\documentclass[a4paper,11pt]{article}
\pdfoutput=1 

\usepackage{jheppub} 

\usepackage{epstopdf} 
\usepackage[T1]{fontenc} 
\usepackage[dvipsnames]{xcolor}

\usepackage{graphics}
\usepackage[utf8]{inputenc}
\usepackage{xspace}
\usepackage{amsmath}
\usepackage{amssymb}
\usepackage{url}
\usepackage{rotating}
\usepackage{enumerate}
\usepackage{graphicx}
\usepackage{pdflscape, afterpage, capt-of} 
\usepackage{slashed} 
\usepackage[normalem]{ulem}
\usepackage{nicefrac}
\usepackage[export]{adjustbox} 
\usepackage{makecell} 
\usepackage{subcaption} 
\usepackage{bbold}
\usepackage{stackengine}

\numberwithin{equation}{subsection}

\usepackage{def} 

\newcommand{\private}[1]{} 

\title{\boldmath One-loop calculations in the chirality-flow formalism}

\author[a]{Andrew Lifson,}
\author[b,c]{Simon Plätzer}
\author[a]{and Malin Sjodahl}

\affiliation[a]{Department of Physics, 
	 Lund University, Box 118, 221 00 Lund, Sweden}
\affiliation[b]{Institute of Physics, NAWI Graz, University of Graz, Universitätsplatz 5, A-8010 Graz, Austria}
\affiliation[c]{Particle Physics, Faculty of Physics, University of Vienna, Boltzmanngasse 5, A-1090 Wien, Austria}

\emailAdd{andrew.lifson@thep.lu.se}
\emailAdd{simon.plaetzer@uni-graz.at}
\emailAdd{malin.sjodahl@thep.lu.se}

\abstract{ In a few recent papers we introduced the chirality-flow
  formalism, which was shown to make calculations
  of tree-level Feynman diagrams simple and transparent.  Chirality flow, 
  which is based on the spinor-helicity formalism,
  allows to often immediately analytically write down a tree-level
  Feynman diagram in terms of spinor inner products.  In this paper,
  we argue that there is also a significant simplification of the Lorentz structure at the one-loop level,
  at least when using the four-dimensional formulation of the
  four-dimensional helicity scheme. Additionally, we find that the
  possible terms in a tensor decomposition of loop integrals are 
  highly constrained, and therefore the tensor reduction procedure is simplified.

}

\begin{document} 
\preprint{LU-TP 23-01, MCNET-23-03}
\maketitle
\flushbottom

\section{Introduction}
\label{sec:introduction}

In a few recent papers we introduced the chirality-flow formalism
\cite{Lifson:2020pai,Alnefjord:2020xqr,Lifson:2022ijv,Platzer:2022nfu}.  This
graphical method builds on the spinor-helicity formalism
\cite{DeCausmaecker:1981jtq, Berends:1981rb, Berends:1981uq,
  DeCausmaecker:1981wzb, Berends:1983ez, Farrar:1983wk, Kleiss:1984dp,
  Berends:1984gf, Gunion:1985bp, Gunion:1985vca, Kleiss:1985yh,
  Hagiwara:1985yu, Kleiss:1986ct, Kleiss:1986qc, Xu:1986xb,
  Dittmaier:1998nn, Schwinn:2005pi, Dixon:1996wi, Elvang:2013cua}, and
relies on splitting the Lorentz algebra into left- and right-chiral
parts.  As for the spinor-helicity formalism, all objects, in
particular polarization vectors, are rewritten in terms of massless
spinors, but the chirality-flow method takes the simplification one
step further by recasting everything into diagrammatic ``flows'', i.e.,
contractions of spinors.

This allows for rewriting Feynman rules and diagrams directly in terms
of these flows and invariants, 
and results in amplitudes that may be directly expressed in terms of
spinor inner products, i.e., the only Lorentz invariant quantities at
hand \cite{Lifson:2020pai,Alnefjord:2020xqr}. Methods which are able
to directly decompose amplitudes in terms of the available group and
spinor structures are of utmost importance to provide reliable input
to resummation approaches at the amplitude level 
\cite{Forshaw:2019ver,DeAngelis:2020rvq,Platzer:2020lbr,Platzer:2022nfu,Platzer:2022jny}.

For tree-level calculations, the chirality-flow formalism leads to significant
simplifications \cite{Lifson:2020pai,Alnefjord:2020xqr}.  In practice,
it is often possible to immediately write down the amplitude
corresponding to a Feynman diagram. On top of this, reference vectors
for external gauge bosons or spin directions can be chosen in such a
way that many Feynman diagrams simplify or vanish.
This --- as well as the
simplified algebra --- led to a significant speed-up for a
tree-level test implementation in \mg{} \cite{Lifson:2022ijv}.

Moving beyond tree-level, it is imperative to use a consistent regularization scheme,
which requires a careful treatment of chirality and $\ga^5$
\cite{tHooft:1972tcz, Breitenlohner:1977hr,Jones:1982zf, Korner:1991sx, Kreimer:1993bh, Larin:1993tq, Trueman:1995ca, Jegerlehner:2000dz,
	Tsai:2009hp, Tsai:2009it, Ferrari:2014jqa, Ferrari:2015mha, Ferrari:2016nea, Viglioni:2016nqc, Gnendiger:2017rfh, Bruque:2018bmy, Belusca-Maito:2022usw}.
A comprehensive summary of different regularization methods can be
found in \cite{Gnendiger:2017pys, Heinrich:2020ybq}.  In conventional
dimensional regularization (CDR) \cite{Bollini:1972ui, tHooft:1972tcz,
  Breitenlohner:1977hr}, all objects are regularized in $d$
dimensions, which (at least naively) would destroy many of the
simplifications brought about by chirality flow.
While in the long term we aim at a full treatment of chirality flow in
CDR (such as to use chirality flow to extract $\epsilon$-dependent
quantities in tree-level calculations), in the present paper we
instead exploit a (partially) four-dimensional regularization scheme,
the 4$d$ formulation (FDF) \cite{Fazio:2014xea} of the 4$d$ helicity
scheme (FDH)\footnote{
		The equivalence of FDH (and therefore FDF) to CDR is shown in \cite{Jack:1993ws, Jack:1994bn}. } 
\cite{Bern:1991aq,Bern:2002tk}, in order to retain the
significant simplifications which chirality flow draws from using charge conjugations,
Fierz and possibly Schouten identities. In FDF, loop numerators are written as purely
four-dimensional objects together with some additional Lorentz
scalars,
while the loop momenta and integrals are in $d$ dimensions.  This
implies that the algebraic manipulations implemented for tree-level
calculations in four dimensions \cite{Lifson:2020pai,
  Alnefjord:2020xqr} can be retained, and Feynman diagrams are
therefore easily ``flowable'' --- as in tree-level
calculations.\footnote{ It is likely also possible to use chirality
flow in other four-dimensional regularization schemes such as
\cite{Battistel:1998sz, BaetaScarpelli:2000zs, BaetaScarpelli:2001ix,
  Catani:2008xa, Bierenbaum:2010cy, Cherchiglia:2010yd, Pittau:2012zd,
  Bierenbaum:2012th, Donati:2013iya, Donati:2013voa, Pittau:2014tva,
  Hernandez-Pinto:2015ysa, Page:2015zca, Sborlini:2016gbr,
  Sborlini:2016hat, Page:2018ljf, TorresBobadilla:2020ekr,
  Pittau:2021jbs}, but we do not explore this here.}  After using
chirality flow to simplify the Lorentz algebra, the integrals are in
$d$ dimensions, so we are able to use all of the standard properties and
results of dimensionally regularized integrals.

The rest of this paper is organized as follows. In
\secref{sec:chirality flow}, we give a brief introduction to chirality
flow and how it is used in Feynman-diagram-based amplitude
calculations. Then, the 4$d$ formulation of the 4$d$ helicity scheme
is introduced in \secref{sec:FDF}. In \secref{sec:flowing loops}, we
illustrate with a few examples how to do one-loop calculations in
chirality flow, describing how chirality flow simplifies the Lorentz
algebra and tensor reduction. Finally, we conclude in \secref{conclusion}.

\section{Introduction to chirality flow}
\label{sec:chirality flow}

In this section, we give a brief introduction to the chirality-flow formalism,
and exemplify how spinors, propagators and vertices are defined. A complete
list of standard model external wave functions, vertices and propagators can be found
in \cite{Alnefjord:2020xqr}, whereas
all structures needed for this paper are contained either in the main text or in \appref{sec:additional rules}.
For details of conventions, we refer the reader to \cite{Lifson:2020pai}.

The basic building blocks of the chirality-flow formalism are the left-
and right-chiral spinors, which we represent graphically in terms of dotted
and undotted lines respectively \cite{Lifson:2020pai},
\begin{alignat}{2}
 \lanSp{i} &= \raisebox{-0.3\height}{\includegraphics[scale=0.4]{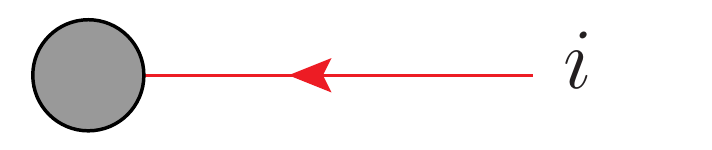}},
 \qquad  
   \sqlSp{i} &&= \raisebox{-0.3\height}{\includegraphics[scale=0.4]{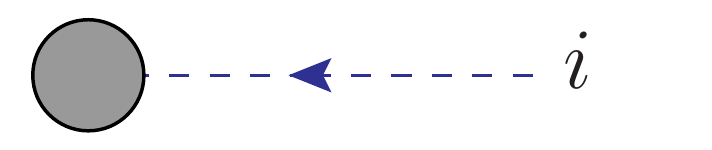}},
 \nonumber \\
 \ranSp{j} &= \raisebox{-0.3\height}{\includegraphics[scale=0.4]{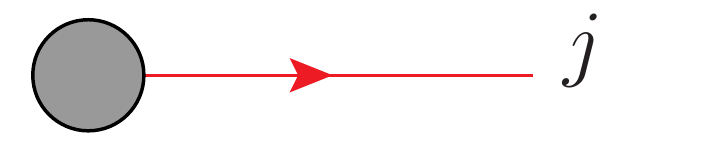}}, 
   \qquad 
   \sqrSp{j} &&= \raisebox{-0.3\height}{\includegraphics[scale=0.4]{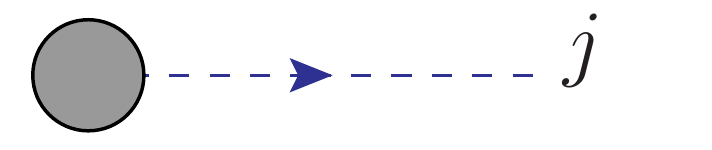}},
\end{alignat}
with $\lanSp{i}=\lanSp{p_i}$ etc.
We represent contractions of these spinors with ``flows'' 
\begin{equation}
  \lan i j \ran 
  \defequal  
   \epsilon^{\al \be}\la_{i,\be}\la_{j,\al}
   = 
   {\includegraphics[scale=0.4]{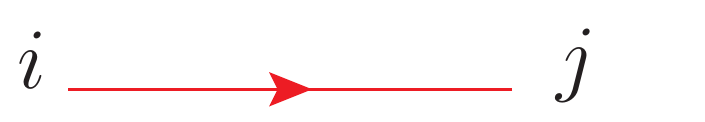}} 
  ~, 
   \;\;\;
  \sql  i j  \sqr 
  \defequal
  \epsilon_{\da \db}\tla_i^{\db}\tla_j^{\da}
  = \raisebox{-0.2\height}{\includegraphics[scale=0.4]{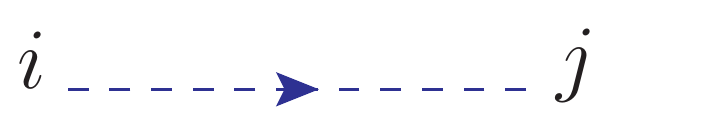}} ~,
    \label{eq:inner products}
\end{equation}
where (up to a phase) $\lan ij \ran \sim \sql ij \sqr \sim \sqrt{2p_i\cdot p_j}$,
and where the antisymmetry of the spinor inner product is obvious.
We read chirality-flow lines along the chirality-flow arrow,
and, because the inner product is antisymmetric, swapping the chirality-flow arrow direction induces a minus sign.

All particles and momenta are then written as (a combination of) massless spinors.\footnote{
	Scalar particles have no Lorentz structure and therefore no flow representation.} 
For example, 
massless polarization vectors are given by
\begin{align}
	\epsilon_{\blue{L}}(p_i,r) &= \frac{1}{\lan r i \ran}\raisebox{-0.2\height}{\includegraphics[scale=0.35]{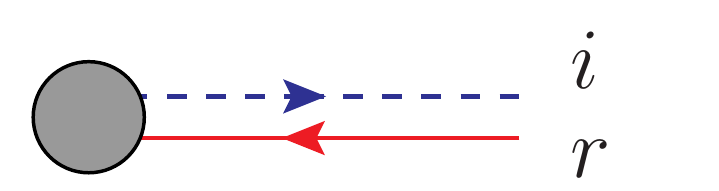}}
	\quad
	\text{or}
	\quad\,\,
	\frac{1}{\lan r i \ran} \raisebox{-0.2\height}{\includegraphics[scale=0.35]{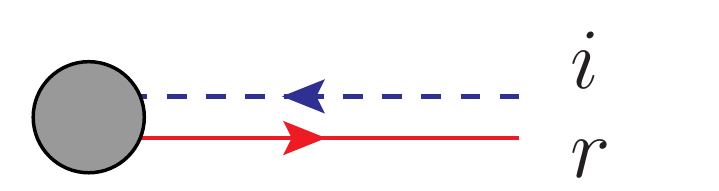}}\hspace*{-2 mm},\nonumber\\
	\epsilon_{\red{R}}(p_i,r) &= \frac{1}{\sql i r \sqr}\raisebox{-0.2\height}{\includegraphics[scale=0.35]{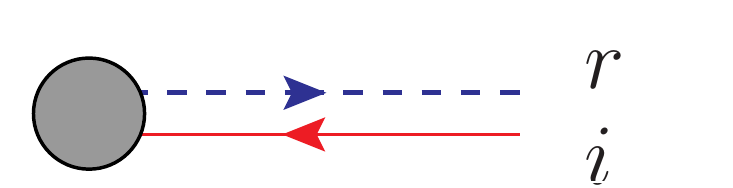}}
	\quad\,\,
	\text{or}
	\quad\,\,
	\frac{1}{\sql i r \sqr} \raisebox{-0.2\height}{\includegraphics[scale=0.35]{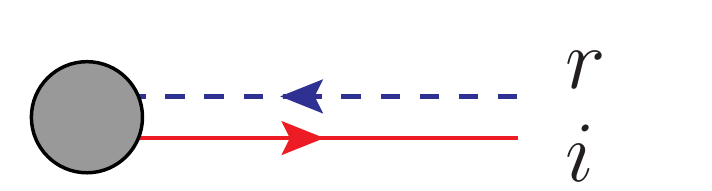}}\hspace*{-2 mm},
	\label{eq:pol vecs massless}
\end{align}
where, similar to chiral spinors, we label the polarization vectors $\blue{L}/\red{R}$
for convenience,\footnote{ 
	The L/R label of a photon is given by the chirality of the spinor in its numerator containing its momentum.  }  
with $\epsilon_{\blue{L}}(p_i,r)$ denoting a negative-helicity incoming or positive-helicity
outgoing photon of momentum $p_i$, and $\epsilon_{\red{R}}(p_i,r)$
denoting a positive-helicity incoming or negative-helicity outgoing photon. 
Here, $r$ is an arbitrary (massless) reference vector with $r\cdot p_i \neq 0$, 
and either set of opposing arrow directions may be used as
long as it matches the rest of the diagram \cite{Lifson:2020pai}.

To describe massive particles and momenta,
a massive momentum $p$ with $p^2=m^2\neq 0$ 
is decomposed into a sum of massless momenta $p^{\flat}$ and $q$ as
\begin{equation}
p^{\mu} = p^{\flat,\mu}+\alpha q^\mu~, \qquad (p^\flat)^2=q^2=0, \qquad p^2=m^2~,
\qquad \alpha = \frac{m^2}{2p^\flat\cdot q} = \frac{m^2}{2p\cdot q}~,
\label{eq:massive momentum decomposition}
\end{equation}
with, for example, an incoming spinor with spin along the axis 
$s^\mu = (p^\mu - 2\alpha q^\mu)/m$ given by\footnote{
	We use the chiral basis for the Dirac $\ga$-matrices. } 
\cite{Alnefjord:2020xqr}
\begin{equation}
u^{+}(p)
    =\begin{pmatrix}
    \frac{m}{\sql p^\flat q \sqr }
    \raisebox{-0.2\height}{\includegraphics[scale=0.40]{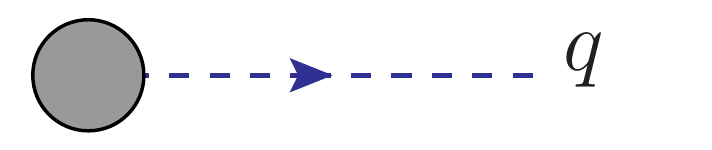}}  \\
    \phantom{\frac{m}{\sql p^\flat q \sqr}}
    \raisebox{-0.2\height}{\includegraphics[scale=0.40]{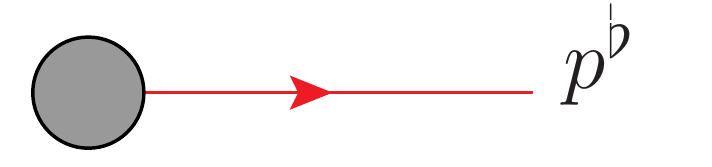}}
    \end{pmatrix}\;.
\end{equation}

While it is natural in chirality flow to measure spin along any direction,
we can of course also measure it along the direction of motion,
in which case we choose
$\alpha=1$,  $q^\mu=\eigvecm{p}$ and $p^{\flat,\mu}=\eigvecp{p}$,
with $\vec{p}_{\plus}$ pointing in the same direction as $\vec{p}$
and $\vec{p}_{\minus}$ pointing in the opposite direction, i.e.,
\begin{align}
\eigvecp{p} &= 
\frac{p^0 +|\vec{p}|}{2} \big(1,\phat\big)~,
&
\eigvecm{p} &= 
\frac{p^0 -|\vec{p}|}{2} \big(1,-\phat\big)~, 
\end{align}
such that the three-vector of the spin $s^\mu = \frac{1}{m}(p_f^{\mu} - p_b^{\mu}) =\frac{1}{m}(|\vp|, p^0 \hat{p})$, is directed along the motion, giving the helicity basis,
\begin{align}
u^+(p)
&=
\begin{pmatrix}
\frac{m}{\sql p_f p_b \sqr}
\raisebox{-5.5pt}{\includegraphics[scale=0.375]{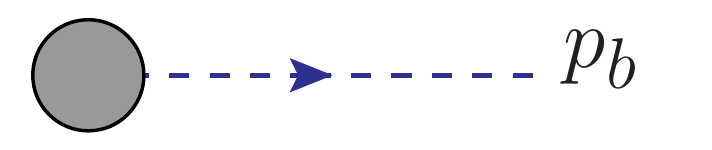}}
\hspace{-1.75ex}
\\
\phantom{\frac{m}{\sql p_f p_b \sqr}}
\raisebox{-5.5pt}{\includegraphics[scale=0.375]{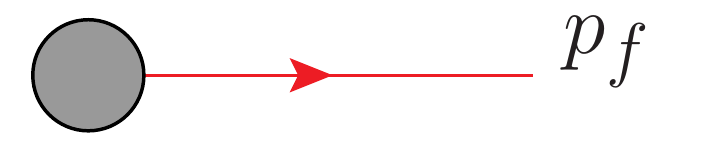}}
\hspace{-1.75ex}
\end{pmatrix}
\stackrel{\hbox{$m\rightarrow 0$}}{\hbox{$\longrightarrow$}}
\begin{pmatrix}
0
\\
\raisebox{-5.5pt}{\includegraphics[scale=0.375]{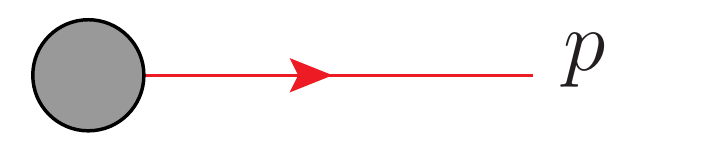}}
\end{pmatrix}
\;. 
\end{align}

Vertices and propagators are also naturally described using chirality flow.
For example, the QED vertex can be translated to
\begin{align}
	\includegraphics[scale=0.45,valign=c]{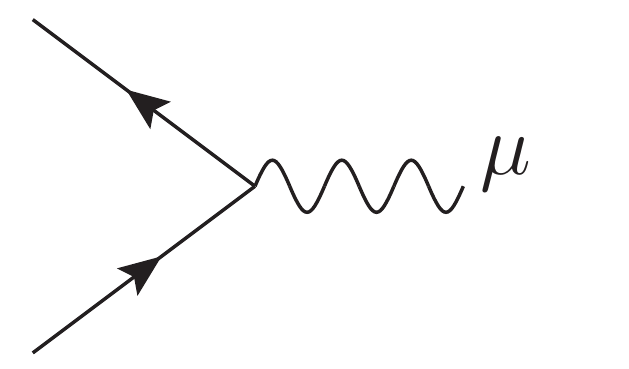}\!\!\!\!\! =
	ie Q_f \gamma^\mu
	&=
	ieQ_f \begin{pmatrix}
		0&   \sqrt{2}\tau^{\mu,\da\be}
		\\
		\sqrt{2}\taubar^{\mu}_{\al\db} & 0
	\end{pmatrix}
	\nonumber\\
    &=
	ieQ_f \sqrt{2}\begin{pmatrix}
		0&   \raisebox{-0.35\height}{\includegraphics[scale=0.35]{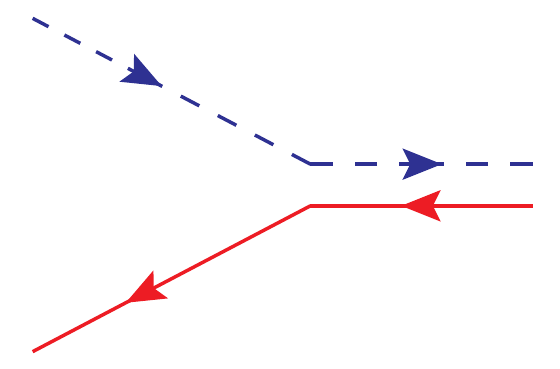}}
		\\
		\raisebox{-0.4\height}{\includegraphics[scale=0.385]{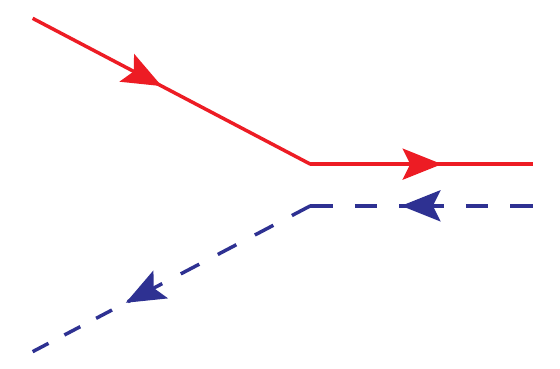}} & 0
	\end{pmatrix}~,
	\label{eq:fermion_photon_vertex}  
\end{align}
where we introduce our differently-normalized Pauli matrices $\tau^\mu \equiv \si^\mu/\sqrt{2}$ 
and $\taubar^\mu \equiv \sibar^\mu/\sqrt{2}$,
and our use of the chiral basis is made explicit.
The propagator for a massless gauge boson in the Feynman gauge contains the chirality-flow rule for the metric,
which is a double line with arrows opposing,
\begin{align}
	\raisebox{-0.25\height}{\includegraphics[scale=0.4]{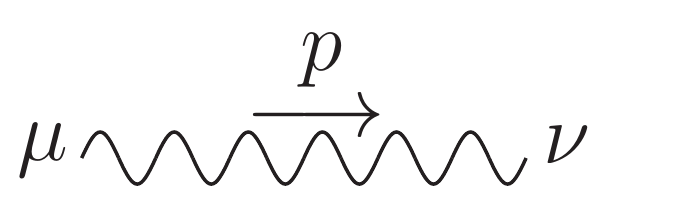}}
	&=  \quad  -i\frac{g_{\mu\nu}}{p^2}
	&
	&\rightarrow
	&
	&-\frac{i}{p^2}\raisebox{-0.25\height}{\includegraphics[scale=0.45]{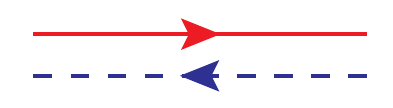}}
	&
	&\mbox{or}
	&
	&-\frac{i}{p^2}\raisebox{-0.25\height}{\includegraphics[scale=0.45]{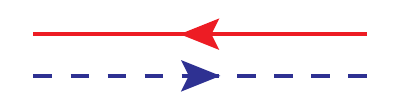}}\;,
	\label{eq:vbpropagator}
\end{align}
while the fermion propagator in the flow picture is
\begin{align}
   \raisebox{-0.1\height}{\includegraphics[scale=0.4]{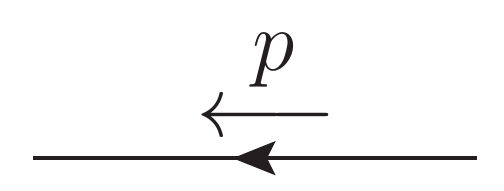}}
   &= \
\frac{i}{p^2-m^2} 
  \begin{pmatrix} 
    m {\delta^{\da}}_{\db} & \sqrt{2}p^{\da \be} \\ 
    \sqrt{2}\bar{p}_{\al \db} &  m {\delta_{\al}}^{\be}
  \end{pmatrix}
= \frac{i}{p^2-m^2} 
  \begin{pmatrix} 
    m\includegraphics[scale=0.34]{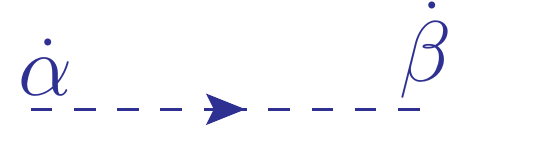}
    & \includegraphics[scale=0.34]{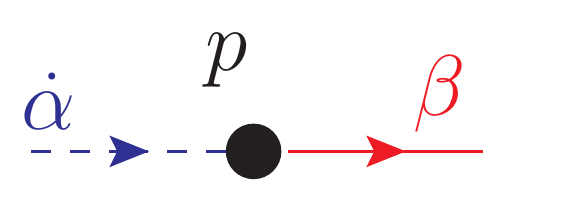} \\ 
    \includegraphics[scale=0.34]{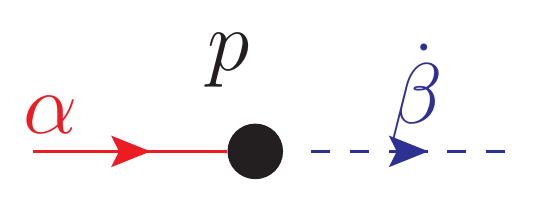}
    & m\includegraphics[scale=0.34]{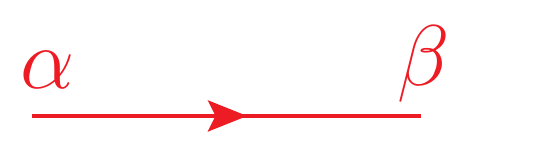}
  \end{pmatrix} \;,
\end{align}
where we have introduced a graphical ``momentum-dot'' notation for momenta slashed
with $\sigma$ or $\sigmabar$.
Note that any momentum in the Feynman rules will be translated to this momentum-dot in the chirality-flow rules using
\begin{align} 
	p^{\mu} &\rightarrow \frac{1}{\sqrt{2}}\raisebox{-0.15\height}{\includegraphics[scale=0.4]{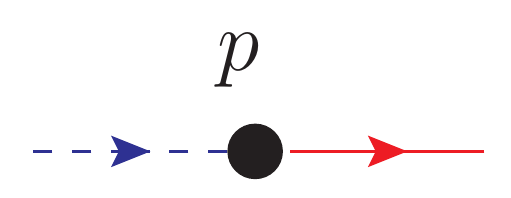}}~,
	& 
	&\text{or} 
	&
	p^{\mu} &\rightarrow \frac{1}{\sqrt{2}}\raisebox{-0.15\height}{\includegraphics[scale=0.4]{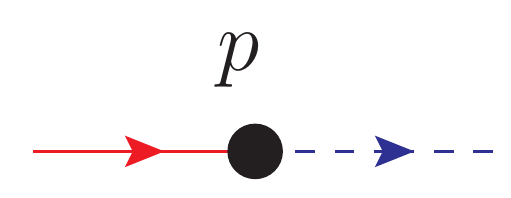}}~,
	\label{eq:mom flow}
\end{align}
where, like with the polarization vectors in \eqref{eq:pol vecs massless},
either arrow direction in \eqsrefa{eq:vbpropagator}{eq:mom flow}
is allowed as long as it matches the rest of the diagram \cite{Lifson:2020pai}.

Using these rules, it is easy to immediately write down the values of Feynman diagrams,
for example, in massless QED,
\begin{align}
  &\raisebox{-0.4\height}{\includegraphics[scale=0.5]{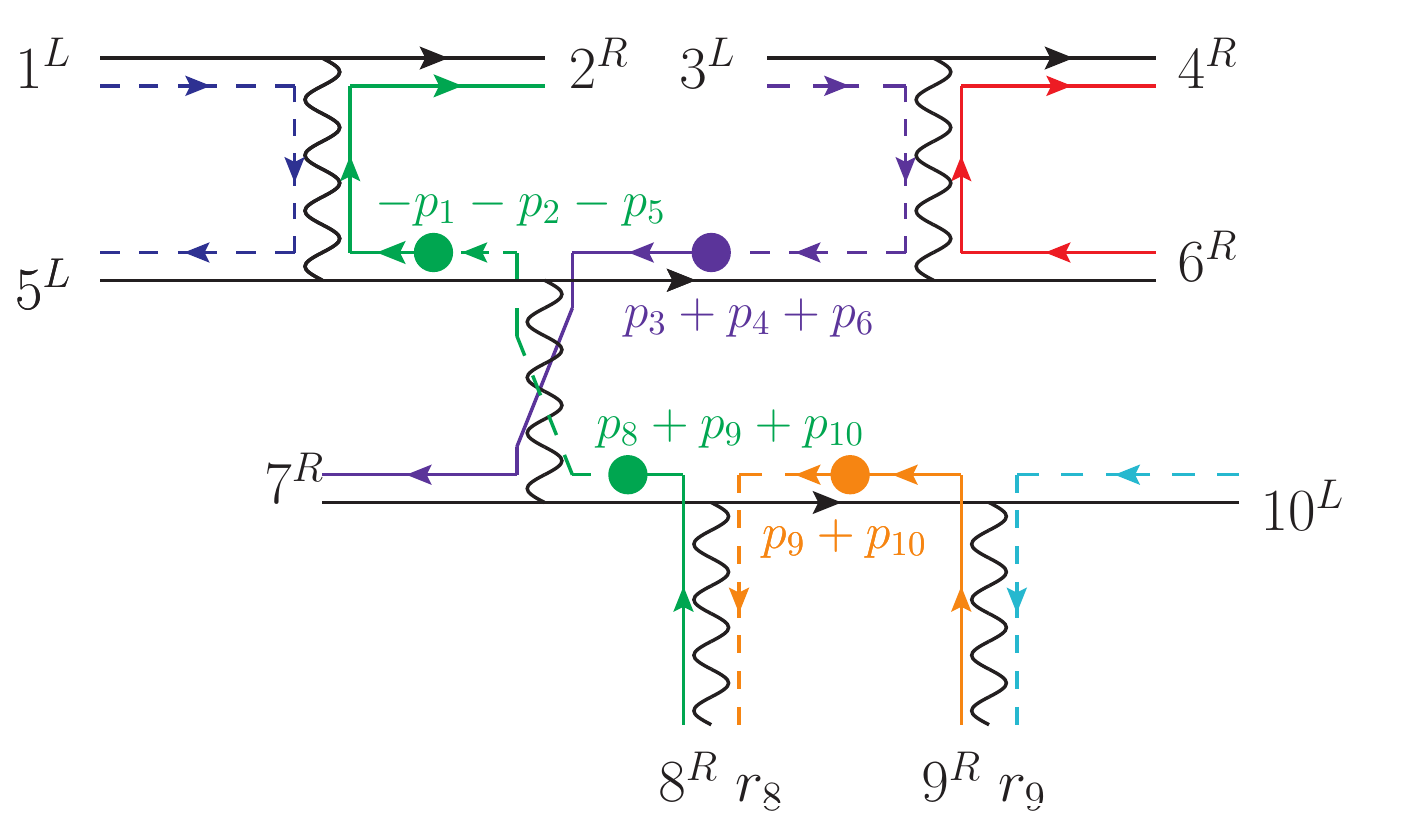}}
  \begin{tabular}{ c c}
    $=$ & $\underbrace{(\sqrt{2} e i)^8}_{\text{vertices}}\underbrace{\frac{(-i)^3}{s_{1\,2}\;s_{3\,4}\;s_{7\,8\,9\,10}}}_{\text{photon propagators}}$\\
    $\times$ & $\underbrace{\frac{(i)^4}{s_{1\,2\,5}\;s_{3\,4\,6}\;s_{8\,9\,10}\;s_{9\,10}}}_{\text{fermion propagators}}$\\
    $\times$ & $\underbrace{\frac{1}{[8\, r_8 ] [ 9 \, r_9 ] }}_{\text{polarization vectors}}$
  \end{tabular}
  \nonumber\\
  &
  \times
  {\textcolor{blue}{[ 1 5 ]}}
  {\textcolor{red}{  \langle 6 4 \rangle} }
  {\textcolor{sky}{ [ 10\,\, r_9]} }   
  {\textcolor{orange} {\Bigg(
    \underbrace{\langle 9 9\rangle}_{0} [9r_8]
    + \langle 9 \,10\rangle [10\,r_8]
    \Bigg)}} 
  {\textcolor{lilac}{\Bigg(
    \underbrace{[ 3 3]}_{0} \langle 3 7 \rangle
    + [3 4 ]\langle 4 7 \rangle
    + [ 3 6 ] \langle  6 7 \rangle
    \Bigg)}}
  \nonumber\\
  & \times
    {\textcolor{jaxoGreen}{\Big(
      -\langle 8 9\rangle [9 1]\langle 1 2\rangle
      -\langle 8 9\rangle [9 5]\langle 5 2\rangle
      -\langle 8 \, 10\rangle [10\,\,1]\langle 1 2\rangle
      -\langle 8 \, 10\rangle [10\,\, 5]\langle 5 2\rangle
      \Big)}}\;,
\end{align}
where we have superimposed a chirality-flow diagram onto a ten-point Feynman diagram. 
For massless QED and QCD, 
any chirality-flow arrow direction which has opposing arrows for bosons and which flows through momentum-dots is equivalent \cite{Lifson:2020pai}.
If masses or scalar particles are involved, then some care is needed to set consistent arrow directions \cite{Alnefjord:2020xqr}.

Finally, we also note that diagrams can be made to vanish by appropriately choosing the reference
vectors of our polarization vectors and massive spinors.  For example, the above diagram
vanishes if we pick $r_9=10$.

In this paper, we will argue that much of the simplification from
chirality flow can be retained at one-loop level. For example, we
may simplify the algebra of the electron self-energy in Feynman gauge as
\begin{equation}
  \raisebox{-0.4\height}{\includegraphics[scale=0.3]{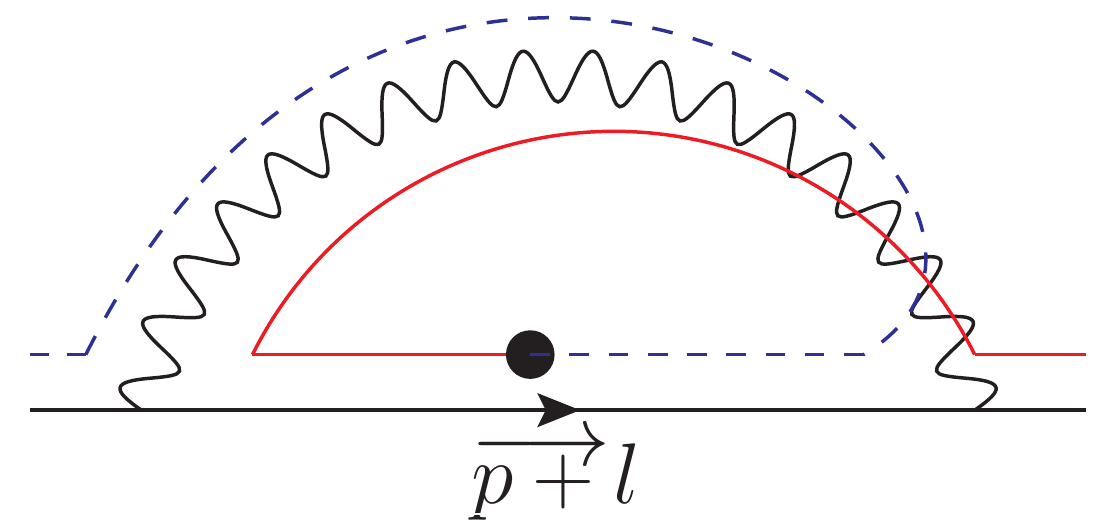}}
  \quad\sim\quad
  \raisebox{-0.4\height}{\includegraphics[scale=0.3]{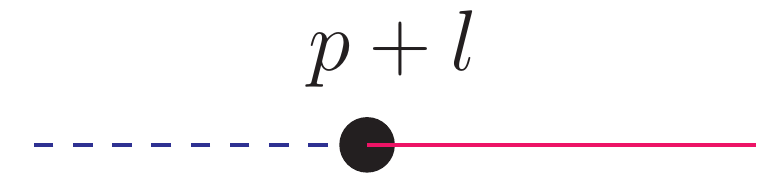}}\;,
  \label{eq:ferm self energy intro}
\end{equation}
where we defer a discussion on the arrow direction until 
\secref{sec:flowing loops}. The diagrammatic identity above is of
course simple, but we will demonstrate that similar diagrammatic 
decompositions can be used together with certain methods of decomposing tensor integrals, 
and thereby significantly simplify calculations.

\section{FDF}
\label{sec:FDF}

In this section we summarize the FDF formalism. 
This formalism was created in 2014 with the aim of providing a truly 4d numerator algebra for loop integrals.
It has also been used in the context of color-kinematics duality \cite{Mastrolia:2015maa, Primo:2016omk}, 
while some example calculations in FDF can be found in \cite{Gnendiger:2017pys, Gnendiger:2017rfh}.
Since the chirality-flow formalism makes use of explicitly 4d relations,
the FDF method is easily converted into a flow formalism
(though we expect that a similar flow formalism holds in CDR).

In FDF we have five vector spaces, 
the standard Minkowski space $\Sfour$ with four integer dimensions,
and the infinite-dimensional spaces $\QSd$, 
$\QSds$, 
$\QStwoe$,
$\QSne$,
and $\QSnetwoe$,
which satisfy \cite{Fazio:2014xea}
\begin{align}
	\Sfour &\subset \QSd \subset \QSds ~,
	&
	\QSd  &= \Sfour \oplus \QStwoe ~,
	\nonumber \\
	\QSnetwoe &= \QStwoe \oplus \QSne ~, 
	&
	\QSds &= \QSd \oplus \QSne~,
	\label{eq:dimensional subspaces}
\end{align}
where $\QSd$ is the vector space of CDR, $\QSds$ with $\ds = 4$ is the vector space of FDH and FDF,
and we follow the notation of \cite{Gnendiger:2017pys,Anger:2018ove}.
Since the subspaces in \eqref{eq:dimensional subspaces} are orthogonal,
we can split up vectors and tensors as e.g.
\begin{align}
	\gads &= \gafour + \ganepstwoeps~,
	&
	\metds &= \metfour + \metnepstwoeps~,
\end{align}
where the subscript in square brackets gives the vector space of the object.
The metric is used to project onto the different subspaces, for example,
\begin{align}
	g^{\mu\nu}_{\dimens{\ds}}g_{\nu\rho\dimens{4}} &= \delta^{\mu}_{\rho\dimens{4}}~,
	&
	g^{\mu\nu}_{\dimens{\ds}}g_{\nu\rho\dimens{d}} &= \delta^{\mu}_{\rho\dimens{d}}~,
	\label{eq:metric projectors FDF}
\end{align}
while the trace of the metric is given by the dimension of its (sub)space, 
and the Dirac gamma matrices have their usual anticommutation relations
\begin{align}
	(g_{\dimens{dim}})^{\mu}_{~\mu} &= dim~,
	&
	\left\lbrace \ga^\mu_{\dimens{dim}},\ga^\nu_{\dimens{dim}}\right\rbrace &= 2g^{\mu\nu}_{\dimens{dim}}~.
	\label{eq:metric trace and gamma algebra}
\end{align}

The main starting point of FDF is to re-write $\QSds$ as
\begin{align}
	\QSds  = \Sfour \oplus \QStwoe \oplus \QSne = \Sfour \oplus \QSnetwoe~,
\end{align}
that is, as the purely 4d Minkowski space plus an extra space $\QSnetwoe$ 
which can be represented by 4d objects multiplying a new algebra called the 
$-2\eps$ selection rules (\mteSRs{}).
The \mteSRs{} are defined using the following replacements 
\begin{align}
	\metnepstwoeps &\rightarrow G^{MN}~,
	&
	\ltwoeps^\mu &\rightarrow i\mu Q^M~,
	&
	\ganepstwoeps &\rightarrow \ga^5\Ga^M~,
	\label{eq:2eps selection rules}
\end{align}
together with the following algebra
\begin{align}
	G^{MN} G^{NP} &= G^{MP}~,
	&
	G^{MM} &= 0~,
	&
	G^{MN} &= G^{NM}~,
	\nonumber \\
	\Ga^M G^{MN} &= \Ga^{M}~,
	&
	\Ga^{M} \Ga^M &= 0~,
	&
	Q^M \Ga^{M} &= 1~,
	\nonumber \\
	Q^M G^{MN} &= Q^{M}~,
	&
	Q^M Q^M &= 1~,
	&
	\left\lbrace \Ga^M,\Ga^N \right\rbrace &= 2G^{MN}~.
	\label{eq:2eps selection rules algebra}
\end{align}
In any given Feynman diagram the \mteSRs{} can be precalculated, 
and contribute an overall multiplicative factor of $0$ or $\pm 1$ to the diagram. 

In the Feynman rules of FDF, the \mteSRs{} 
imply that fermion propagators within loops are rewritten to contain only 4d terms and a new mass\footnote{
	This mass should not be confused with the `t Hooft mass,
	but signifies the extra-dimensional part of the loop momentum, 
	see \eqref{eq:d-dim mom mu^2}.} 
$\mu$ \cite{Fazio:2014xea}
\begin{align}
	 \raisebox{-0.1\height}{\includegraphics[scale=0.4]{./Jaxodraw/FermionProp}} \sim \
	\slashed{l}_{\dimens{\ds}} + m
	&= \slashed{l}_{\dimens{4}} + i\mu\ga^5 + m
	\nonumber \\
	&=   \begin{pmatrix} 
		(m-i\mu) \includegraphics[scale=0.34]{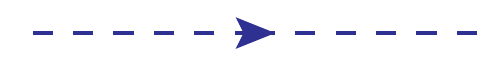} \;\;\;\;\;\; 
		& \phantom{(m-i\mu)}\includegraphics[scale=0.34]{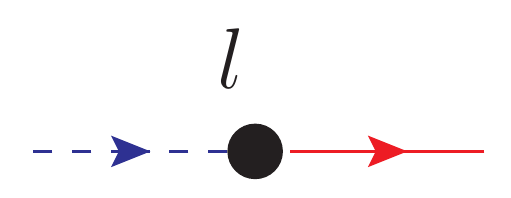} \\ 
		\phantom{(m+i\mu)}\includegraphics[scale=0.34]{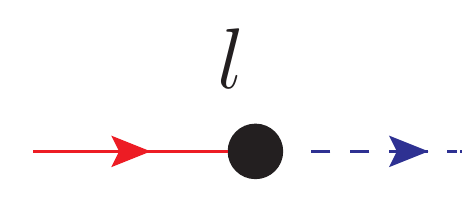}\;\;\;\;\;\; 
		& (m+i\mu)\includegraphics[scale=0.34]{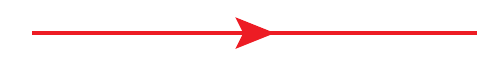}
	\end{pmatrix} ~,
\label{eq:fermion prop FDF}
\end{align}
and that vector boson propagators are split into two pieces,
with numerator structure
\begin{align}
		\raisebox{-0.25\height}{\includegraphics[scale=0.4]{Jaxodraw/PhotonProp}}
	\!\!\sim \
	-i\metds = -i\metfour  -iG^{MN}
	= 	-i\raisebox{-0.25\height}{\includegraphics[scale=0.45]{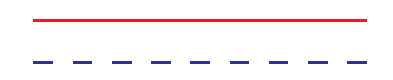}}
	-iG^{MN}
	~,
\end{align}
in Feynman gauge.
Here, the first term is the usual 4d propagator with suppressed chirality-flow arrows 
(recall that these arrows should be opposing, see \secref{sec:chirality flow}), 
and the second is called an FDF scalar and has trivial Lorentz structure.\footnote{
	Note that although FDF is designed to reproduce FDH results, 
	the FDF scalar is fundamentally different from the $\eps$-scalar which appears in FDH \cite{Gnendiger:2017pys}.}

To build the correct Feynman diagrams with FDF, we must use the FDF
Feynman rules, which can be found in \cite{Fazio:2014xea}.  These
rules include new terms not present at tree level in chirality flow.
However, these new terms are all either Lorentz scalars and therefore
have no flow representation, or contain additional momenta which have
the chirality-flow representation in \eqref{eq:mom flow},
the metric which has the chirality-flow representation in \eqref{eq:vbpropagator},
or $\ga^5$ which only affects the overall sign of a chirality-flow graph.  Therefore,
we do not show them explicitly here.

All loop integrals $I^d_{i_1\dots i_k}[N^{\mu_1\dots\mu_n}]$ in FDF, 
and therefore all loop momenta $\ld^\mu$, are in $d$ dimensions.
The loop integral is defined in the usual way
\begin{align}
	I^d_{i_1\dots i_k}[N^{\mu_1\dots\mu_n}] \equiv \lint \frac{N^{\mu_1\dots\mu_n}}{D_{i_1}\dots D_{i_k}}~,
\end{align}
for $N^{\mu_1\dots\mu_n}$ some numerator which may be a scalar, vector, or general tensor in Lorentz space,
and $D_i$ a propagator momentum of the usual form $p^2_{\dimens{d}}-m^2$.
When squaring a loop momentum we use the \mteSRs{}, 
\eqsrefa{eq:2eps selection rules}{eq:2eps selection rules algebra},
to obtain
\begin{align}
	\ld^\mu &= \lfour^\mu + \ltwoeps^\mu~,
	&
	\ld^2 &= \lfour^2 + \ltwoeps^2 = \lfour^2 -\mu^2 ~,
	\label{eq:d-dim mom mu^2}
\end{align}
where we identify the space-like mass $ \ltwoeps^2 = -\mu^2$.
A defining feature of FDF is that only even powers of $\mu$ 
are allowed to contribute to the amplitude.
Integrals involving $\mu^2$ are reduced to integrals without $\mu^2$ using \cite{Bern:1995db}
\begin{align}
	I^d_{i_1\dots i_k}[(\mu^2)^r] = (2\pi)^r I^{d+2r}_{i_1\dots i_k}[1]\prod_{j=0}^{r-1}(d-4-2j)~.
	\label{eq:mu integral relation}
\end{align}
The above identity is essentially related to dimensional shift
relations originating from powers of the loop momentum in the
numerator, as {\it e.g.} also employed in the decomposition of tensor
integrals \cite{Davydychev:1991va}.

From the above, we see that all of the Lorentz
algebra is done in four dimensions and is easily ``flowable'', while all
loop integrals are conveniently performed in d-dimensions.

\section{Flowing loops}
\label{sec:flowing loops}

\begin{figure}
	\centering
	\includegraphics[scale=0.35]{Jaxodraw/QEDAxialAnomalyFeyn1}
	\hspace{1cm}
	\includegraphics[scale=0.35]{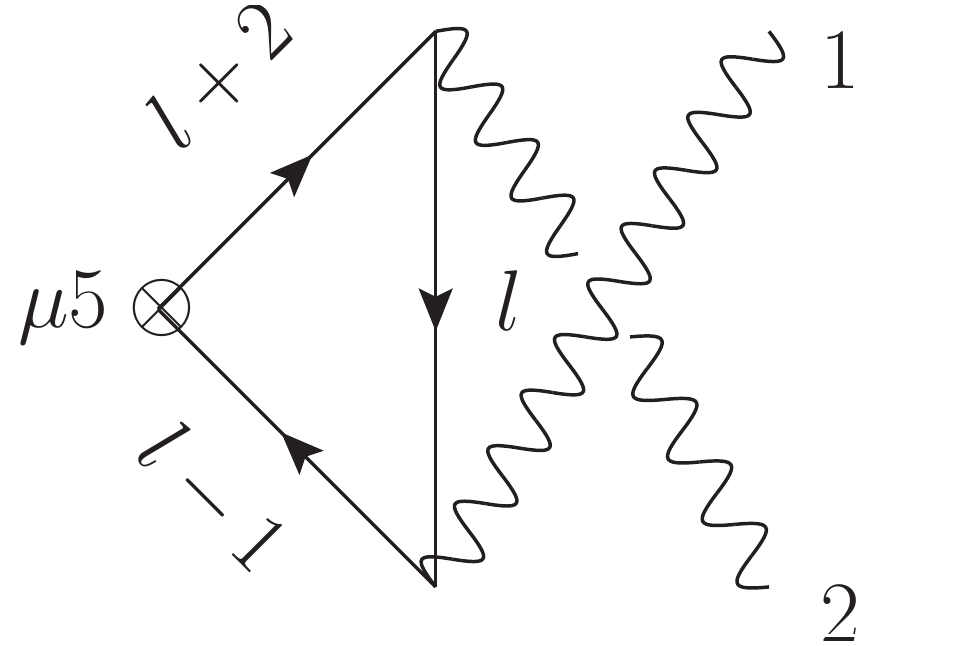}
	\caption{The two Feynman diagrams for the axial anomaly. 
		The inserted axial current operator is labeled $\mu 5$ $\otimes$.}
	\label{fig:QED axial anomaly}
\end{figure}

In this section we show how to turn the FDF formalism into a chirality-flow formalism.
To understand how this works, it is useful to first go through an example,
for which we choose the axial anomaly of massless QED.
To calculate the anomaly, we consider the axial current $j^{\mu 5} = \bar{\psi}\ga^\mu\ga^5\psi$,
and the divergence $\partial_\mu j^{\mu 5}$ of its matrix element to create two photons.
The matrix element is calculated using the two diagrams in \figref{fig:QED axial anomaly},
and is well-known (see e.g.\ \cite{Adler:1969gk, Bell:1969ts, Adler:1969er, Peskin:1995ev, Gnendiger:2017rfh}) to equal
\begin{align}
		i(p_1+ p_2)_\mu \includegraphics[scale=0.3,valign=c]{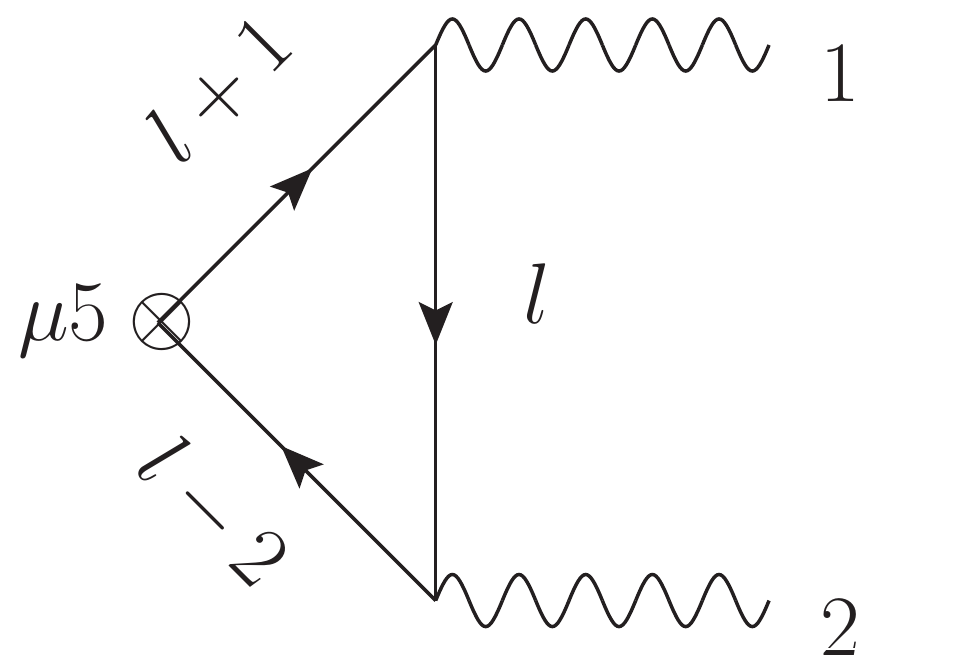} \hspace{-0.4cm}= 
	\,\,i(p_1+ p_2)_\mu  \includegraphics[scale=0.3,valign=c]{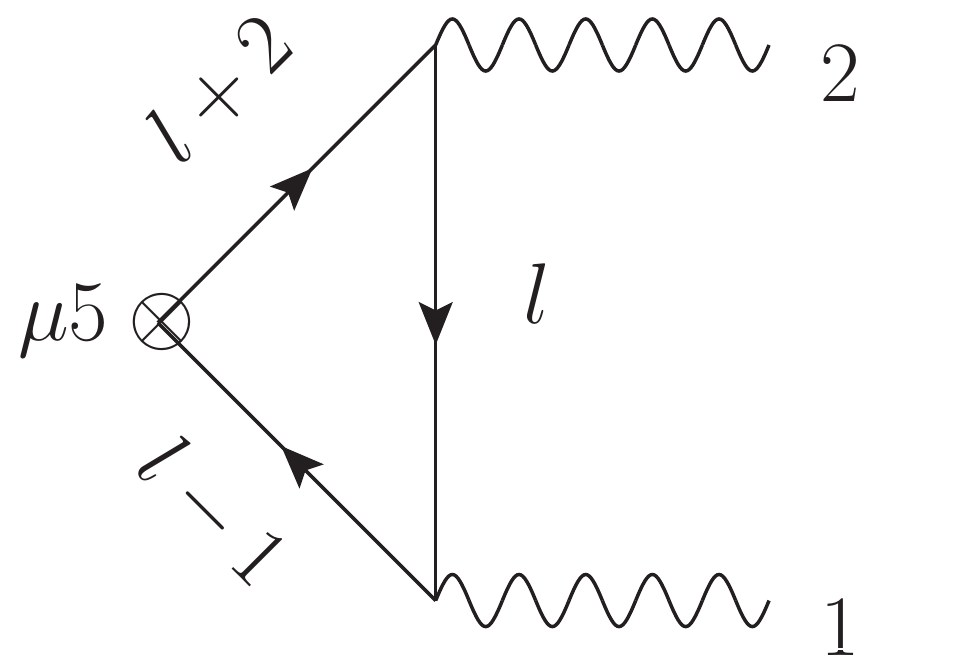}\hspace{-0.4cm}
	= \,\,\frac{ie^2}{(4\pi)^2}\Tr\Big[\ga^5 \slashed{1}\slashed{\eps}^*_1 \slashed{2}\slashed{\eps}^*_2\Big]~,
	\label{eq:axial anomaly Peskin}
\end{align}
where $1,2 \equiv p_1,p_2$ are the (outgoing) momenta of the two photons with polarization vectors $\eps_1^*,\eps_2^*$.

To obtain a comparison to chirality flow, we choose a set of polarizations, 
then use \eqsrefa{eq:mom outer products}{eq:pol vecs massless outer prods} 
to express the polarization vectors and momenta in terms of spinors, obtaining
\begin{align}
	\frac{ie^2}{(4\pi)^2}\Tr\Big[\ga^5 \slashed{1}\slashed{\eps}_{1,L} \slashed{2}\slashed{\eps}_{2_R}\Big] &= 
	\frac{ie^2}{(4\pi)^2}\frac{2}{\cAngle{r_1 1}\cSquare{2 r_2}} 
	\Big[ -\cAngle{1 r_1} \cSquare{12} \underbrace{\cAngle{22}}_{0} \cSquare{r_2 1} 
	+ \underbrace{\cSquare{11}}_{0} \cAngle{r_1 2} \cSquare{2 r_2} \cAngle{21} \Big]\nonumber \\
	&= 0~, \nonumber \\
	\frac{ie^2}{(4\pi)^2}\Tr\Big[\ga^5 \slashed{1}\slashed{\eps}_{1,L} \slashed{2}\slashed{\eps}_{2_L}\Big] &= 
	\frac{ie^2}{(4\pi)^2}\frac{2}{\cAngle{r_1 1}\cAngle{r_2 2}} 
	\big[ -\cAngle{1 r_1} \cSquare{12} \cAngle{2 r_2} \cSquare{2 1} 
	+ \underbrace{\cSquare{11}}_{0} \cAngle{r_1 2} \underbrace{\cSquare{2 2}}_{0} \cAngle{r_2 1} \big] \nonumber \\
	&= \frac{-2ie^2}{(4\pi)^2}\cSquare{12}\cSquare{21} ~,
	\label{eq:axial anomaly as spinor products}
\end{align}
where we used the explicit representation of $\ga^5$ in the chiral basis to separate the trace into two terms,
and the cyclicity of the trace to write everything in terms of spinor inner products.
Notice that the axial anomaly vanishes if the photons have opposite helicity.

We now go through the calculation leading up to \eqsrefa{eq:axial anomaly Peskin}{eq:axial anomaly as spinor products} 
in chirality flow.\footnote{
	To see this calculation in the pure FDF formalism, see \cite{Gnendiger:2017rfh}.} 
It is easiest to set the polarizations and the reference vectors of the photons at the beginning.
We choose $\eps_{1,\blue{L}}$ with $r_1 = 2$ and $\eps_{2,\red{R}}$ with $r_2 = 1$.
Using the QED vertex \eqref{eq:fermion_photon_vertex}, fermion propagator \eqref{eq:fermion prop FDF},
and splitting the trace of $\ga$-matrices into two two-component traces, we have
\begin{align}
	&i(p_1+ p_2)_\mu  \includegraphics[scale=0.3,valign=c]{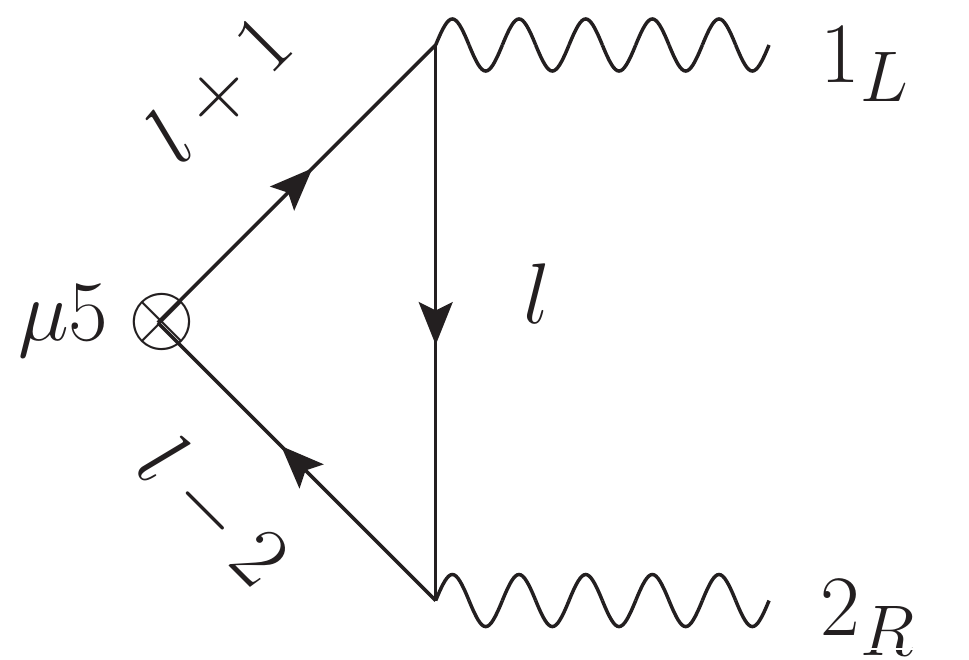} \hspace{-0.4cm}= 
	\frac{2e^2}{\cAngle{21}\cSquare{21}}\lint \left\lbrace 
	\includegraphics[scale=0.27,valign=c]{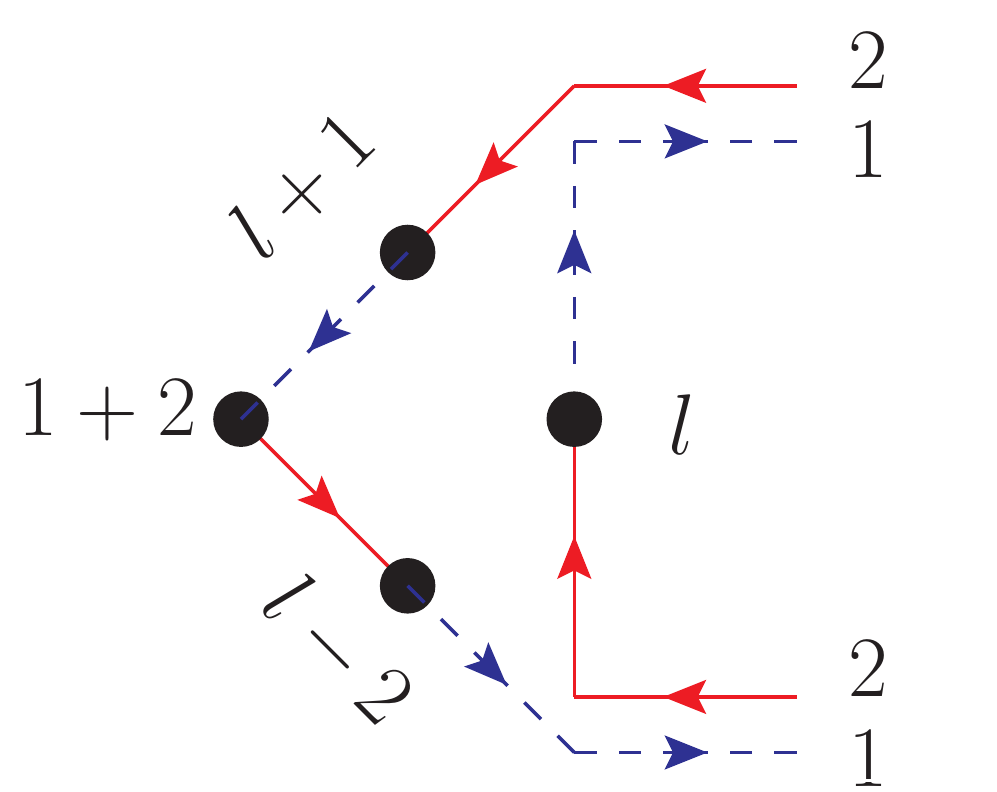}\hspace{-0.2cm}
	-\,\, \includegraphics[scale=0.27,valign=c]{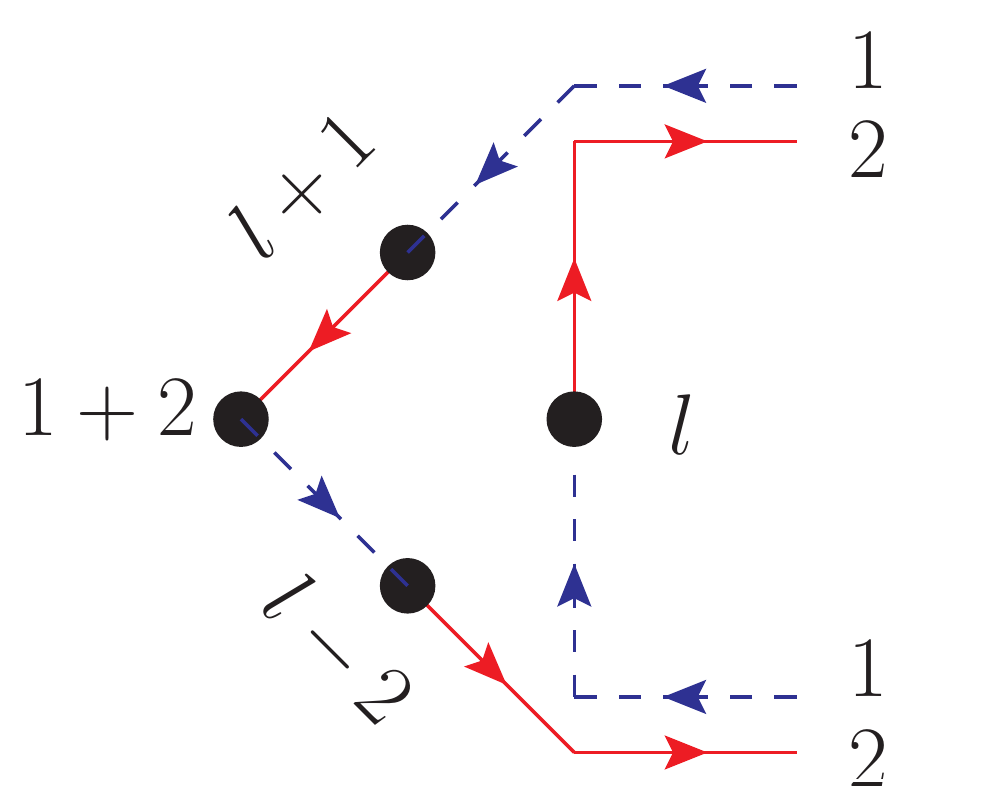} \right.
	\nonumber \\
	&+\mu^2\includegraphics[scale=0.27,valign=c]{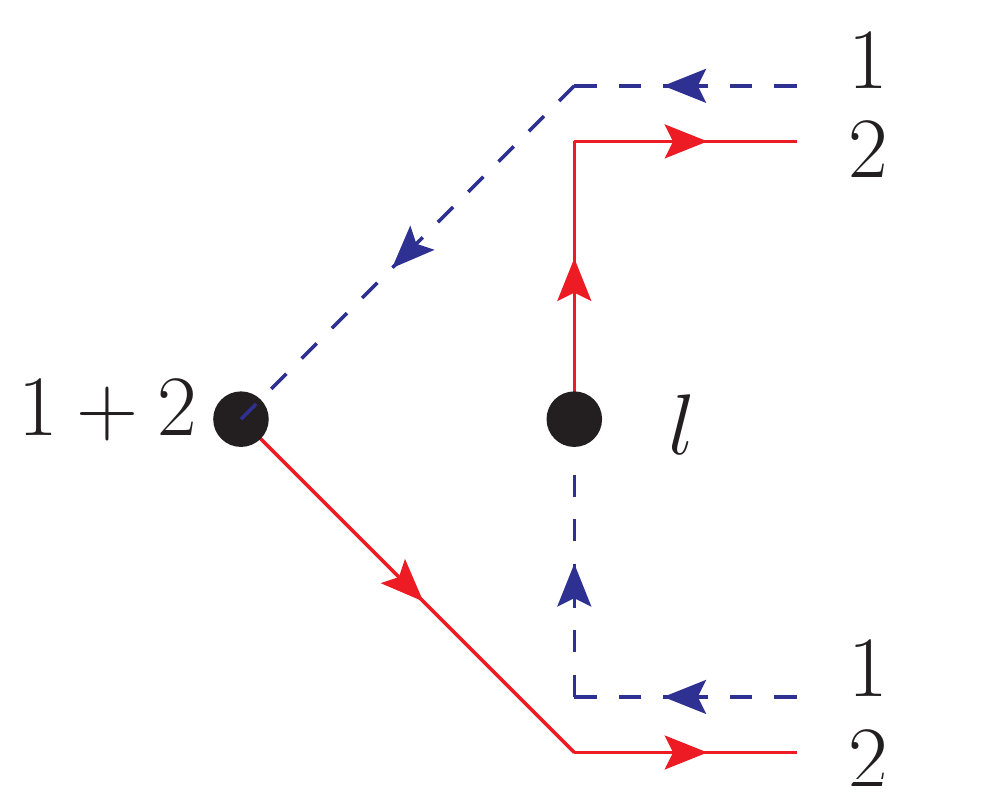}\hspace{-0.2cm}
	-\,\, \mu^2\includegraphics[scale=0.27,valign=c]{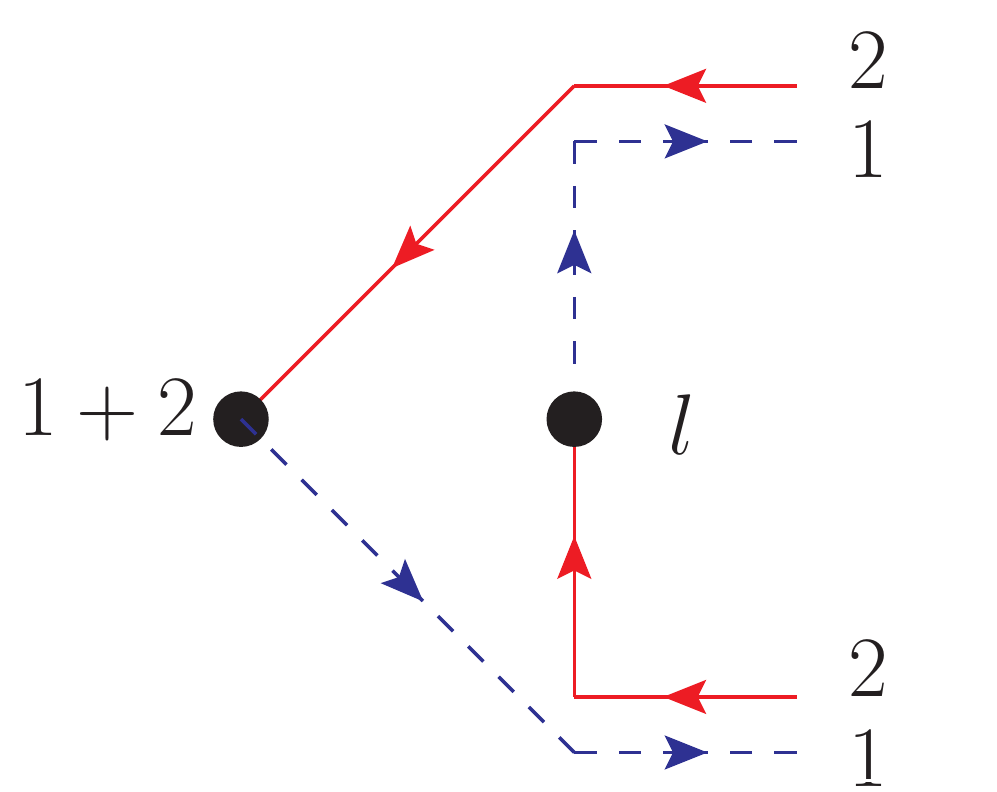} 
		+\mu^2\includegraphics[scale=0.27,valign=c]{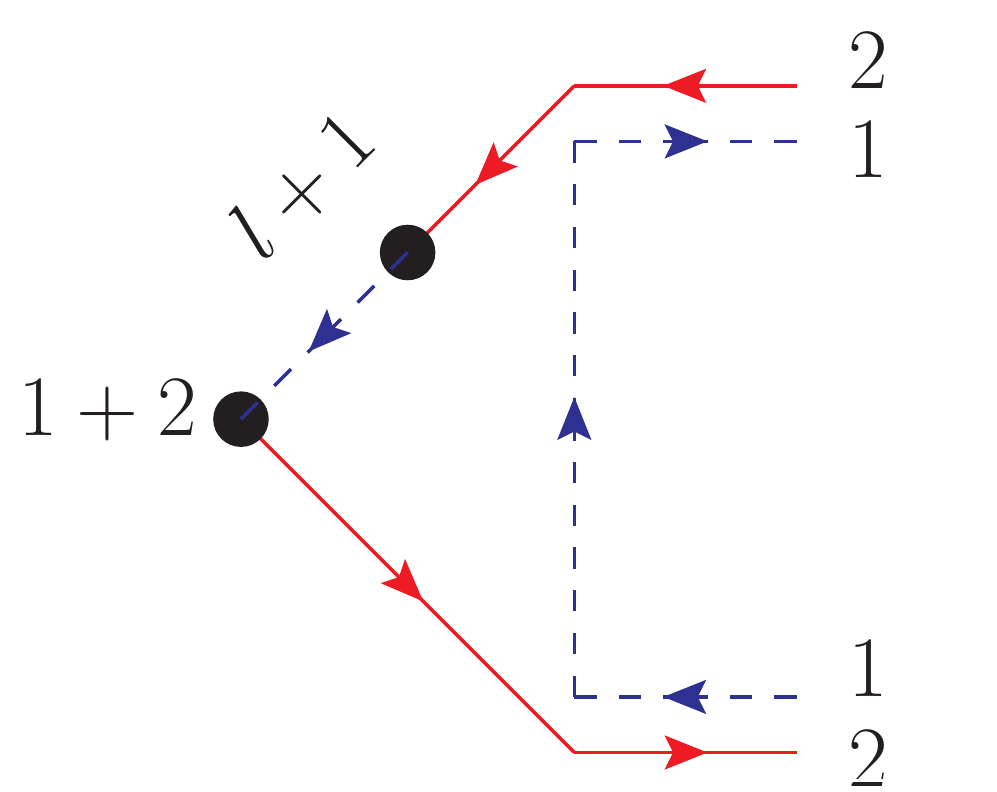}\hspace{-0.2cm}
	-\,\, \mu^2\includegraphics[scale=0.27,valign=c]{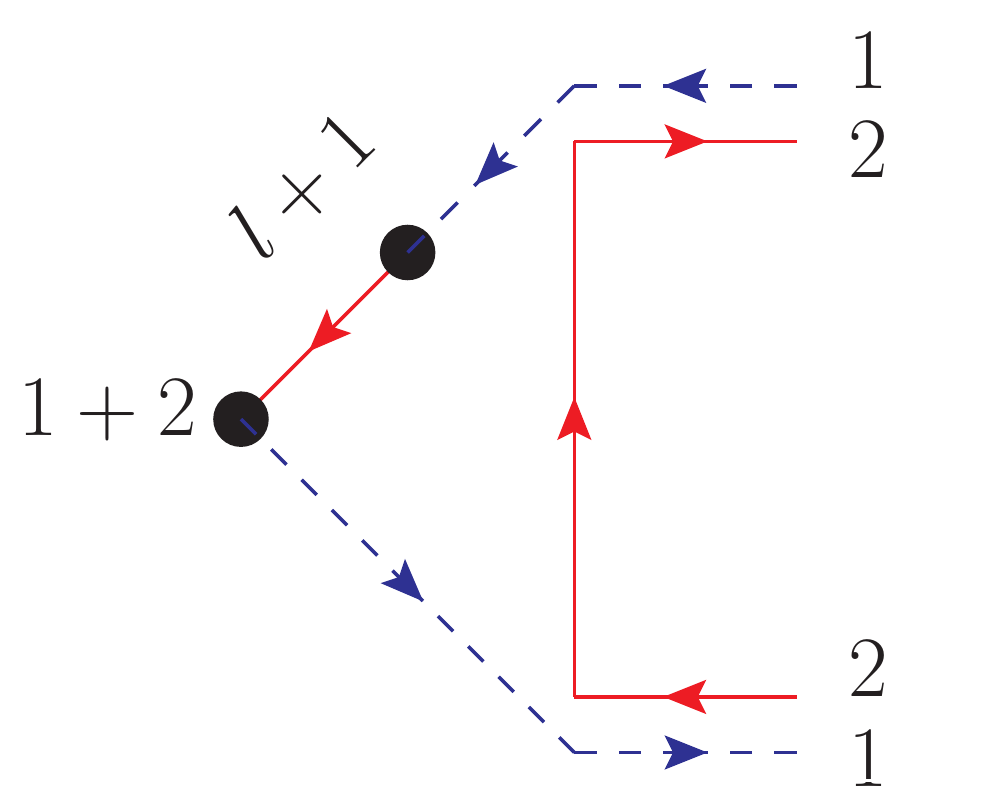} 
	\nonumber \\
		&+\left.\mu^2\includegraphics[scale=0.27,valign=c]{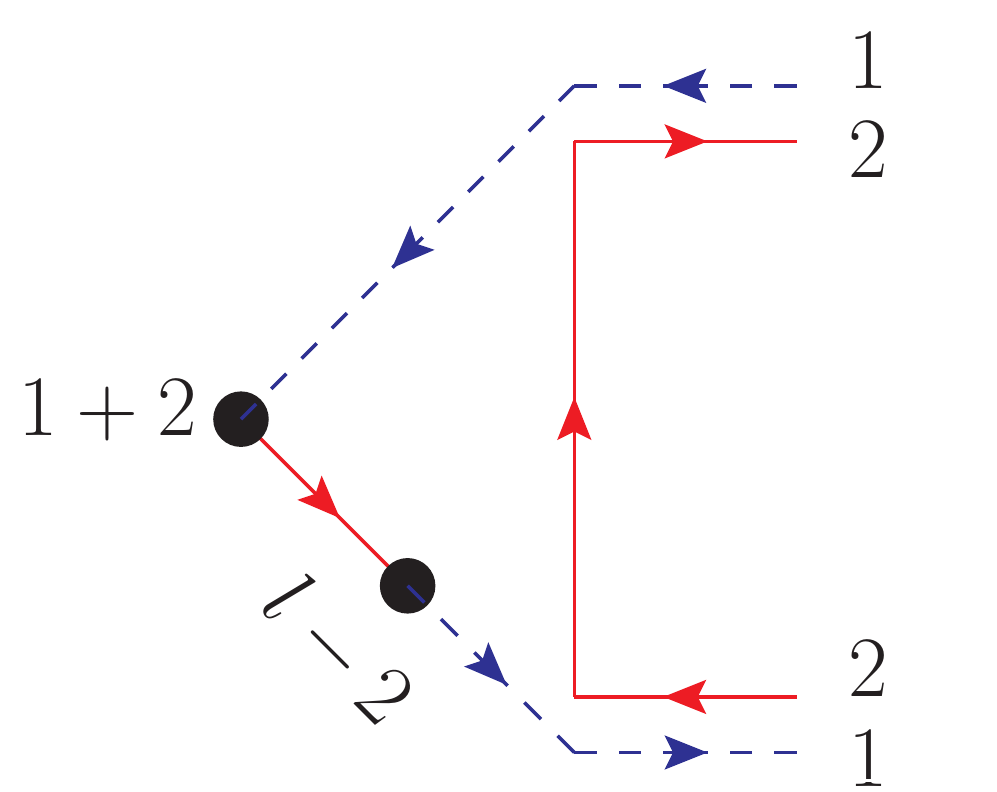}\hspace{-0.2cm}
	-\,\, \mu^2\includegraphics[scale=0.27,valign=c]{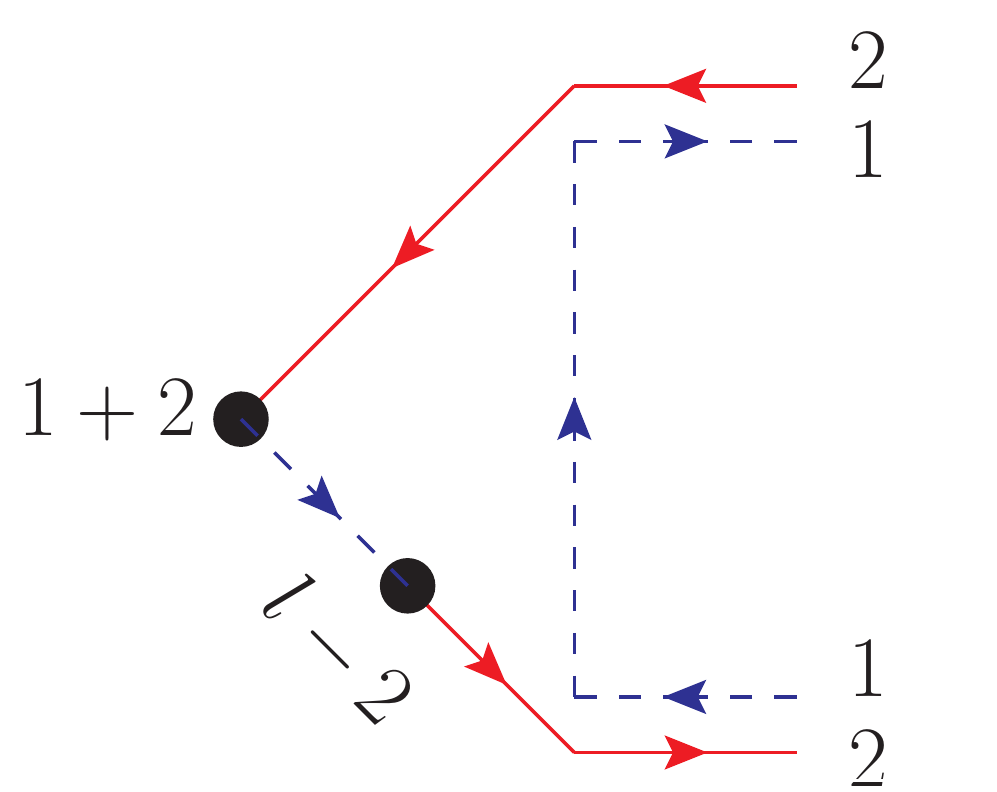} \right\rbrace
	\frac{1}{\ld^2(l+1)_{\dimens{d}}^2 (l-2)_{\dimens{d}}^2}~,
	\label{eq:axial anomaly LR first}
\end{align}
which is immediately simplified since all terms proportional to $\mu^2$ 
and parts of the first two chirality-flow diagrams vanish, due to 
$\cAngle{ii} = \cSquare{jj} = 0$ and/or due to the Weyl equation, e.g. $\sla{2}\ranSp{2} = 0$
(here, and in the rest of the paper, we use the slash notation to refer to contraction with a Pauli matrix, see \eqref{eq:mom outer products}).
Note how the clever choice of reference vectors helps remove many terms, as at tree level.

Next, we swap the arrow directions on all flow lines in the second chirality-flow diagram, 
and separate the momenta in the $l+1$ and $l-2$ momentum-dots of the first chirality-flow diagram to obtain
\begin{align}
	i(p_1+ p_2)_\mu \includegraphics[scale=0.3,valign=c]{Jaxodraw/QEDAxialAnomalyFeyn1LR} \hspace{-0.4cm}&=
	\frac{2e^2}{\cAngle{21}\cSquare{21}}\lint 
	\left\lbrace \includegraphics[scale=0.27,valign=c]{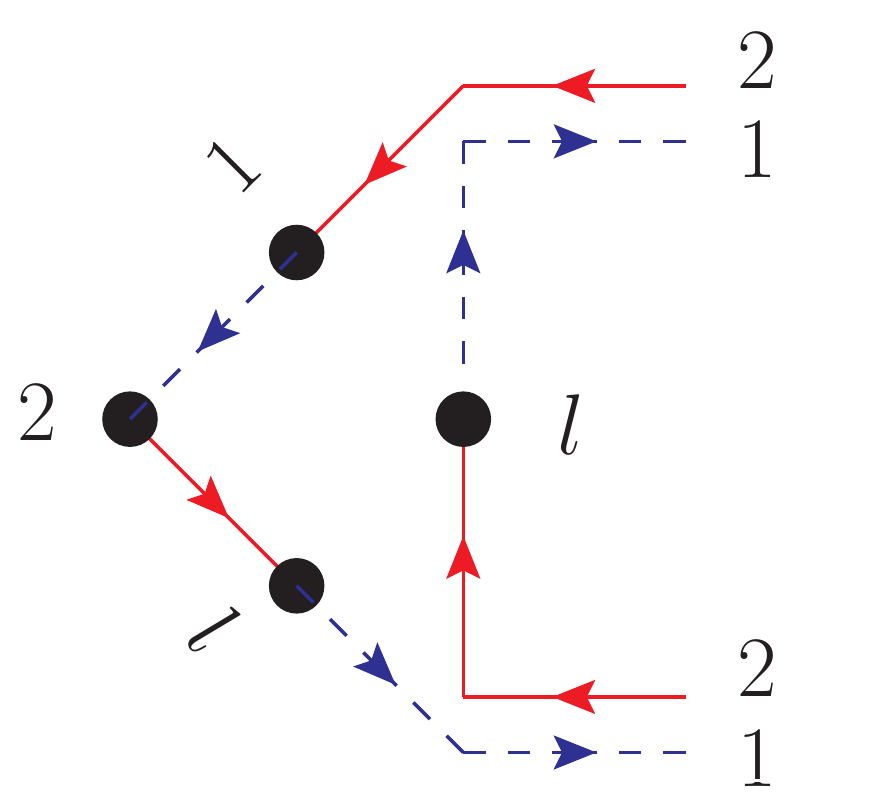} 
	+ \includegraphics[scale=0.27,valign=c]{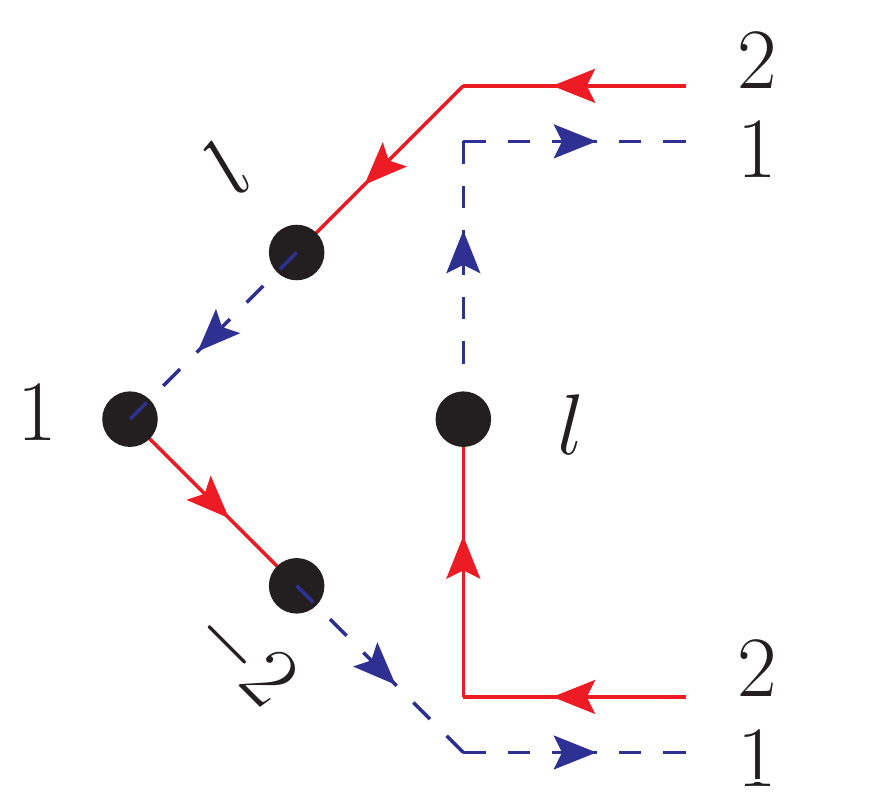} \right.\nonumber \\ 
	&\left.+ \includegraphics[scale=0.27,valign=c]{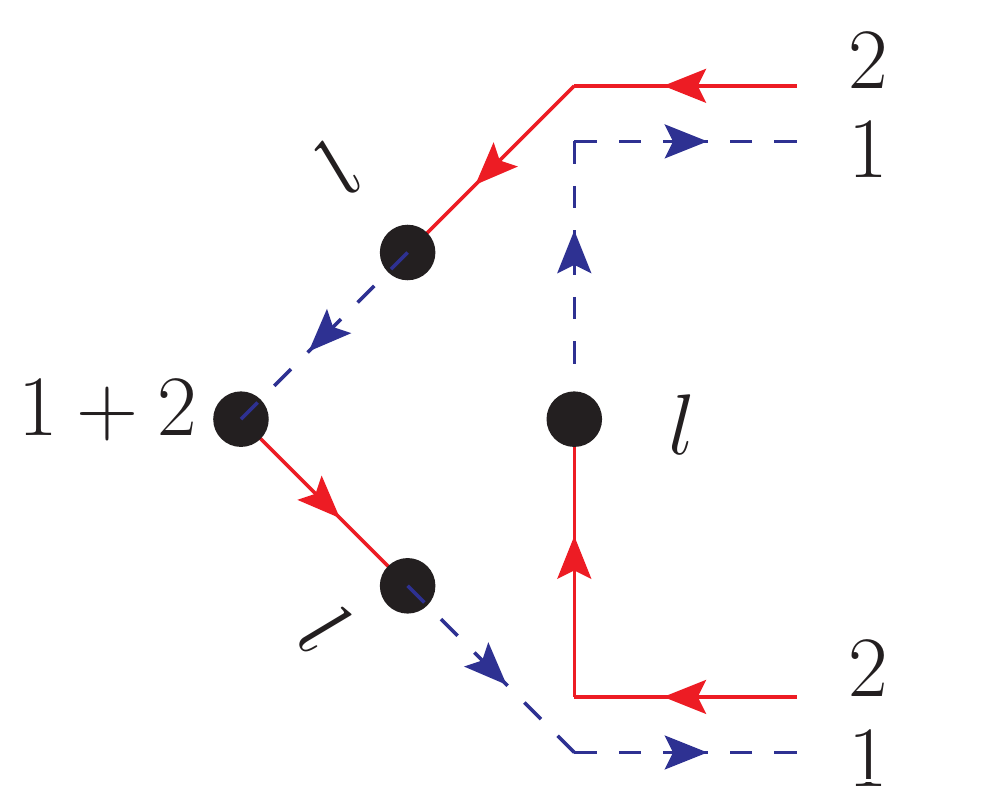}\hspace{-0.2cm}
		-\,\, \includegraphics[scale=0.27,valign=c]{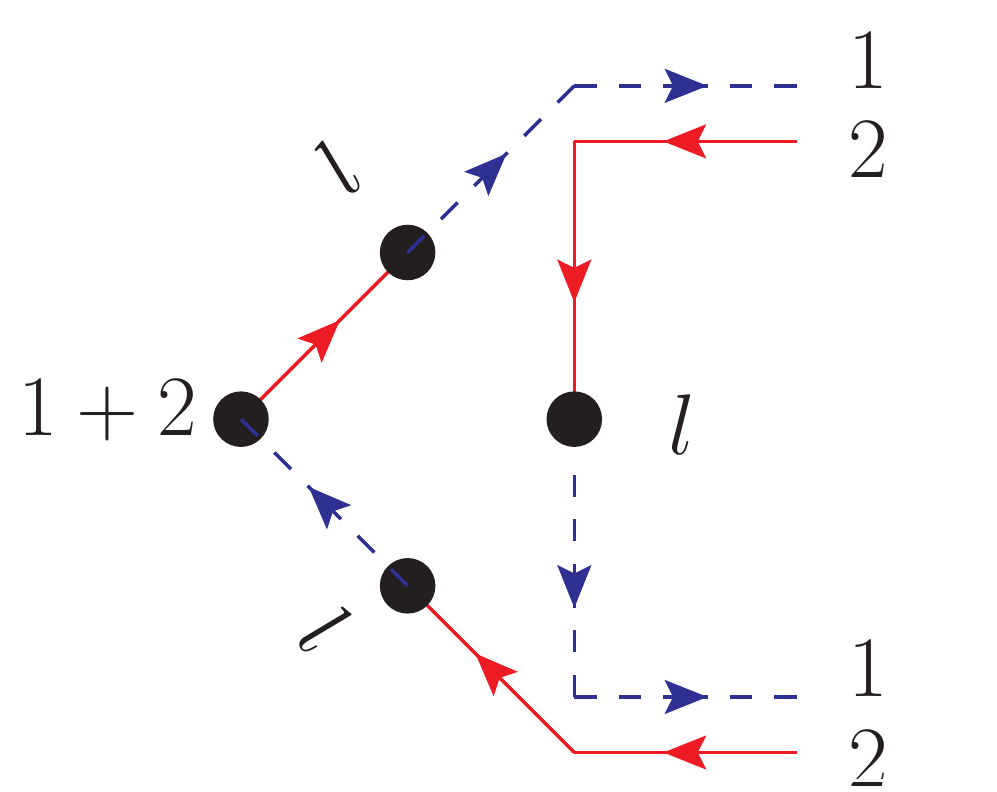}\right\rbrace
	\frac{1}{\left[\ld^2(l+1)_{\dimens{d}}^2 (l-2)_{\dimens{d}}^2\right]}
	\nonumber \\
	&= 0
\end{align}
where, reading out the inner products, 
we see that the first two chirality-flow diagrams cancel each other, as do the two last two chirality-flow diagrams. 
Therefore, in chirality flow, the axial anomaly for opposite-helicity photons can be made to vanish 
\textit{before} having to do the integration,
with --- similar to tree level --- a large simplification coming from our choice of reference vectors.

Note that if the terms did not cancel, they would anyway individually vanish after integration.
For example, the first chirality-flow diagram can be rewritten as
\begin{align}
	\lint \includegraphics[scale=0.27,valign=c]{Jaxodraw/QEDAxialAnomalyRchir4dSimp1Diag1LR} 
	\frac{1}{\left[\ld^2(l+1)_{\dimens{d}}^2 (l-2)_{\dimens{d}}^2\right]}
	&= \cAngle{21}\cSquare{12}\lint\frac{\lanSp{2}\slabar{l}\rsqSp{1}\lanSp{2} \slabar{l} \rsqSp{1}}{\left[\ld^2(l+1)_{\dimens{d}}^2 (l-2)_{\dimens{d}}^2\right]}~,
\end{align}
which is a rank-two tensor integral. 
This integral can be solved using standard (d-dimensional) tensor analysis
\begin{align}
	\lint \frac{\ld^\mu\ld^\nu}{\ld^2(l+1)_{\dimens{d}}^2 (l-2)_{\dimens{d}}^2} &= 
	C_{00}g^{\mu\nu} +  C_{11} 1^{\mu}1^{\nu} + C_{22} 2^{\mu}2^{\nu} + C_{12}\left(1^\mu 2^\nu + 1^\nu 2^\mu\right)~,
\end{align}
where we used that $\slabar{l} \equiv \lfour^\mu \sibar_{\dimens{4}\mu} = \ld^\mu \sibar_{\dimens{4}\mu}$ 
due to the projective nature of the metric, \eqref{eq:metric projectors FDF}.
However, every coefficient $C_{ij}$ multiplies a vanishing contribution,
either due to the Weyl equation, e.g.\ $C_{11} \lanSp{2}\slabar{1}\rsqSp{1}\lanSp{2} \slabar{1} \rsqSp{1} = 0$,
or due to the Fierz identity, 
$C_{00} \lanSp{2}\sibar^\mu\rsqSp{1}\lanSp{2} \sibar_\mu \rsqSp{1} = 2C_{00} \cAngle{22}\cSquare{11} = 0$.
This exemplifies one of the major advantages of chirality flow for loop diagrams. 

If we instead choose photon $2$ to be ``left chiral'' (have positive helicity), i.e.,\ 
$\eps_{1,\blue{L}}$ with $r_1 = 2$ and $\eps_{2,\blue{L}}$ with $r_2 = 1$,
we get
\begin{align}
	&i(p_1+ p_2)_\mu \includegraphics[scale=0.3,valign=c]{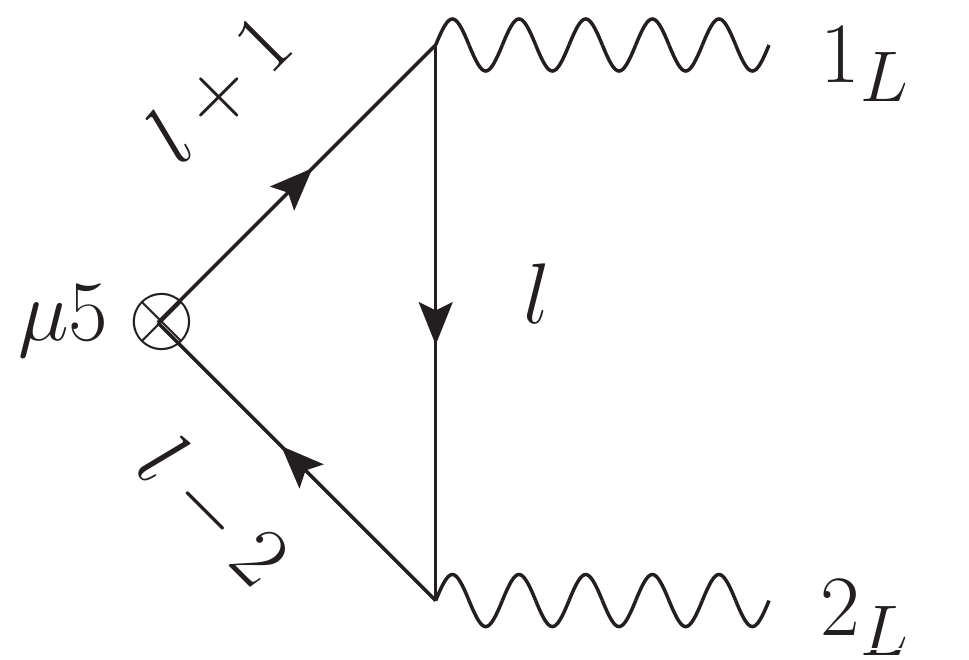} \hspace{-0.4cm}= 
	\frac{2e^2}{\cAngle{21}\cAngle{12}}\lint \left\lbrace 
	\includegraphics[scale=0.27,valign=c]{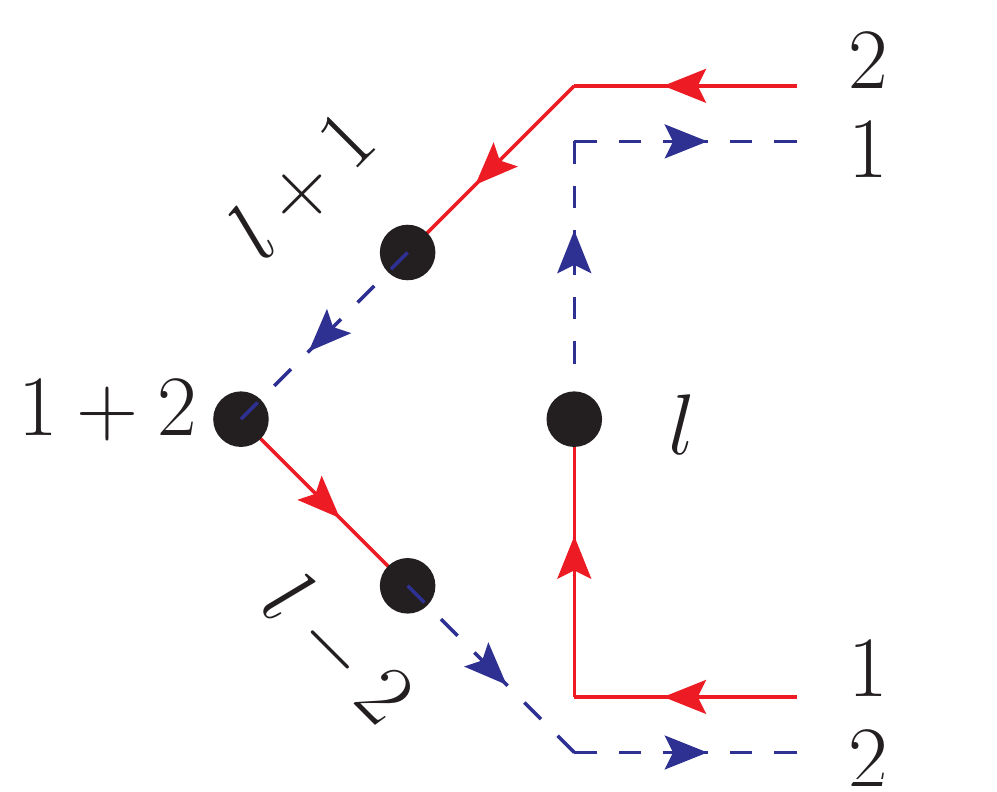}\hspace{-0.2cm}
	-\,\, \includegraphics[scale=0.27,valign=c]{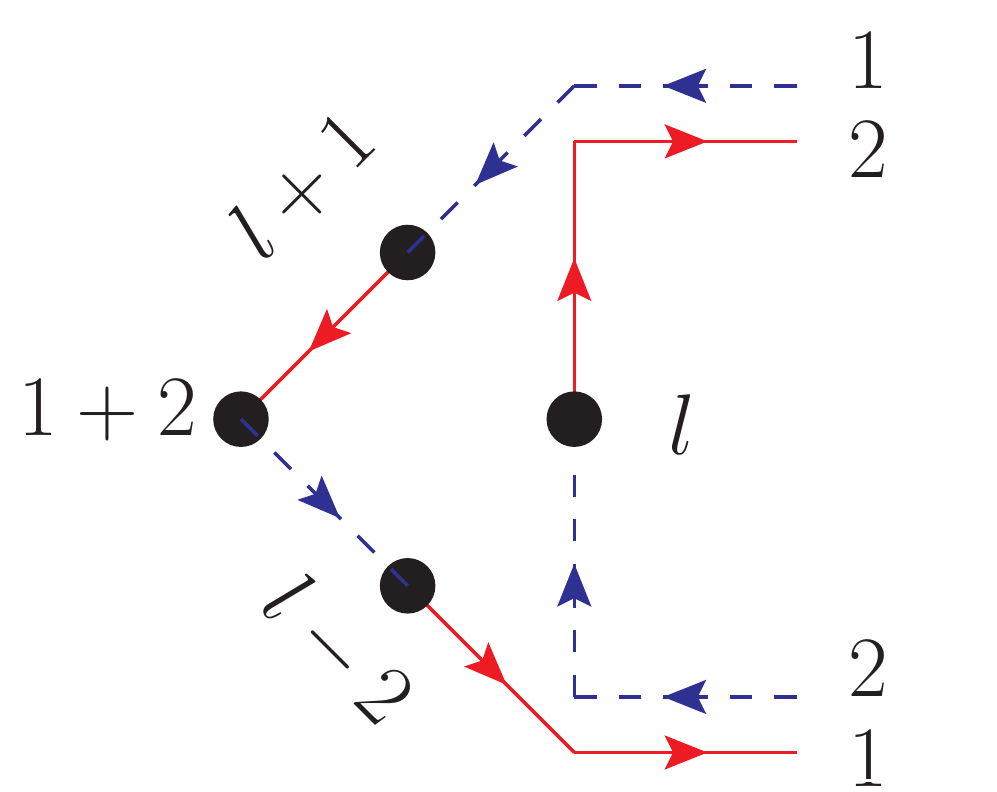} \right.
	\nonumber \\
	&+\mu^2\includegraphics[scale=0.27,valign=c]{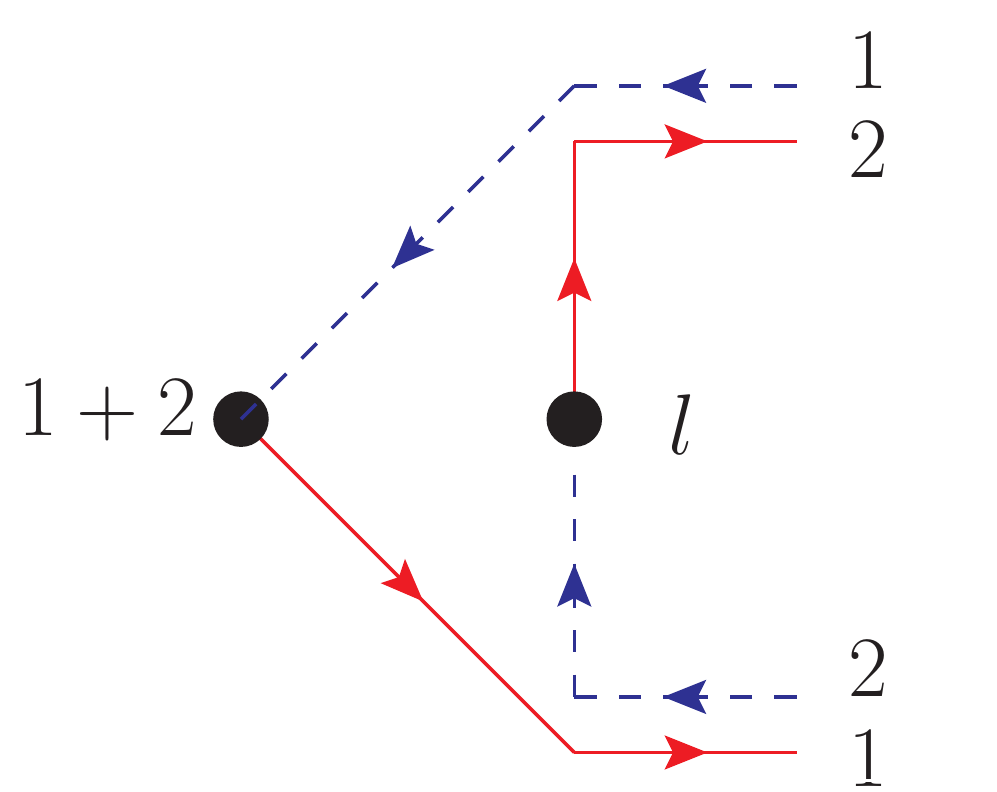}\hspace{-0.2cm}
	-\,\, \mu^2\includegraphics[scale=0.27,valign=c]{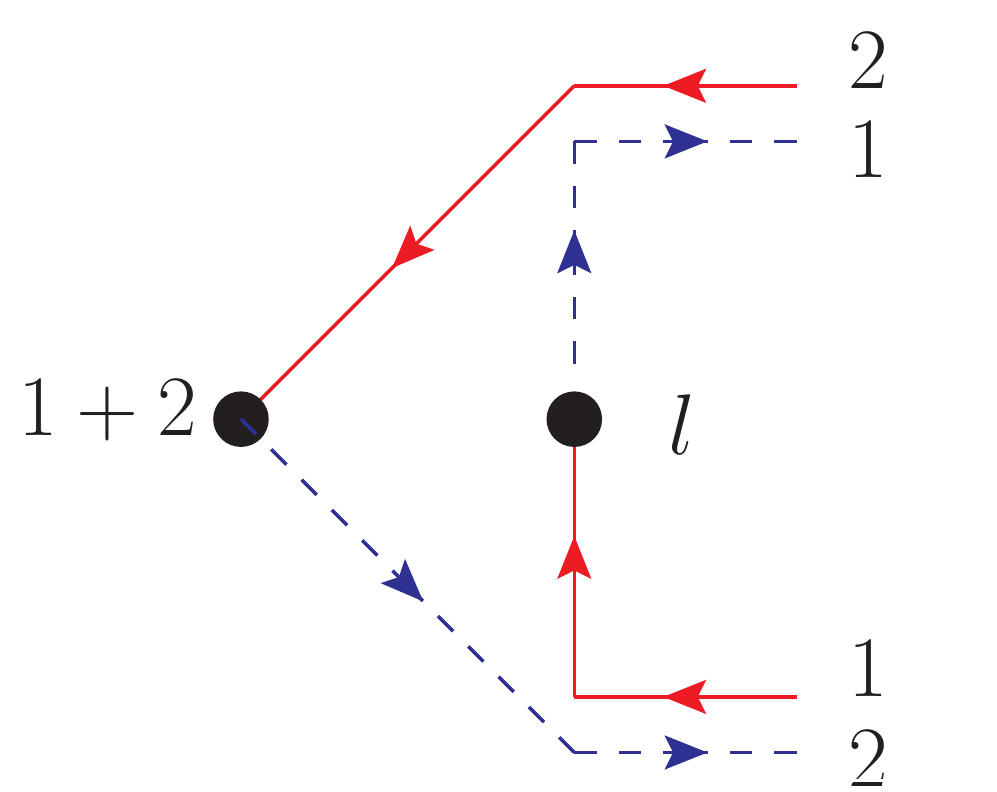} 
	+\mu^2\includegraphics[scale=0.27,valign=c]{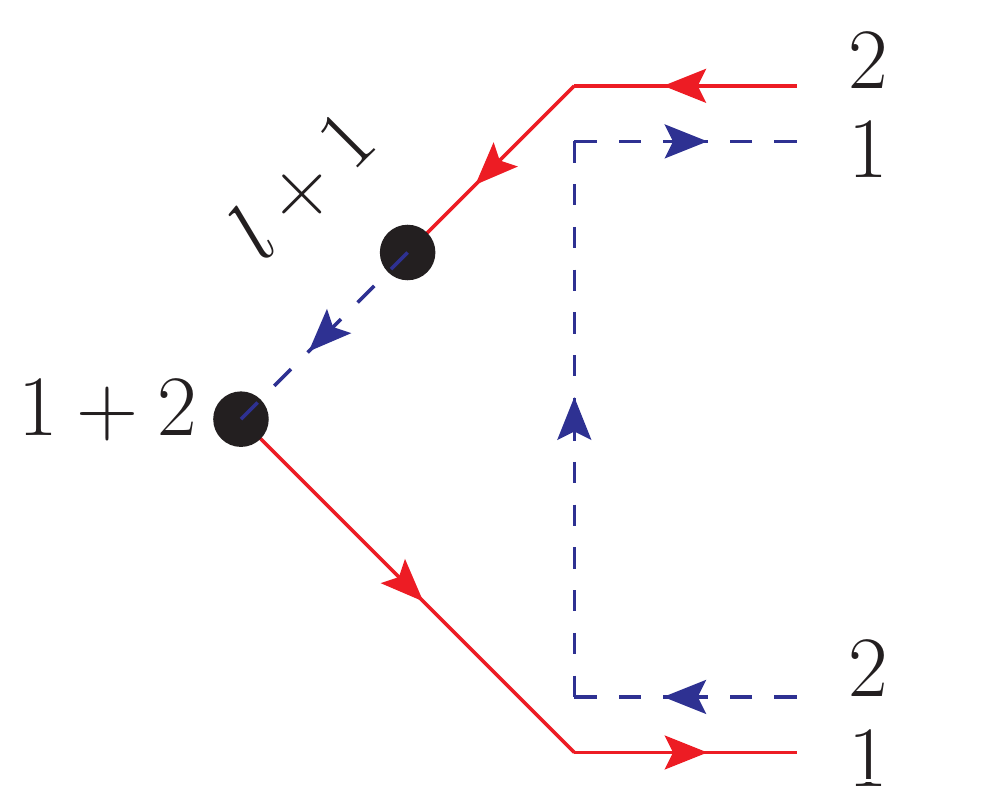}\hspace{-0.2cm}
	-\,\, \mu^2\includegraphics[scale=0.27,valign=c]{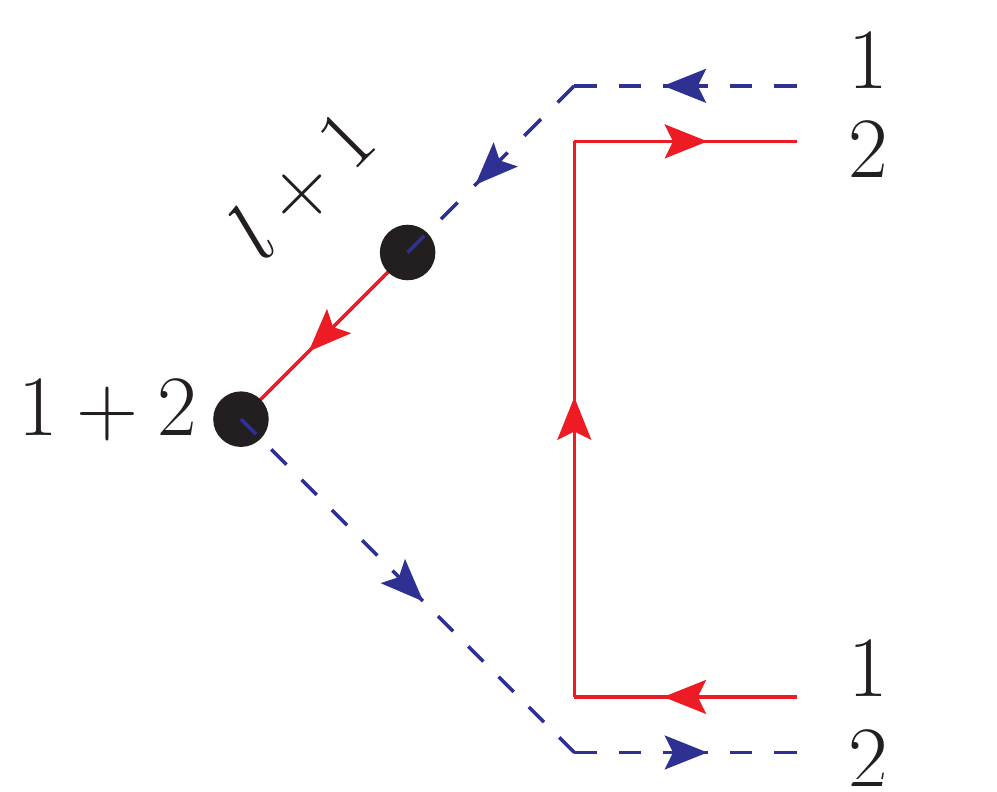} 
	\nonumber \\
	&+\left.\mu^2\includegraphics[scale=0.27,valign=c]{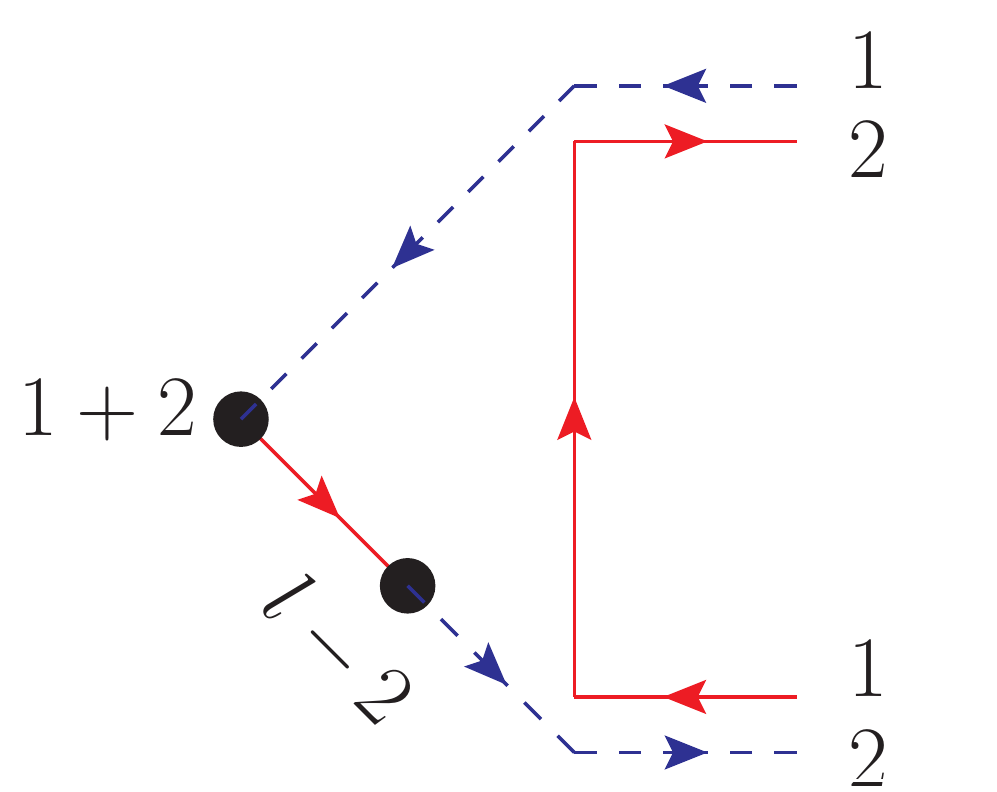}\hspace{-0.2cm}
	-\,\, \mu^2\includegraphics[scale=0.27,valign=c]{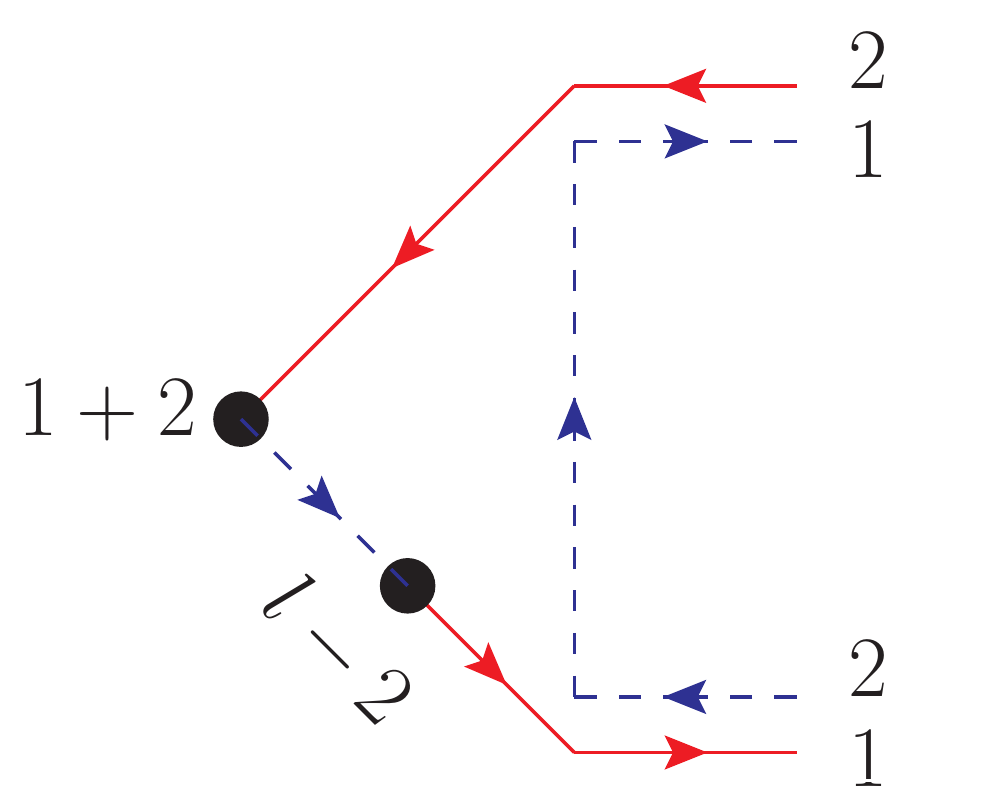} \right\rbrace
	\frac{1}{\ld^2(l+1)_{\dimens{d}}^2 (l-2)_{\dimens{d}}^2}~.
\end{align}
Tensor integrals with three loop momenta, $I_{i_1i_2i_3}^d[l^\mu l^\nu
  l^\rho]$, can be reduced by using \cite{Platzer:2022nfu}
\begin{align}
    \raisebox{-0.2\height}{\includegraphics[scale=0.3]{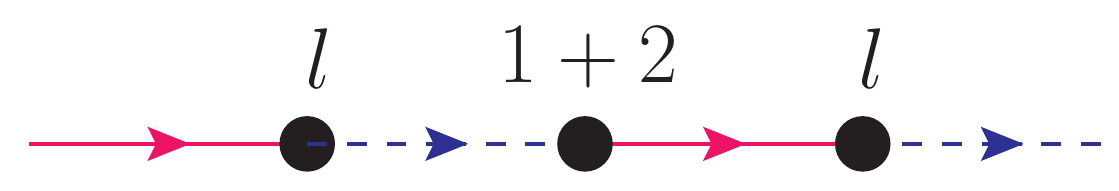}}&= 
	2 \lfour\cdot(1+2)_{\dimens{4}} \raisebox{-0.2\height}{\includegraphics[scale=0.3]{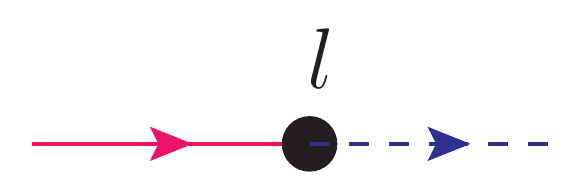}}\,
	- \,\lfour^2 \raisebox{-0.2\height}{\includegraphics[scale=0.3]{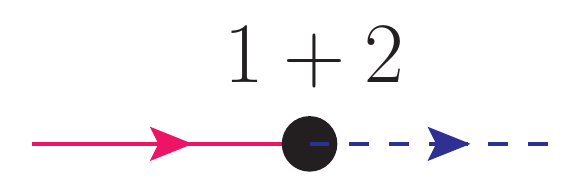}}
	+ i\underbrace{\eps^{\mu\nu\rho\eta}l_{\mu}(1_\nu+2_{\nu})l_\rho\sibar_{\eta}}_{0}~,
	\label{eq:reduce rank 3 to rank 1}
\end{align}
and we can then write the Lorentz structure as spinor products and strings to
obtain (after cancellation of some terms)
\begin{align}
	i(p_1+ p_2)_\mu \includegraphics[scale=0.3,valign=c]{Jaxodraw/QEDAxialAnomalyFeyn1LL} \hspace{-0.4cm}&= 
	\frac{2e^2\cSquare{12}}{\cAngle{12}}\lint \Bigg\lbrace \left(\lfour^2 - \mu^2\right)\left(\lanSp{2}\slabar{l}\rsqSp{2} - \lanSp{1}\slabar{l}\rsqSp{1} \right)
	+ 2 \lanSp{1}\slabar{l}\rsqSp{1} \lanSp{2}\slabar{l}\rsqSp{2}
	\nonumber \\
	&\hphantom{= 
		\frac{2e^2\cSquare{12}}{\cAngle{12}}\lint }  -2\mu^2\cAngle{12}\cSquare{21}  \Bigg\rbrace
	\frac{1}{\ld^2(l+1)_{\dimens{d}}^2 (l-2)_{\dimens{d}}^2}
	\nonumber \\
	&= \frac{2e^2\cSquare{12}}{\cAngle{12}}\left\lbrace\lint \frac{\lanSp{2}\slabar{l}\rsqSp{2} - \lanSp{1}\slabar{l}\rsqSp{1} }{(l+1)_{\dimens{d}}^2 (l-2)_{\dimens{d}}^2} \right.
	\nonumber \\
	&\hphantom{=\frac{2e^2\cSquare{12}}{\cAngle{12}}\Bigg\lbrace} \left.+ 2\lint \frac{\lanSp{1}\slabar{l}\rsqSp{1} \lanSp{2}\slabar{l}\rsqSp{2} - \mu^2\cAngle{12}\cSquare{21}}{\ld^2(l+1)_{\dimens{d}}^2 (l-2)_{\dimens{d}}^2}\right\rbrace~,
\end{align}
where we used that $\ld^2 = \lfour^2 - \mu^2$ from \eqref{eq:d-dim mom mu^2}.
To solve this, we have to solve three integrals,
which, using standard tensor reduction, we write as
\begin{align}
	\lint \frac{\ld^\mu}{(l+1)_{\dimens{d}}^2 (l-2)_{\dimens{d}}^2} &= C_1^{[0]} 1^{\mu} + C_2^{[0]} 2^{\mu}~,
	\nonumber \\
	\lint \frac{\ld^\mu\ld^\nu}{\ld^2(l+1)_{\dimens{d}}^2 (l-2)_{\dimens{d}}^2} &= 
	C_{00}g^{\mu\nu} +  C_{11} 1^{\mu}1^{\nu} + C_{22} 2^{\mu}2^{\nu} + C_{12}\left(1^\mu 2^\nu + 1^\nu 2^\mu\right)~, 
    \nonumber \\
	\lint \frac{\mu^2}{\ld^2(l+1)_{\dimens{d}}^2 (l-2)_{\dimens{d}}^2} &= C_{[\mu^2]}~,
\end{align}
such that
\begin{align}
	i(p_1+ p_2)_\mu \includegraphics[scale=0.3,valign=c]{Jaxodraw/QEDAxialAnomalyFeyn1LL} \hspace{-0.4cm}&= 
	2e^2\cSquare{12}\cSquare{21} \left( C_1^{[0]} - C_2^{[0]}  + 4C_{00} + 2C_{12}\cAngle{12}\cSquare{21} - 2C_{[\mu^2]} \right)~,
\end{align}
where it was again obvious due to the Weyl equation that not all
tensor coefficients were required.  Using \eqref{eq:mu integral
  relation} for the integral with $\mu^2$, and calculating the tensor
coefficients using \cite{Davydychev:1991va, Giele:2004iy,
  Platzer:2012oneloop, Patel:2015tea, Patel:2016fam,
  Shtabovenko:2016whf} we find
\begin{align}
	C_1^{[0]} - C_2^{[0]}  + 4C_{00} + 2C_{12}\cAngle{12}\cSquare{21}  &= 0~, 
	& 
	C_{[\mu^2]} &= \frac{1}{2}\frac{i}{(4\pi)^2}~,
\end{align}
and therefore we obtain the known result for the anomaly, 
\eqref{eq:axial anomaly as spinor products}
\begin{align}
	i(p_1+ p_2)_\mu \includegraphics[scale=0.3,valign=c]{Jaxodraw/QEDAxialAnomalyFeyn1LL} \hspace{-0.4cm}&= 
	-\frac{2ie^2}{(4\pi)^2}\cSquare{12}\cSquare{21}~.
\end{align}
This example shows many of the features of chirality flow at one loop.
We saw that many terms vanished due to the choice of reference
vector, as well as a transparent simplification of the tensor
reduction.

Though it was not required in this example,
we may also need to take some care with chirality-flow arrows and minus signs at one loop
(see the discussion of the fermion self-energy in \secref{sec:abelian theories}). 
This, along with other details of one-loop chirality-flow calculations, 
will be discussed below.

\subsection{Reduction of tensor integrals}
\label{sec:tensor reduction}

The method of performing one-loop calculations with chirality flow explored here is
traditional in the sense that we do not exploit unitarity-based
approaches, but rely on separating the numerator algebra from a tensor
integral, which is then decomposed into the various tensor structures,
and subsequently reduced to master integrals. The chirality-flow method
allows to use spinor identities and equations of motion directly, such
as to directly identify those tensor structures which will not
contribute to an amplitude.

As we saw in the axial anomaly above, it is
easy to identify which contributions of a tensor integral in chirality
flow will vanish.  A typical rank-$n$ tensor integral will occur as
\begin{align}
	&\lanSp{i_1}\sibar_{\mu_1}\rsqSp{j_1}\dots\lanSp{i_n}\sibar_{\mu_n}\rsqSp{j_n} I^d_{d_1\dots d_k}[l^{\mu_1}\dots l^{\mu_n}]~,
	\nonumber \\
	&\lanSp{i_1}\sibar_{\mu_1}\si_{\mu_2}\ranSp{j_1}\dots\lanSp{i_n}\sibar_{\mu_n}\rsqSp{j_n} I^d_{d_1\dots d_k}[l^{\mu_1}\dots l^{\mu_n}]~,
		\nonumber \\
	&\lanSp{i_1}\sibar_{\mu_1}\si_{\mu_2}\ranSp{j_1}\dots\lsqSp{i_n}\si_{\mu_{n-1}}\sibar_{\mu_n}\rsqSp{j_n} I^d_{d_1\dots d_k}[l^{\mu_1}\dots l^{\mu_n}]~,
		\nonumber \\
	&\text{etc.}~,
\end{align}
and coefficients will vanish for one of three reasons.
Either, they will vanish due to the Weyl equation,
e.g.,\ $\slabar{j}_1\rsqSp{j_1}= 0$,
due to massless momenta being contracted with consecutive Pauli matrices, e.g.\ $\sla{j}_1\slabar{j}_1 = j_1^2 = 0$,
or from the Fierz identity contracting two spinor strings with a common spinor, e.g.
\begin{align}
	\lanSp{i_1}\sibar^\mu\rsqSp{j_1} \lsqSp{j_1} \si_{\mu}\ranSp{i_2} = 2\cAngle{i_1i_2}\cSquare{j_1j_1} = 0~,
\end{align} 
where we define a string of spinors to be a sequence like $\lanSp{i_1}\sibar^\mu\rsqSp{j_1}$ which starts and ends with spinors, 
possibly with (many) Pauli matrices in between. 

In general, the number of contributing structures in a tensor integral
is dependent on the choice of gauge or spin reference momenta, and,
like at tree level, choosing these wisely can significantly reduce the
amount of work required to do the calculation.

Additionally, in chirality flow we are sometimes able to reduce the rank of a tensor integral by using equations such as 
\eqref{eq:reduce rank 3 to rank 1}.
In \cite{Platzer:2022nfu}, 
we showed how to reduce strings of multiple momentum-dots into simpler building blocks,
using e.g.
\begin{align}
	\raisebox{-0.2\height}{\includegraphics[scale=0.3]{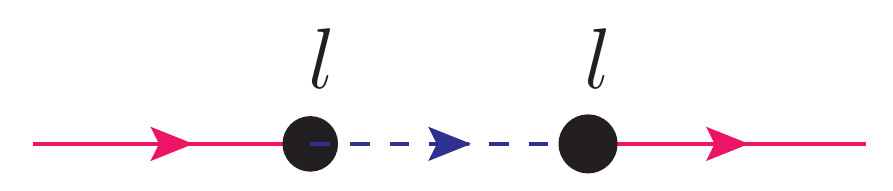}}&= 
	\lfour^2 \raisebox{-0.2\height}{\includegraphics[scale=0.3]{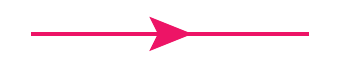}}~, 
	\quad \text{and}
	\nonumber \\
	\raisebox{-0.2\height}{\includegraphics[scale=0.3]{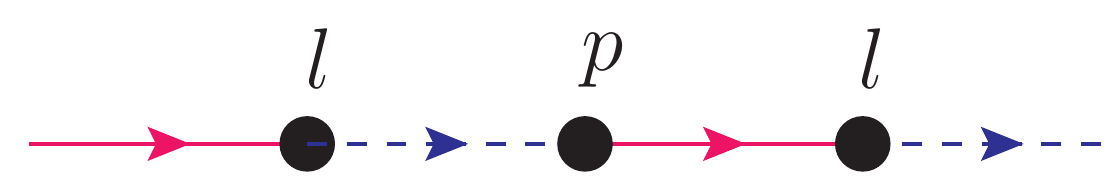}}
	&= 	2 \lfour\cdot p_{\dimens{4}} \raisebox{-0.2\height}{\includegraphics[scale=0.3]{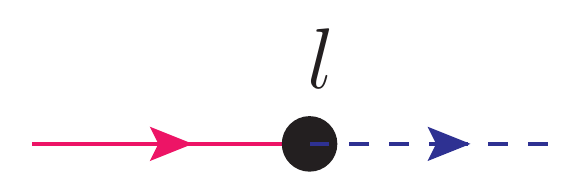}}\,
	- \,\lfour^2 \raisebox{-0.2\height}{\includegraphics[scale=0.3]{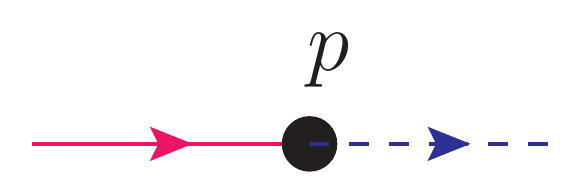}}~,
	\label{eq:example tensor rank reductions}
\end{align}
where the former occurs in diagrams like 
\begin{align*}
	\includegraphics[scale=0.3,valign=c]{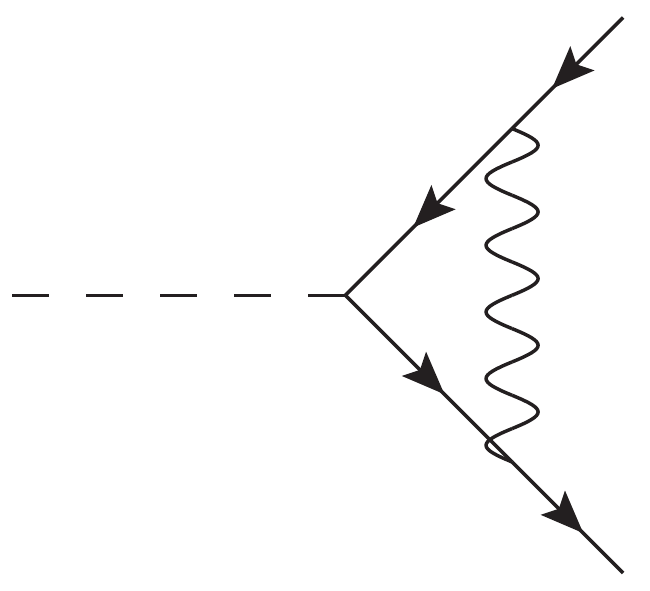}~,
\end{align*}
while the latter appears in e.g.\ the axial anomaly calculation of the previous section. 
To use these relations, recall that the momenta on the denominator are d-dimensional, 
so we have to convert the four-dimensional dot products of \eqref{eq:example tensor rank reductions} to d-dimensional ones using
\begin{align}
	\lfour^2 &= \ld^2 + \mu^2~,
	&
	2 \lfour\cdot p_{\dimens{4}} &= 2 \ld\cdot p_{\dimens{d}} = (l+p)^2_{\dimens{d}} - \ld^2 - p^2_{\dimens{d}}
\end{align}
where we used that $p$ is strictly four-dimensional, so
$p_{\dimens{4}} = p_{\dimens{d}}$. For a full set of relations which
can reduce the rank of a tensor integral, see
\cite{Platzer:2022nfu}. Reduction methods of tensor integrals along
the lines of \cite{Giele:2004iy,Platzer:2012oneloop} then allow us to
directly reduce coefficients of individual tensor structures.

\subsection{Abelian gauge theories}
\label{sec:abelian theories}

One-loop diagrams in abelian theories can be separated into two categories,
diagrams like the axial anomaly in \secref{sec:flowing loops} which have a purely fermionic loop,
and diagrams which have a mixture of fermions and bosons in the loop.

For the former case, the procedure is simple,
draw the chirality-flow diagram as at tree level \cite{Lifson:2020pai,Alnefjord:2020xqr}, 
putting arrows in their natural position, 
i.e.,\ against the fermion flow.
Then perform the integral, making use of the simplifications from \secref{sec:tensor reduction}.
The rest of this section deals with the second class of loops, 
those with bosons and fermions,
and show that these can also be handled.

\paragraph*{Loops with a single fermion line:}
The simplest example of such a diagram is the self-energy of a massless fermion in the Feynman gauge.
There are two diagrams for this process, 
one in which the fermion emits and reabsorbs the gauge boson 
(usually just called a photon below for convenience),
and another in which it emits and reabsorbs the FDF scalar. 
However, the FDF-scalar contribution is proportional to 
$\Ga^M\Ga^M = 0$, so we do not need to consider it. 

If we consider just the Lorentz structure of the self-energy, 
and consider only a single chirality, we have e.g.
\begin{align}
	\includegraphics[scale=0.3,valign=c]{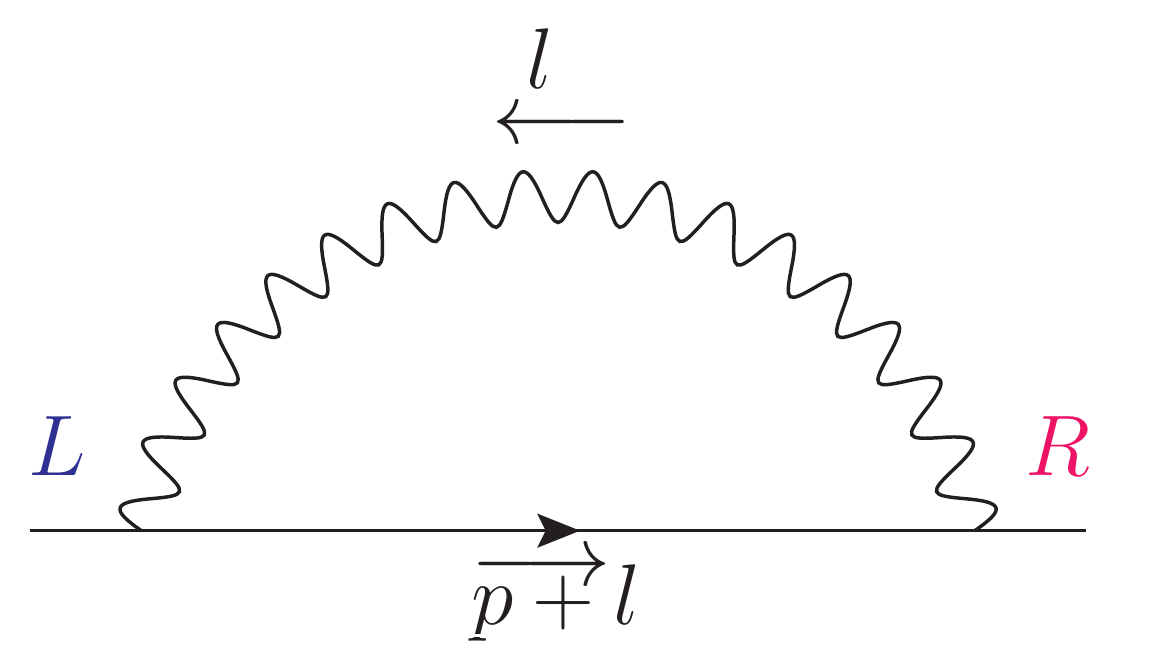} &\sim 
	\includegraphics[scale=0.3,valign=c]{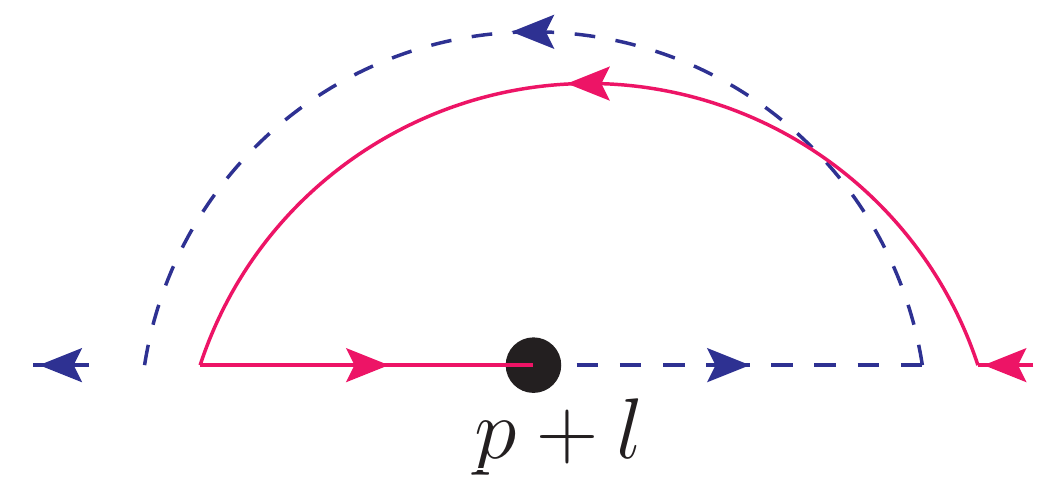}
	\sim \raisebox{-0.2\height}{\includegraphics[scale=0.3]{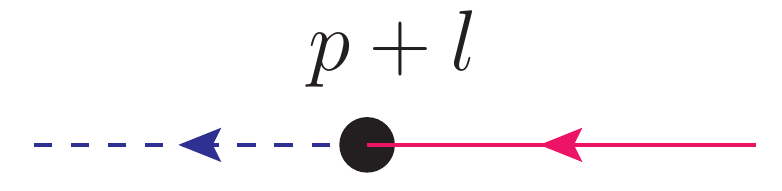}}~,
	\label{eq:fermion self energy simplified}
\end{align}
for the numerator structure.
In the middle diagram, drawn to follow the Feynman diagram, 
we see that naively applying chirality flow to the diagram leads to either the photon having arrows in the same direction or the momentum-dot not having a continuous flow,
both of which are avoided at tree level \cite{Lifson:2020pai}. 
As we will see below, this simply introduces a minus sign into equations. 
In the right-hand diagram,
we see that chirality flow quickly reduces the loop-level Lorentz structure to a simple tree-level chirality structure
(i.e.\ a Lorentz structure which occurs in a tree-level chirality-flow diagram),
without having to apply any anticommutation relations.

We now consider the general case of a single massless fermion emitting a virtual photon,
possibly emitting more photons, and then reabsorbing the first virtual photon.
The simplest version of this is the fermion self-energy, 
schematically given in \eqref{eq:fermion self energy simplified}.\footnote{
	The full calculation of the FDF (and therefore chirality flow) fermion self-energy in QED,
	as well as counterterms, are given in \cite{Gnendiger:2017pys}.}
Considering just the Lorentz algebra, we have for example
\begin{align}
		\includegraphics[scale=0.3,valign=c]{Jaxodraw/fermionSelfEnergyPhotonLR} &\rightarrow \,\,
		\taubar^\nu\Big(\sla{p} + \sla{l}\Big)\taubar_{\nu} = - \Big(\slabar{p} + \slabar{l}\Big)~,
		\label{eq:fermion self energy taus}
\end{align}
where the Pauli matrices with repeated
index are removed using \eqref{eq:taufierz}, and the remaining Lorentz
algebra is calculated using \eqsrefa{eq:tau taubar}{eq:eps relations}. 
Note that our normalization of the Pauli matrices in the fermion-photon vertex, \eqref{eq:fermion_photon_vertex}, 
gives the unusual prefactor of one on the right hand side (which would have read $2$ in
the case of Dirac matrices or more traditionally-normalized Pauli matrices). 
In chirality flow, \eqref{eq:fermion self energy taus} is drawn as
\begin{align}
	\includegraphics[scale=0.3,valign=c]{Jaxodraw/fermionSelfEnergyPhotonLR} &\rightarrow \,
	-\includegraphics[scale=0.3,valign=c]{Jaxodraw/fermionSelfEnergyPhotonFlowLR}
	= -\raisebox{-0.2\height}{\includegraphics[scale=0.3]{Jaxodraw/fermionSelfEnergyFlowLR}}~.
	\label{eq:fermion self energy flow}
\end{align}
If the massless fermion emitted a photon before reabsorption, 
we instead obtain
\begin{align}
	\includegraphics[scale=0.3,valign=c]{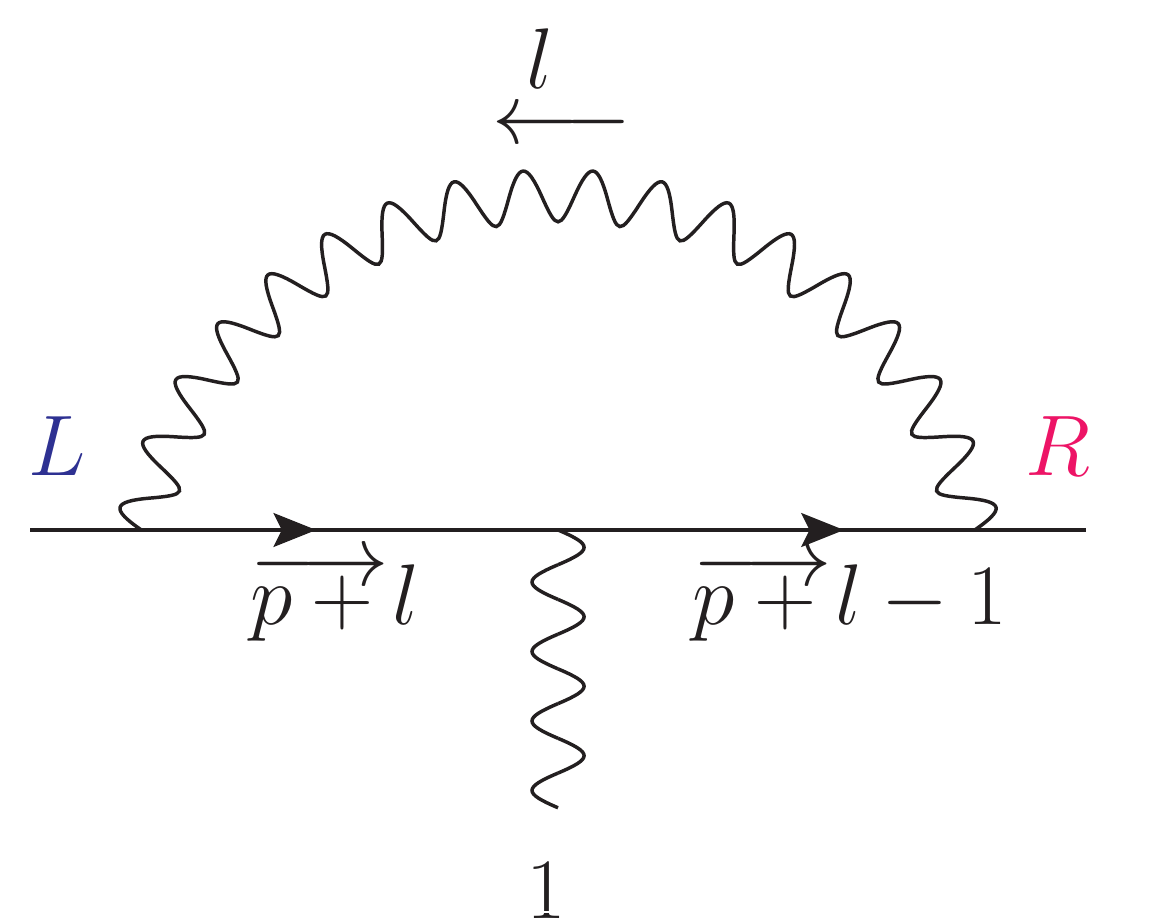} &\rightarrow \,\,
	\taubar^\nu\Big(\sla{p} + \sla{l}-\sla{1}\Big)\taubar^{\nu_1}\Big(\sla{p} + \sla{l}\Big)\taubar_{\nu} + 
	\mu^2 \taubar^\nu \tau^{\nu_1}\taubar_{\nu}
	\nonumber \\
	&=  	- \Big(\slabar{p} + \slabar{l}\Big)\tau^{\nu_1}\Big(\slabar{p} + \slabar{l}-\slabar{1}\Big) -\mu^2\taubar^{\nu_1}~,
\end{align}
using the same method as in \eqref{eq:fermion self energy taus} for the Lorentz algebra. 
Note that we now have a $\mu^2$ term from the fermion propagator, 
and that in the first line we obtain $\mu^2$ instead of $-\mu^2$ due to the $\ga^5$ matrices in the propagator.
In chirality flow, this relation is
\begin{align}
	\includegraphics[scale=0.3,valign=c]{Jaxodraw/fermionSelfEnergyExtPhotonLR} &\rightarrow \,
	-\includegraphics[scale=0.3,valign=c]{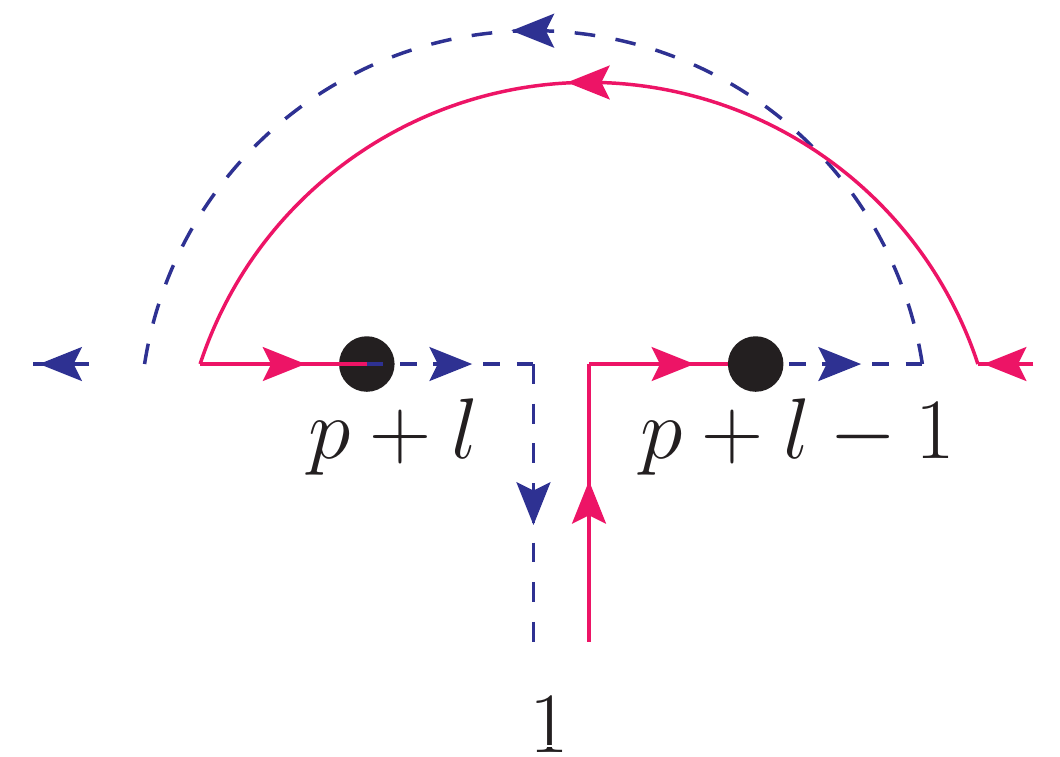} 
	-\mu^2\includegraphics[scale=0.3,valign=c]{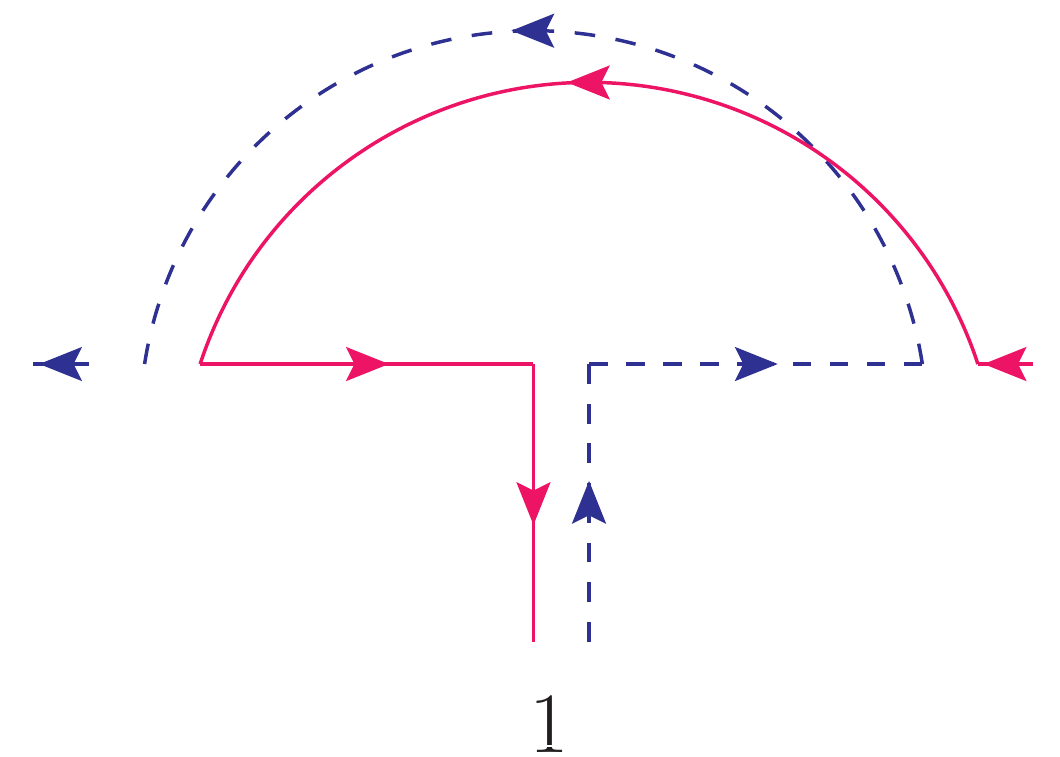}
	\nonumber \\
	 &= -\,\includegraphics[scale=0.3,valign=c]{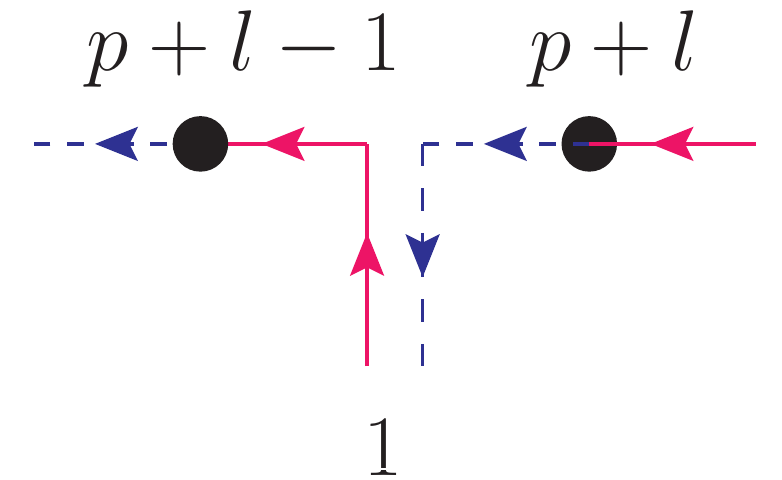} \,
	 - \,\mu^2 \includegraphics[scale=0.3,valign=c]{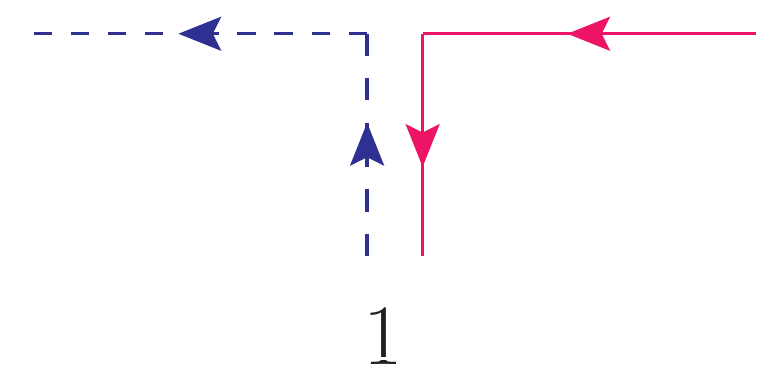}	~,
	 \label{eq:fermion self energy one photon chirality flow}
\end{align}
where we see that for chirality flow to look like a tree-level structure,
we must reverse the order of the momentum-dots in the diagram.

In general, for a massless fermion emitting $n$ photons between emission and absorption of the virtual photon,
the Lorentz structure will be sums of terms with the following form 
\begin{align}
	\includegraphics[scale=0.3,valign=c]{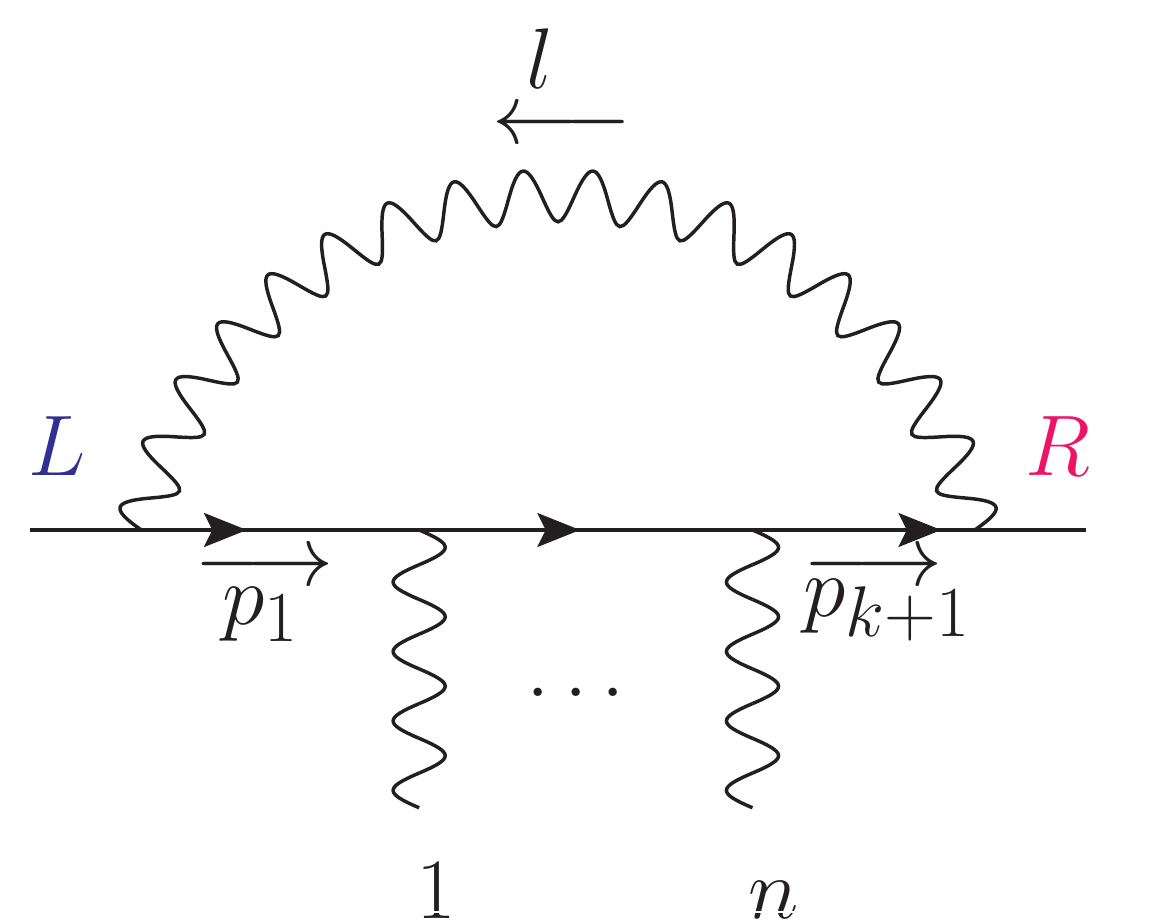} &\rightarrow \,\,
	(\mu^2)^{n-k} \,\taubar^\nu\tau^{\nu_1}\dots\tau^{\nu_{2k+1}}\taubar_{\nu} = 
	- (\mu^2)^{n-k}\,\tau^{\nu_{2k+1}}\dots\tau^{\nu_{1}}~,
	\label{eq:odd jellyfish}
\end{align}
where $k$ is an integer satisfying $\text{max}\left(0, \frac{n-1}{2}\right) \leq k \leq n$, 
we suppress the propagator momenta, 
and the sequence of Pauli matrices always goes $\tau$, $\tau\taubar\tau$,  $\tau\taubar\tau\taubar\tau$ etc.
Note that this relation only holds for an odd number of Pauli matrices,
and that, like in \eqref{eq:fermion self energy one photon chirality flow},
the order of the Pauli matrices on the right-hand side of \eqref{eq:odd jellyfish} is reversed.
If the fermion in \eqref{eq:odd jellyfish} is massive, 
then some powers of $\mu^2$ may be replaced by $m^2$,
and the overall sign in this equation is no longer fixed
and is calculated using \eqref{eq:fermion prop FDF}.

The other thing which could happen if the fermion is massive,
is that the external fermions could have the same chirality, say left-left.
(This may also happen if we for example replace one of the photons $1\dots n$ 
in diagrams like \eqref{eq:odd jellyfish} with a scalar particle.)
In this case, 
there will be an even number of Pauli matrices between emission and absorption,
it is easy to remove the repeated vector index using \eqref{eq:taufierz},
and the Feynman diagram will have Lorentz structures of the form
\begin{align}
	\includegraphics[scale=0.3,valign=c]{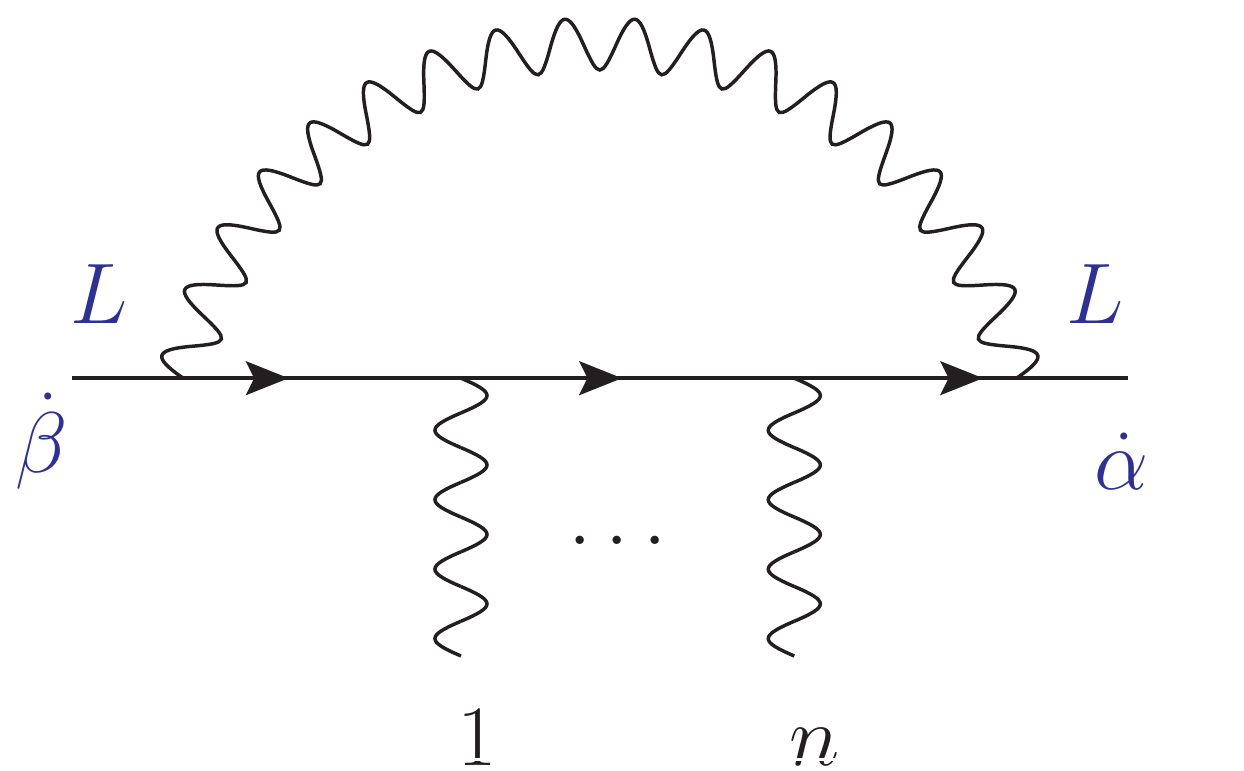} &\sim \,\,
	m^{2r+1}(\mu^2)^{q} 
	\,\tau^\nu\taubar^{\nu_1}\dots\tau^{\nu_{2k}}\taubar_{\nu} = 
	 m^{2r+1}(\mu^2)^{q}
	 \,\de^{\da}_{~\db}\Tr\left(\taubar^{\nu_{1}}\dots\tau^{\nu_{2k}}\right)~,
	 \label{eq:even jellyfish}
\end{align}
where $k,r$, and $q$ are non-negative integers satisfying $ n/2 \leq k \leq n$,
$r = n-k-q \leq n/2$, and 
$q \leq n/2$, 
and the overall sign is determined using \eqref{eq:fermion prop FDF}.

As a simple example of \eqref{eq:even jellyfish} we consider the diagram from 
\eqref{eq:fermion self energy one photon chirality flow},
but for the case of left-left chiralities coming from the replacement of a $\sla{p}$ with a mass
\begin{align}
	\includegraphics[scale=0.3,valign=c]{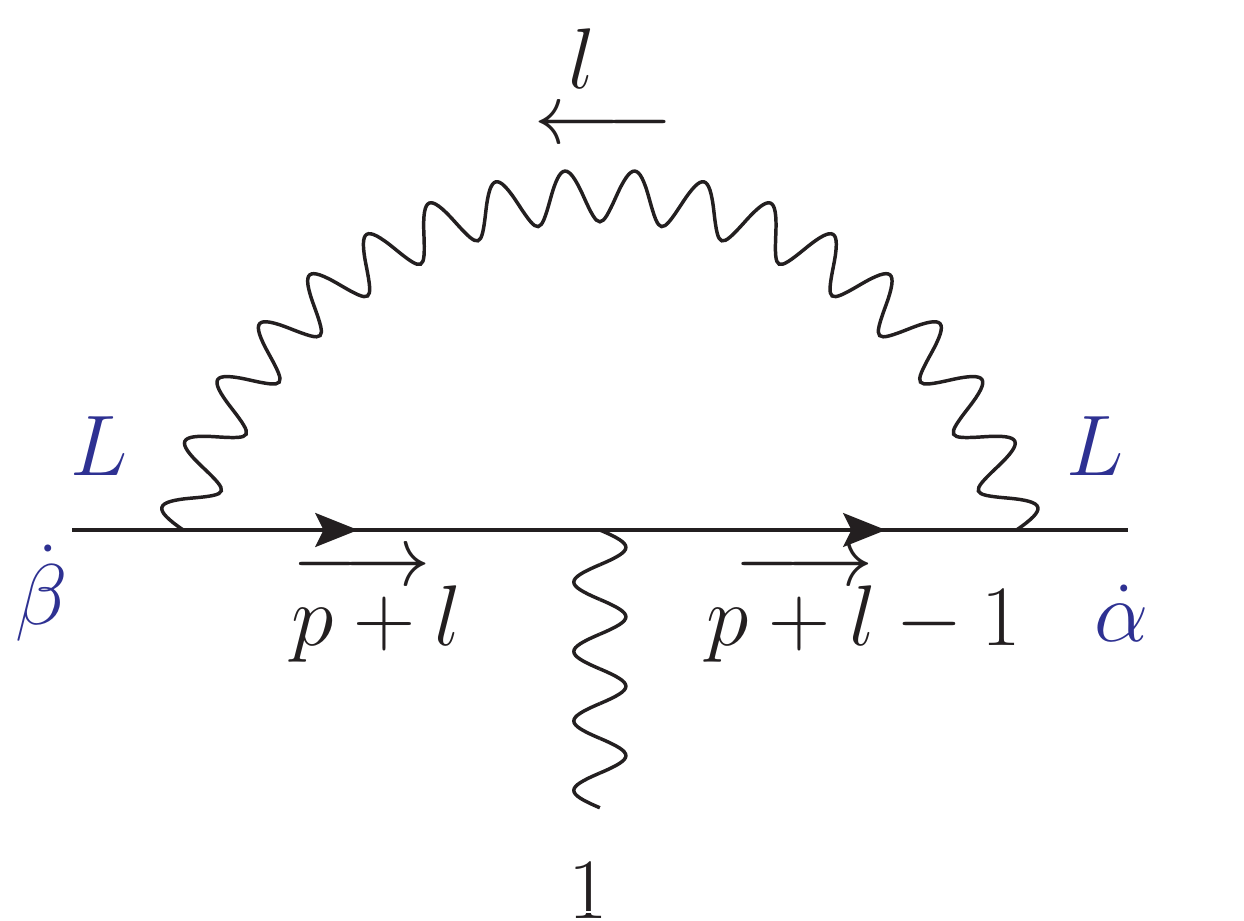} &\rightarrow \,
	m \de^{\da}_{~\db} \Tr\Big[ \taubar^{\nu_1} \Big(\sla{p} + \sla{l}\Big)  \Big]
	+ m \de^{\da}_{~\db} \Tr\Big[ \Big(\slabar{p} + \slabar{l} - \slabar{1}\Big) \tau^{\nu_1}   \Big]~,
\end{align}
or in chirality flow
\begin{align}
	\includegraphics[scale=0.3,valign=c]{Jaxodraw/fermionSelfEnergyExtPhotonLL} &\rightarrow \,
	m \includegraphics[scale=0.3,valign=c]{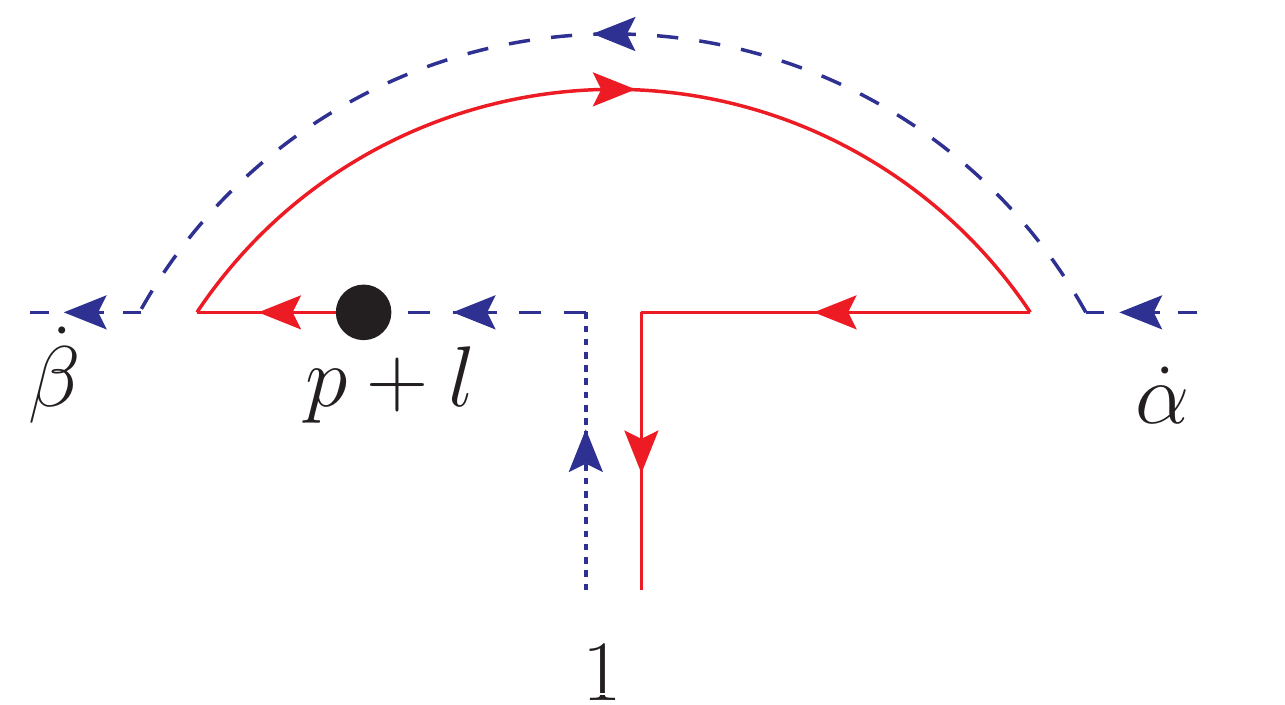}
	+ m \includegraphics[scale=0.3,valign=c]{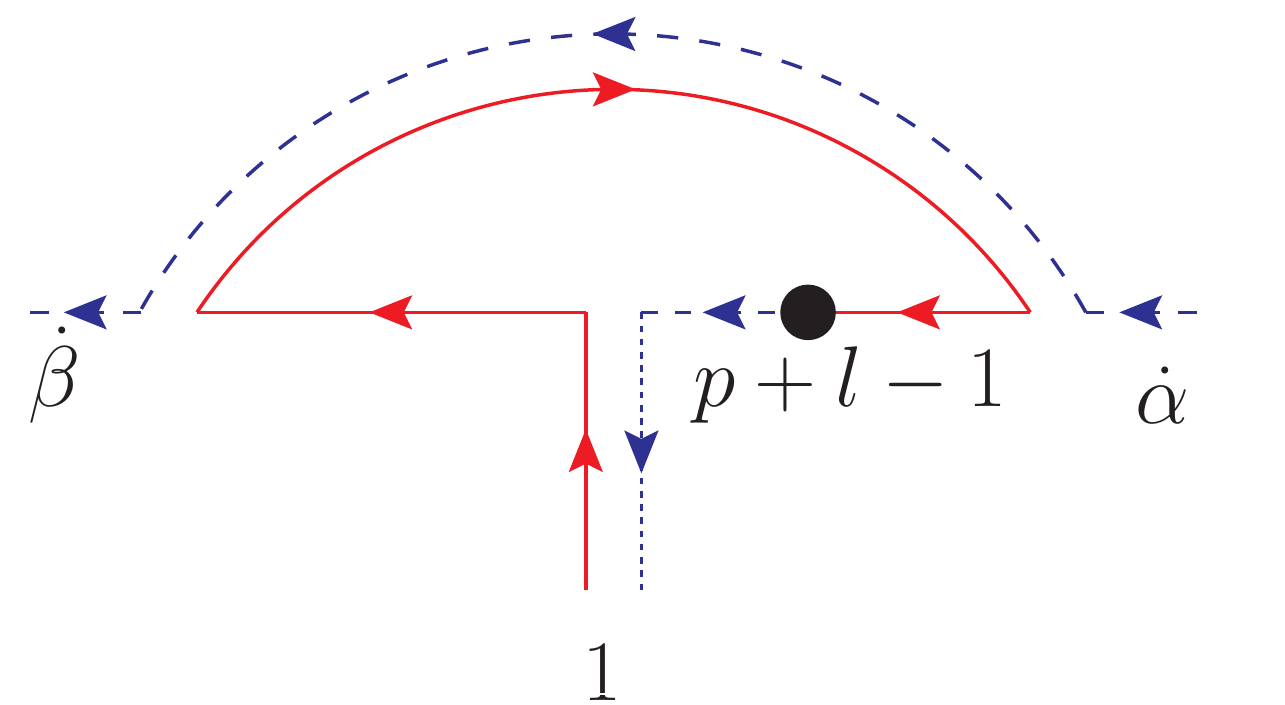}~,
	\label{eq:even jellyfish example}
\end{align}
where the chirality-flow arrows go against the fermion-flow arrows in order to get the correct sign,
as in massive chirality flow \cite{Alnefjord:2020xqr}.

While we have given only half of the chiralities explicitly, 
swapping the chiralities of the fermions in the above examples is trivial,
since it is equivalent to swapping all solid lines for dotted ones and vice versa.\footnote{
	If the theory is chiral then the chiral couplings must also be exchanged.}

\paragraph*{Loops with two or more fermion lines:}
We now consider loops with two or more fermion lines. 
To set the chirality-flow lines consistently in these examples is analogous to the procedure in \cite{Alnefjord:2020xqr},
and is perhaps easiest seen by example (we will give a systematic treatment afterwards).
Consider the following box diagram, redrawn as a chirality-flow diagram,
\begin{align}
	\includegraphics[scale=0.35,valign=c]{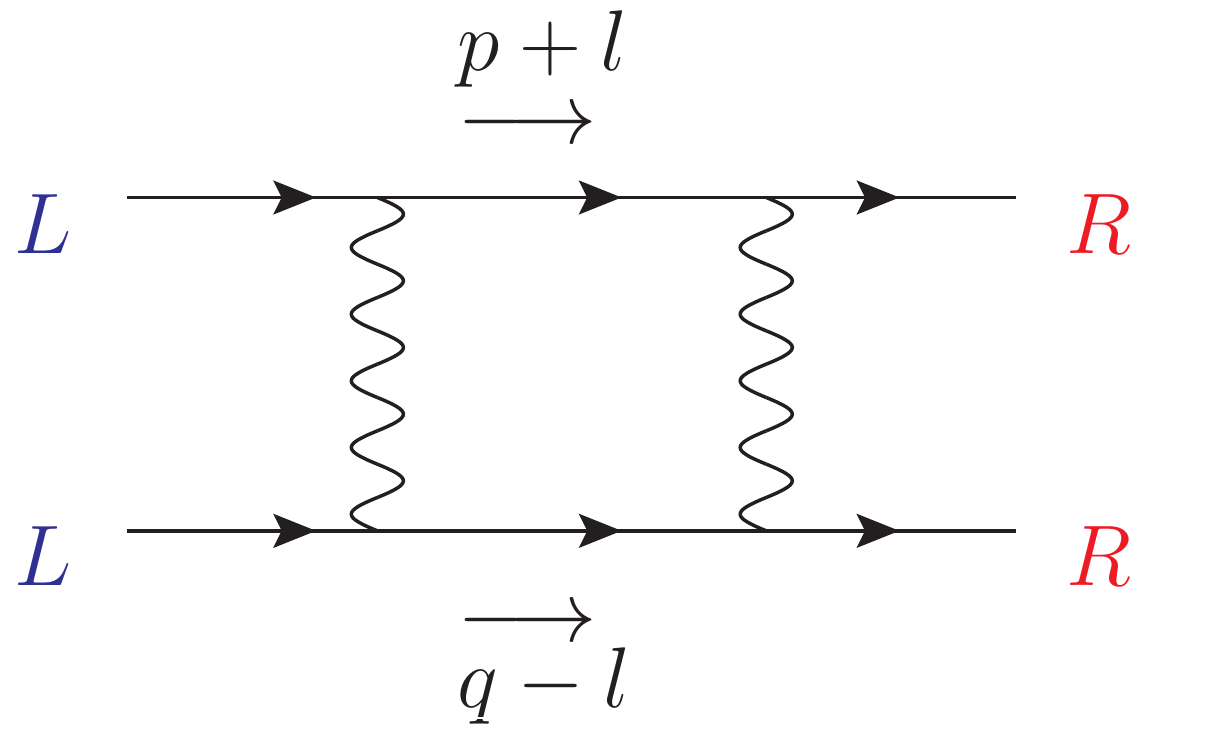} &\rightarrow \,
	\includegraphics[scale=0.35,valign=c]{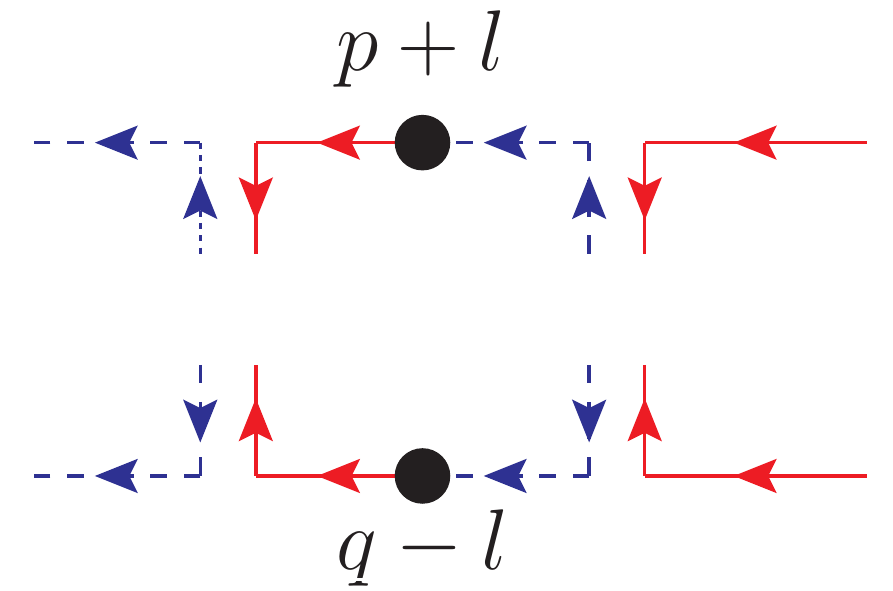}~,
\end{align}
where the chirality-flow lines have chirality-flow arrows opposing fermion-flow arrows,
and photons are drawn as double lines which are not yet connected.
We note that we cannot connect the photons' flow lines yet, since their arrow directions do not match.
To fix this, we use that the chirality-flow lines will always (at least eventually) end with spinors.
Labeling the momenta of these end spinors $i$ and $j$,
we can use
\begin{align}
	\lanSp{i}\taubar^\mu \Big(\sla{q}-\sla{l}\Big)\taubar^\nu \rsqSp{j} = 
	\lsqSp{j}\tau^\nu \Big(\slabar{q}-\slabar{l}\Big)\tau^\mu \ranSp{i}~, 
\end{align}
an example of \eqref{eq:arrow swaps}, 
to swap the chirality-flow arrows on the bottom line, obtaining
\begin{align}
	\includegraphics[scale=0.35,valign=c]{Jaxodraw/bubbleFeynLLRR} &\rightarrow \,
	\includegraphics[scale=0.35,valign=c]{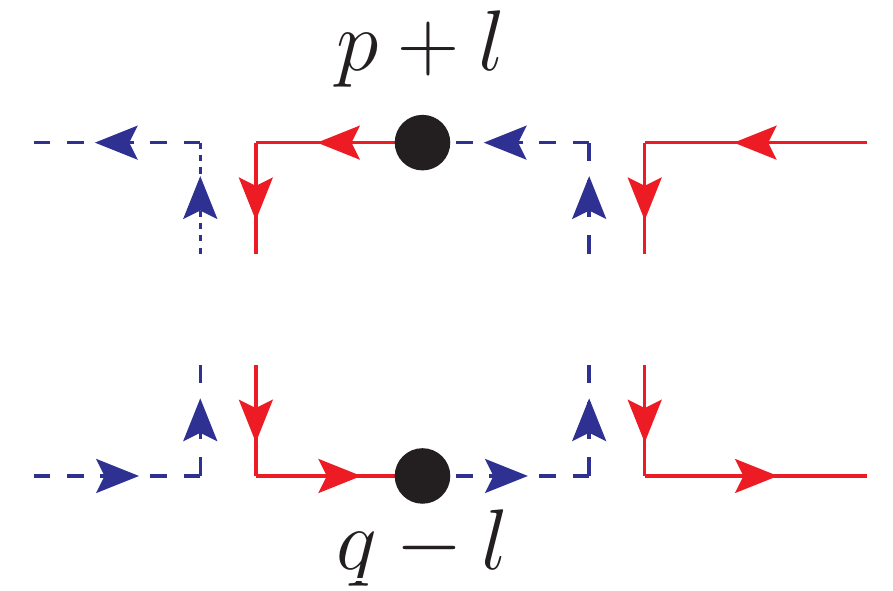}
	= \includegraphics[scale=0.35,valign=c]{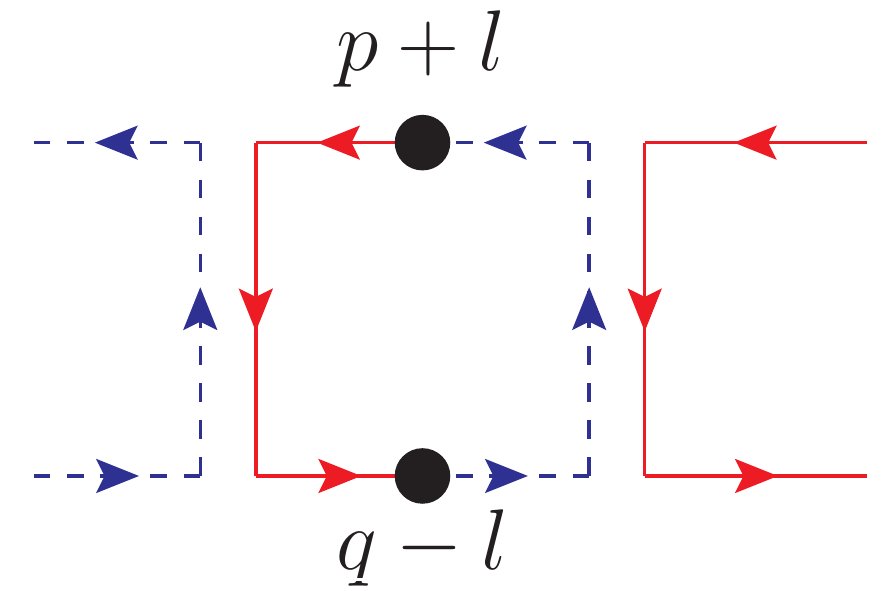}~,
	\label{eq:bubble simple completed}
\end{align}
where we first swapped the arrows, and then connected the photons using the Fierz identity,
leading to a completed chirality-flow diagram. 
Note that we have, for the first time in this paper, a closed chirality-flow string.\footnote{
	This concept is not new to loop calculations. For example, 
	tree-level QCD diagrams also contained traces of two Pauli matrices from the contraction of two momentum-dots from gluon vertices.}
Such chirality-flow strings are simply traces of Pauli matrices, 
or equivalently, strings of inner products.

If the top fermion line of the box diagram is massive, 
for a given helicity configuration we will also have to calculate chirality-flow diagrams like
\begin{align}
	\includegraphics[scale=0.35,valign=c]{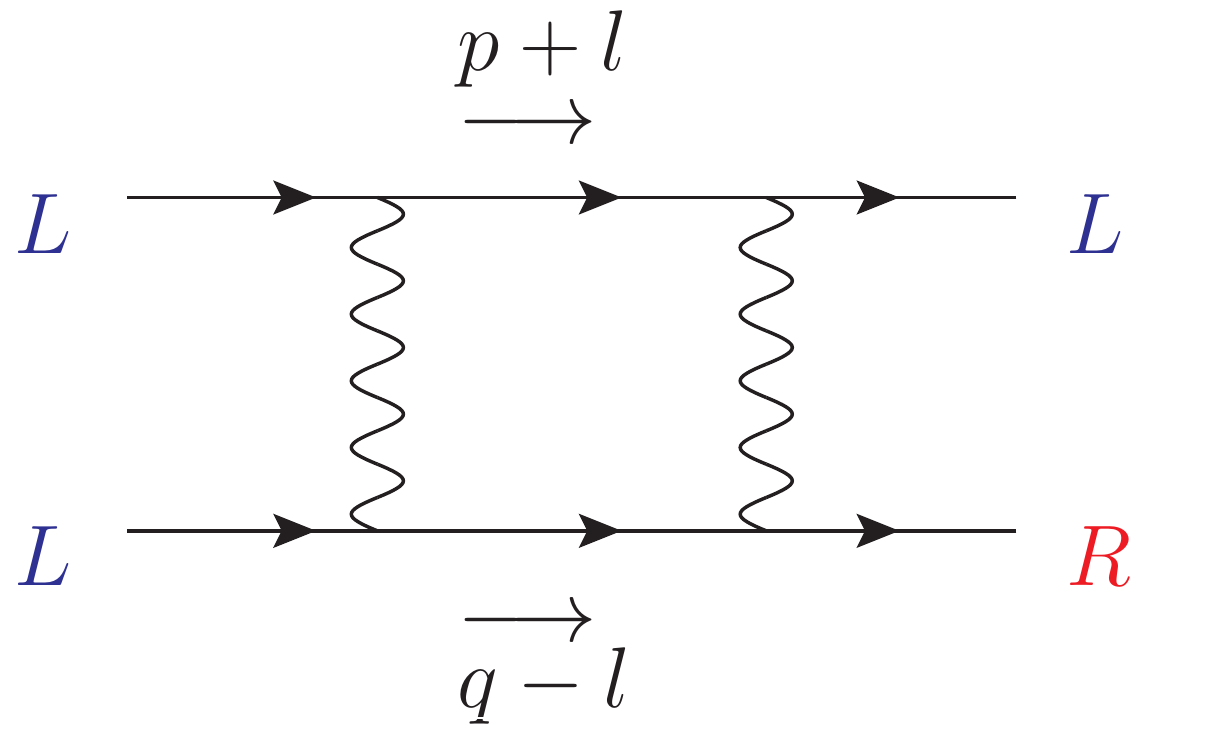} &\rightarrow \,
	\raisebox{-0.53\height}{\includegraphics[scale=0.35]{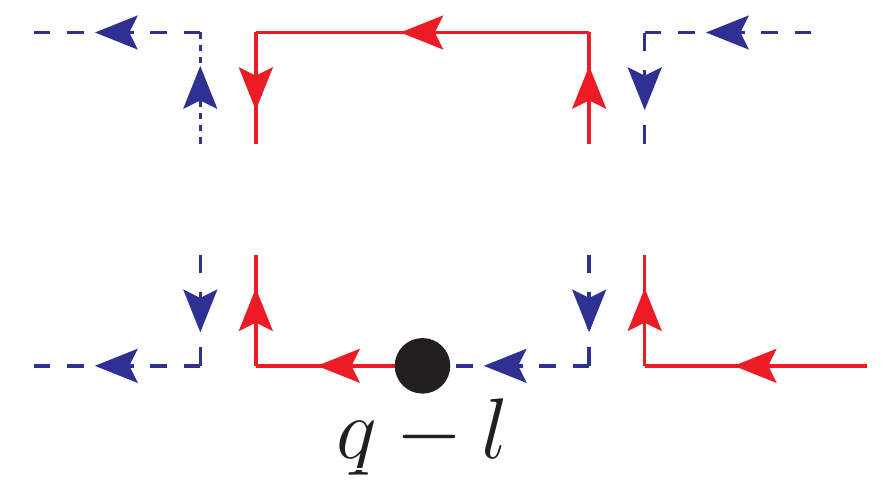}}~.
\end{align}
In this case, we can already use the Fierz identity to join the rightmost photon, obtaining
\begin{align}
	\includegraphics[scale=0.35,valign=c]{Jaxodraw/bubbleFeynLLLR} &\rightarrow \,
	\raisebox{-0.53\height}{\includegraphics[scale=0.35]{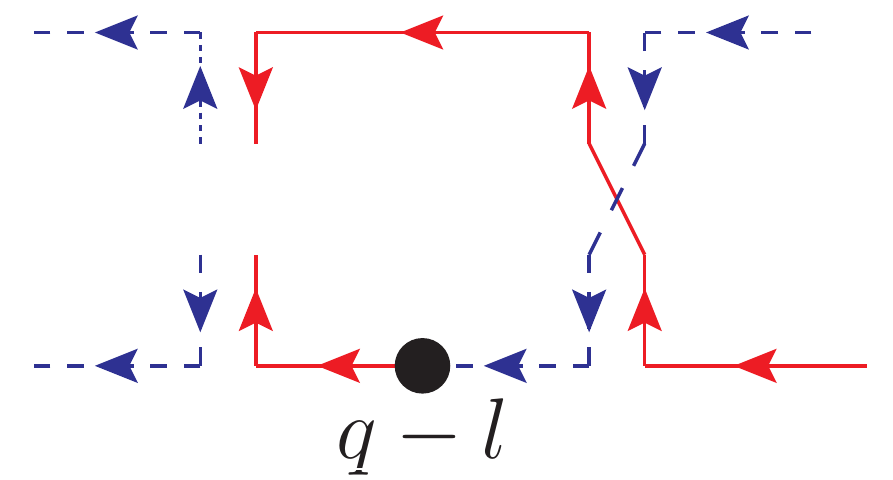}}~,
\end{align}
which has two strings of chirality-flow lines, 
one from the bottom right to the top left (string 1), 
and the other from the top right to the bottom left (string 2). 
To connect these two together, we need to use \eqref{eq:arrow swaps} to flip the arrows on one string or the other. 
If we flip it on the string of lines containing the momentum-dot, there is an odd number of inner products, 
so a minus sign will be introduced. 
Alternatively, we can flip the arrows on string 1 which has two inner products, 
and therefore will not introduce a minus sign, obtaining
\begin{align}
	\includegraphics[scale=0.35,valign=c]{Jaxodraw/bubbleFeynLLLR} &\rightarrow \,
	\raisebox{-0.53\height}{\includegraphics[scale=0.35]{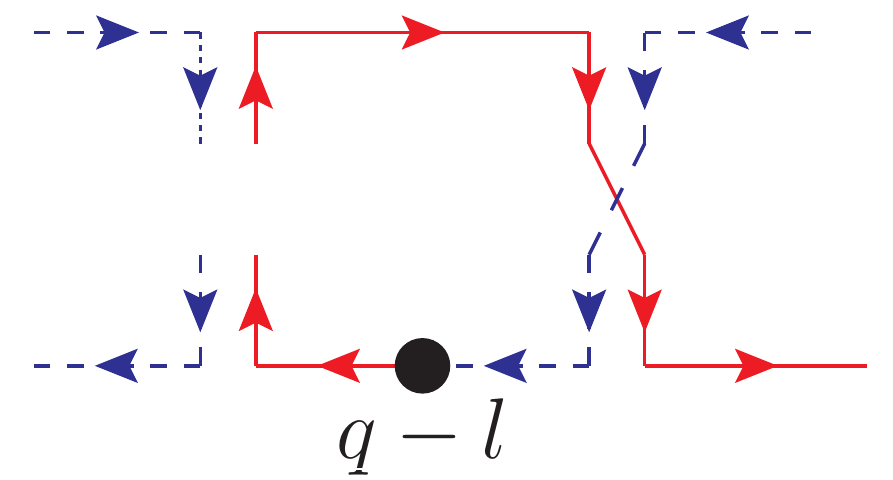}} = 
	\raisebox{-0.53\height}{\includegraphics[scale=0.35]{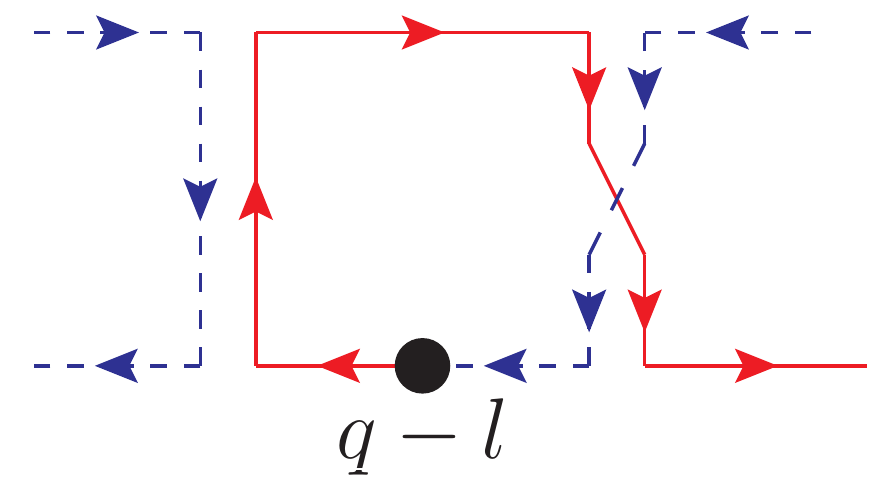}}~.
\end{align}
Note that the final result is independent of where we swap the arrows.
\pagebreak

In a general one-loop abelian Feynman diagram with multiple fermion lines in the loop,
the general procedure to set the arrow directions is:
\begin{enumerate}
	\item Draw the chirality-flow lines for each fermion line with chirality flow opposing fermion flow.
	(Do not use the Fierz identity to attach fermion lines together yet.)
	\item Choose a photon to be Fierzed first. 
	If needed, use \eqref{eq:arrow swaps}
	to swap the arrow directions of one of the fermion lines containing this photon,
	and then use the Fierz identity to connect the flow lines.
	Note that the chirality-flow lines will always, eventually, end in spinors, 
	even if these spinor ends are not explicitly included in the loop calculation,
	so \eqref{eq:arrow swaps} will always hold.
	\label{step:photon fierz}
	\item To Fierz further photons, either repeat step \ref{step:photon fierz} 
	with the fermion lines replaced by the strings of chirality-flow lines,
	or use one of \eqsref{eq:odd jellyfish} or (\ref{eq:even jellyfish}),
	whichever is appropriate.
	\label{step:later photon fierz}
	\item Repeat step \ref{step:later photon fierz} until all photons are replaced with joined chirality-flow lines. 
	This will give a completed one-loop chirality-flow diagram with fully simplified Lorentz structure.
\end{enumerate}

\paragraph*{General $\Rxi$ gauge:}
Since the gauge-parameter (in)dependence is an important cross-check of any perturbative calculation,
we here comment on the general \Rxi{} gauge, for which the photon propagator is 
\begin{align}
	\raisebox{-0.2\height}{\includegraphics[scale=0.3]{Jaxodraw/PhotonProp}} = -\frac{i}{p^2_{\dimens{4}}}\left( \metfour - (1-\xi)\frac{\pfour p^\nu_{\dimens{4}}}{p^2_{\dimens{4}}} \right)~,
	\label{eq:photon propagator Rxi 4d}
\end{align}
in four dimensions. 
This is straightforwardly translated into chirality flow by recalling that 
$p^\mu$ will always be contracted with a Pauli matrix, 
thus becoming a momentum-dot
(see \eqref{eq:mom flow})
\begin{align}
	\raisebox{-0.2\height}{\includegraphics[scale=0.3]{Jaxodraw/PhotonProp}} &\rightarrow \,\,
	-\frac{i}{p^2_{\dimens{4}}}\left(\raisebox{-0.2\height}{\includegraphics[scale=0.3]{Jaxodraw/PhotonPropFlowa}} \ - \ (1-\xi)\frac{1}{2p^2_{\dimens{4}}} 
	\raisebox{-0.2\height}{\includegraphics[scale=0.3]{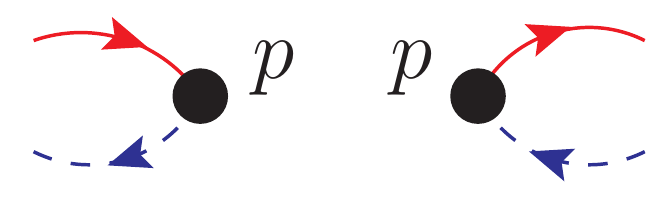}}\right) ~, \quad \text{or}
	\nonumber \\
	&\rightarrow \,\, -\frac{i}{p^2_{\dimens{4}}}\left(\raisebox{-0.2\height}{\includegraphics[scale=0.3]{Jaxodraw/PhotonPropFlowb}} \ - \ (1-\xi)\frac{1}{2p^2_{\dimens{4}}} 
	\raisebox{-0.2\height}{\includegraphics[scale=0.3]{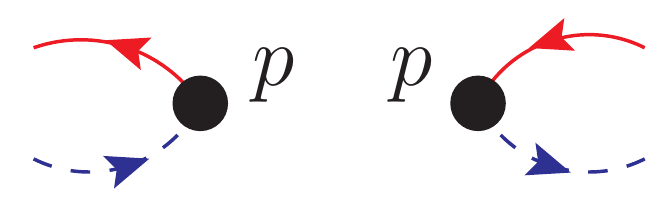}}\right) ~.
	\label{eq:photon propagator Rxi flow}
\end{align}

If the internal photon is part of a loop, 
\eqref{eq:photon propagator Rxi 4d}  is modified using the \mteSRs{}, 
\eqsrefa{eq:2eps selection rules}{eq:2eps selection rules algebra},
to also add the propagator of the FDF scalar
\begin{align}
	\raisebox{-0.2\height}{\includegraphics[scale=0.3]{Jaxodraw/PhotonProp}} &= -\frac{i}{p^2_{\dimens{d}}}\left( \metds - (1-\xi)\frac{\pd p^\nu_{\dimens{d}}}{p^2_{\dimens{d}}} \right)
	\nonumber \\
	&= \underbrace{-\frac{i}{p^2_{\dimens{d}}}\left( \metfour - (1-\xi)\frac{\pfour p^\nu_{\dimens{4}}}{p^2_{\dimens{d}}} \right)}_{\text{4d photon}} \quad \, 
	 \underbrace{-\frac{i}{p^2_{\dimens{d}}}\left(G^{MN} + \mu^2(1-\xi)\frac{Q^M Q^N}{p^2_{\dimens{d}}} \right)}_{\text{FDF scalar}}~.
\end{align}
Treating the FDF-scalar as a separate particle to the photon, we obtain \eqref{eq:photon propagator Rxi flow}
as the flow representation of the photon,
and a trivial prefactor as the flow representation of the FDF-scalar. 

\paragraph*{FDF scalars:}
One key feature of FDF is the appearance of FDF scalars, 
which have their own Feynman rules (see \cite{Fazio:2014xea}).
In abelian theories, there are two new rules,
the propagator for the FDF scalar just discussed,
and the interaction vertex of the FDF-scalar with a fermion
\begin{align}
	\includegraphics[scale=0.3,valign=c]{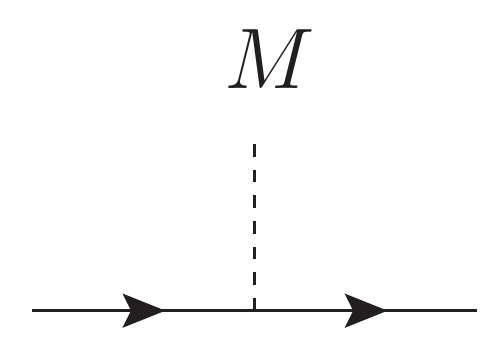} &= ie\ga^5\Ga^M 
	=  \, ie\Ga^M \begin{pmatrix}
		 -\includegraphics[scale=0.3,valign=c]{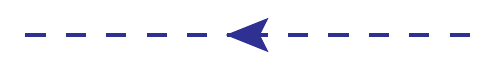} & 0 \\
		 0 & \includegraphics[scale=0.3,valign=c]{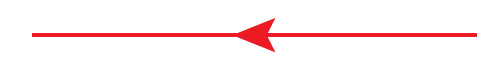}
	\end{pmatrix} ~,
\end{align}
where the flow rule is the same as any scalar (up to the coupling) and the sign on the left-chiral part coming from $\ga^5$.

Looking ahead to non-abelian theories, we find that the Feynman rules for
FDF-scalars only add ingredients we already know how to treat, such as
four-dimensional metrics (\eqref{eq:vbpropagator}) and
four-dimensional momenta (\eqref{eq:mom flow}), together with some of
the \mteSRs{} which will contract to either $0$ or $\pm 1$.
Therefore, FDF scalars are a non-complication from the perspective of
chirality flow. We also note that counterterms do not introduce any
new structure which the chirality-flow formalism would not be able to
handle.

\subsection{Non-abelian gauge theories}
\label{sec:non-abelian theories}
To go from loops in an abelian theory to those in a non-abelian theory is straightforward in chirality flow. 
As we will see below, there is essentially nothing new to the non-abelian case compared to the abelian case, 
only more terms to keep track of.

\subsubsection{QCD}
The aim of this section is to explore what chirality-flow structures we obtain in QCD,
and argue that we can always consistently set chirality-flow arrows.
We will see that all non-abelian Feynman diagrams have a Lorentz structure which is either composed of the tree-level structures already described in 
\cite{Lifson:2020pai, Alnefjord:2020xqr}, 
or of the same structures as in an abelian gauge theory like QED, discussed in the previous section.

Compared to abelian gauge theories like QED, there are two new features in QCD:
the addition of ghosts, and the non-abelian vertices. 
The ghosts have a scalar propagator and couple to a gluon with a momentum which is easily flowed
(see \eqref{eq:mom flow}),
so they are not a complication.

The non-abelian vertices are built from Lorentz structures we have already treated.
This can be seen using a simple illustrative example like
\begin{align}
	\includegraphics[scale=0.3,valign=c]{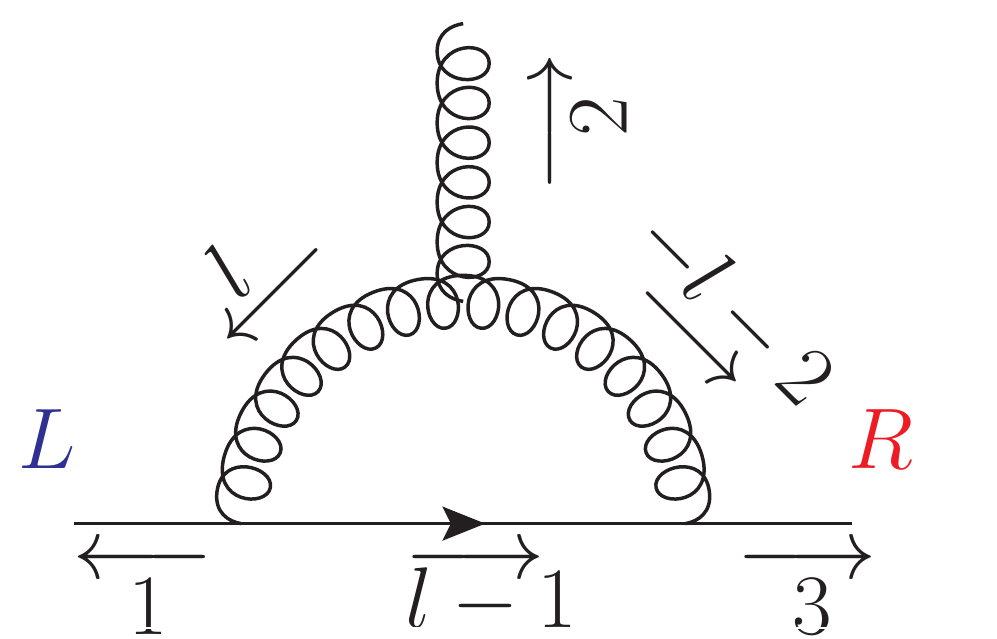}~.
	\label{eq:fermion-gluon triangle diag}
\end{align}
To understand the flow structure of this diagram, 
we first recall the Lorentz structure of the triple gluon vertex
\begin{align}
	\includegraphics[scale=0.35,valign=c]{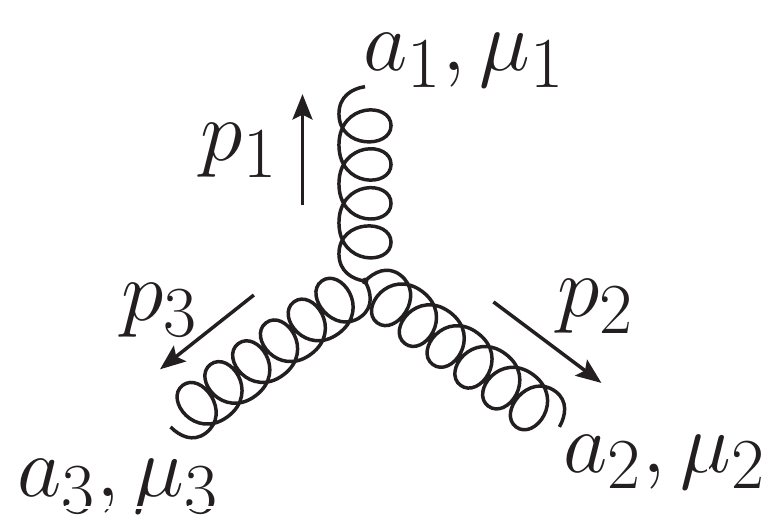}
\!\!\!
&=
\;
i\frac{g_s}{\sqrt{2}} 
if^{a_1a_2a_3}
	\Big(
	g^{\mu_1\mu_2}(p_1-p_2)^{\mu_3}
	+g^{\mu_2\mu_3}(p_2-p_3)^{\mu_1}
	+g^{\mu_3\mu_1}(p_3-p_1)^{\mu_2}
	\Big),
	\nonumber \\
&\rightarrow\;
i\frac{g_s}{\sqrt{2}} 
if^{a_1a_2a_3}
	\left(
\includegraphics[scale=0.325,valign=c]{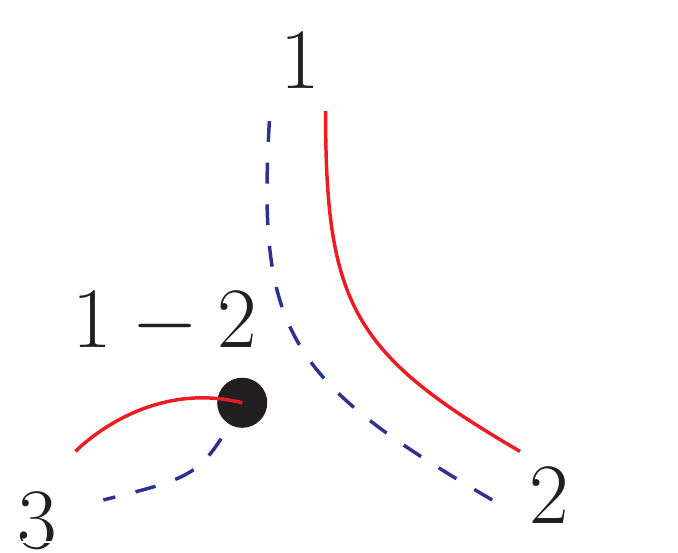} \!\!\!
+ \includegraphics[scale=0.325,valign=c]{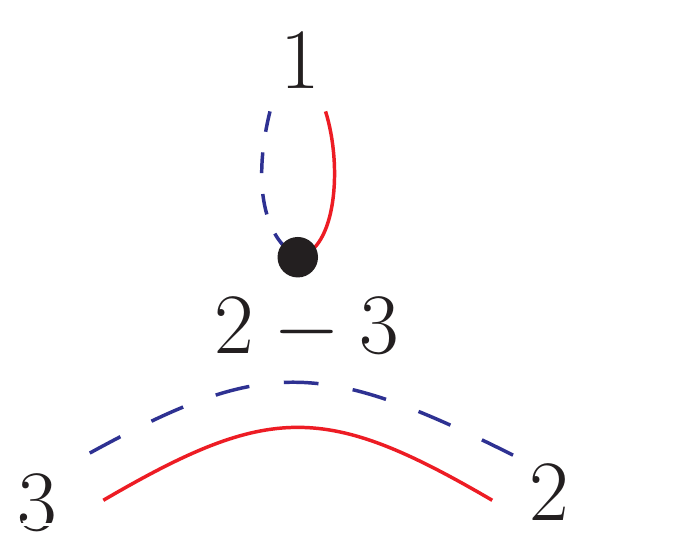} \!\!\!
+ \includegraphics[scale=0.325,valign=c]{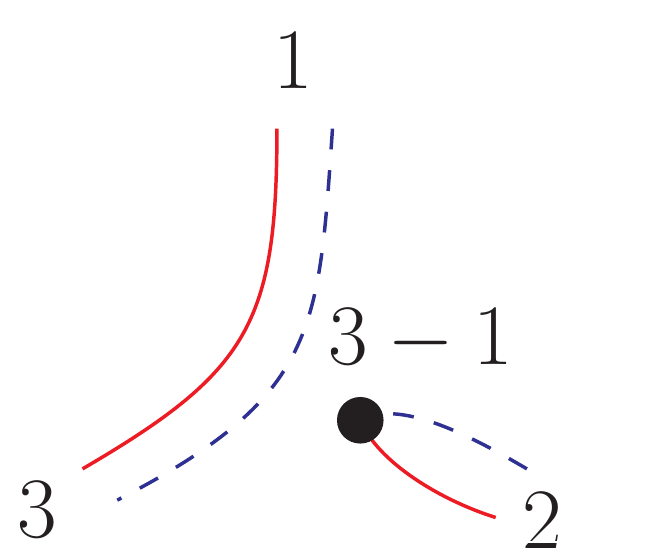}\!\!\!
\right)~.
\label{eq:trpgluvrt}	
\end{align}
Applying this vertex to the loop diagram in \eqref{eq:fermion-gluon triangle diag},
we find (using Feynman gauge and a massless fermion for simplicity)
\begin{align}
	\includegraphics[scale=0.3,valign=c]{Jaxodraw/FermionExtGluonLoopLR} &\rightarrow \,\,
	\includegraphics[scale=0.3,valign=c]{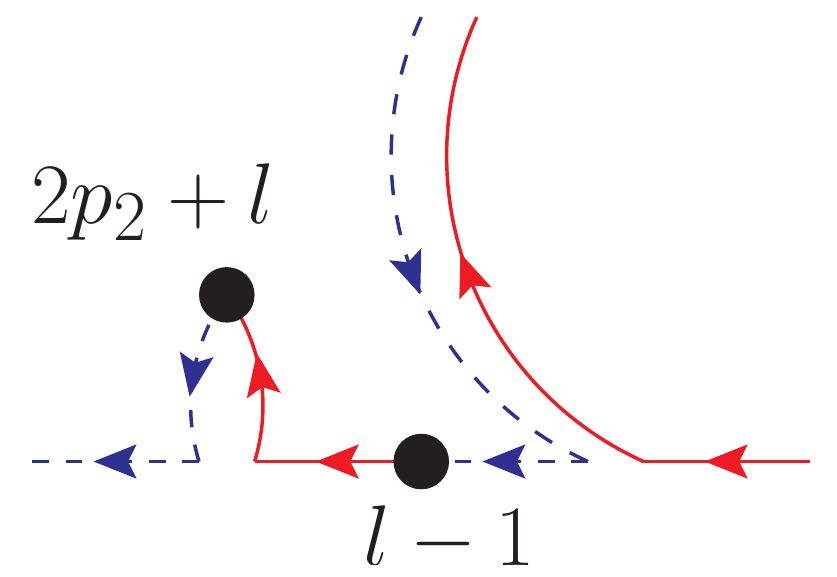}
	- \includegraphics[scale=0.3,valign=c]{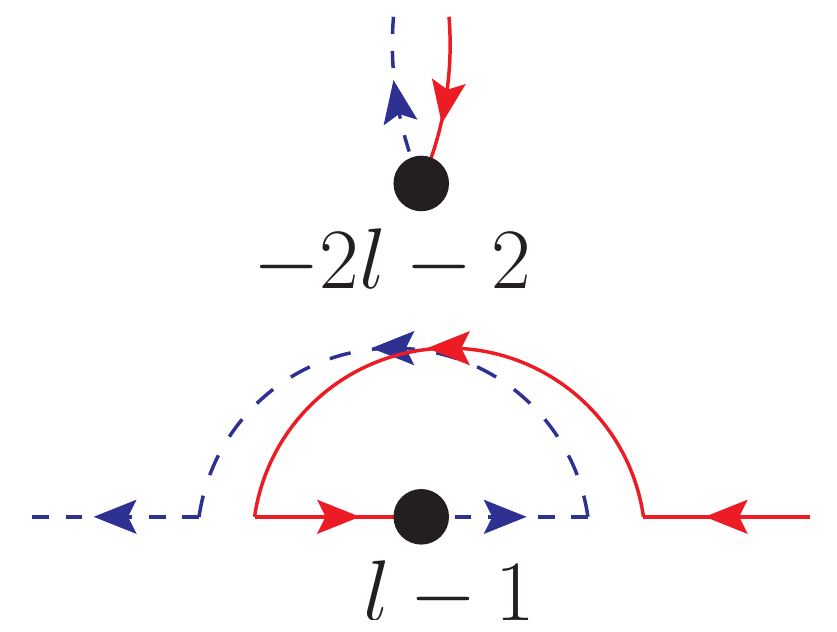}
	+ \includegraphics[scale=0.3,valign=c]{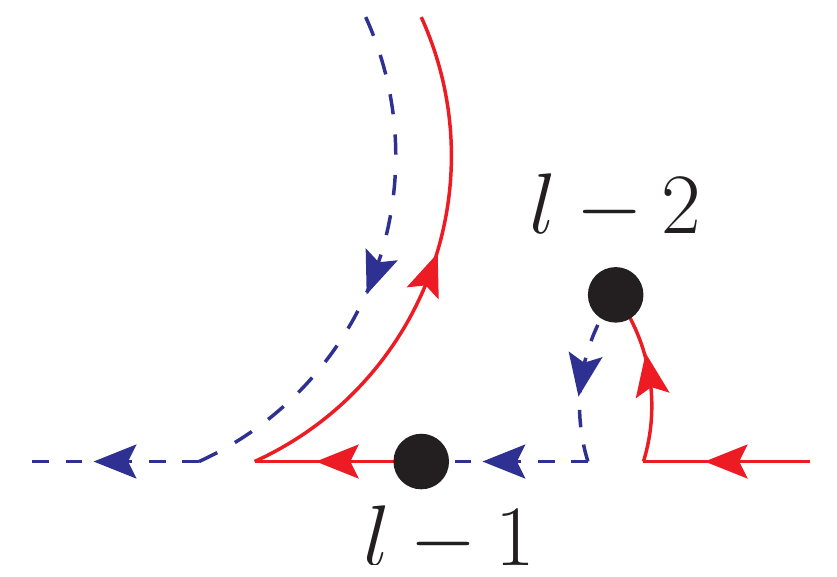} ~,
\end{align}
for the Lorentz structure. 
Here we see that the first and third terms already have tree-level Lorentz structure, 
while the middle term has the ``abelian'' Lorentz structure from \eqref{eq:fermion self energy flow}
(hence the minus sign),
multiplied by a disconnected structure\footnote{
	Recall that the arrow directions of disconnected chirality-flow structures can be set independently \cite{Lifson:2020pai}.}.
Therefore, all of the Lorentz structures in this QCD diagram are either tree-level structures from \cite{Lifson:2020pai},
or built from the Lorentz structures already encountered in abelian gauge theories,
and setting a consistent arrow direction is straightforward.

This conclusion holds for any general QCD one-loop diagram.
This is because the three- and four-gluon vertices, \eqsrefa{eq:trpgluvrt}{eq:four_gluon_vertex},
break up the Lorentz structure into simpler pieces which either:
break the loop Lorentz structure into a tree-level Lorentz structure,
or create loop Lorentz structures of the form of \eqsrefa{eq:odd jellyfish}{eq:even jellyfish}.
Further, the ghosts and FDF scalars only add 4d metrics and momenta multiplied by \mteSRs{} objects,
so again break up the Lorentz structure into either trees or structures like \eqsrefa{eq:odd jellyfish}{eq:even jellyfish}.

Finally, since the external gluons can have either set of arrow directions without introducing a minus sign,
setting arrows is either the same as at tree level,
or follows the abelian case in \secref{sec:abelian theories}.

\subsubsection{Other non-abelian theories}
\label{sec:other non-abelian theories}
Other non-abelian theories will similarly be easily flowable as long as they contain Feynman rules with 4d objects like scalars, Weyl and Dirac spinors,
polarization vectors, momenta, the metric, $\ga^\mu$, and $\ga^5$, together with the \mteSRs{}. 
In the Standard Model, the other non-abelian theory is the electroweak theory,
which behaves like QCD but has the addition of massive polarization vectors, chiral vertices, 
and more (loop-level) scalars.

Chiral vertices and massive polarization vectors were both discussed in \cite{Alnefjord:2020xqr}.
The chiral vertices are easiest understood by drawing the chirality-flow arrow opposing the fermion-flow arrow,
giving
\begin{align}
	\ga^\mu P_{\blue{L}} &\rightarrow\,\, \sqrt{2} \includegraphics[scale=0.3,valign=c]{Jaxodraw/FermionVectorVertexL}~,
	&
	\ga^\mu P_{\red{R}} &\rightarrow\,\, \sqrt{2} \includegraphics[scale=0.3,valign=c]{Jaxodraw/FermionVectorVertexR}~,
\end{align}
for $P_{\red{R}/\blue{L}} = (1\cpm\ga^5)/2$.
After having made this assignment, all arrow swaps can be done as previously described. 

The massive polarization vectors are given in \eqref{eq:pol vecs massive flow},
with the transverse polarization vectors having the same structure as massless polarization vectors,
while the longitudinal polarization vector corresponds to a momentum-dot.
Therefore, one consequence of longitudinal polarization vectors is that we can obtain closed chirality-flow strings.
For example, the axial anomaly with longitudinally-polarized $W$ bosons has the following Lorentz structure
\begin{align}
	i(1_\mu + 2_\mu) \includegraphics[scale=0.26,valign=c]{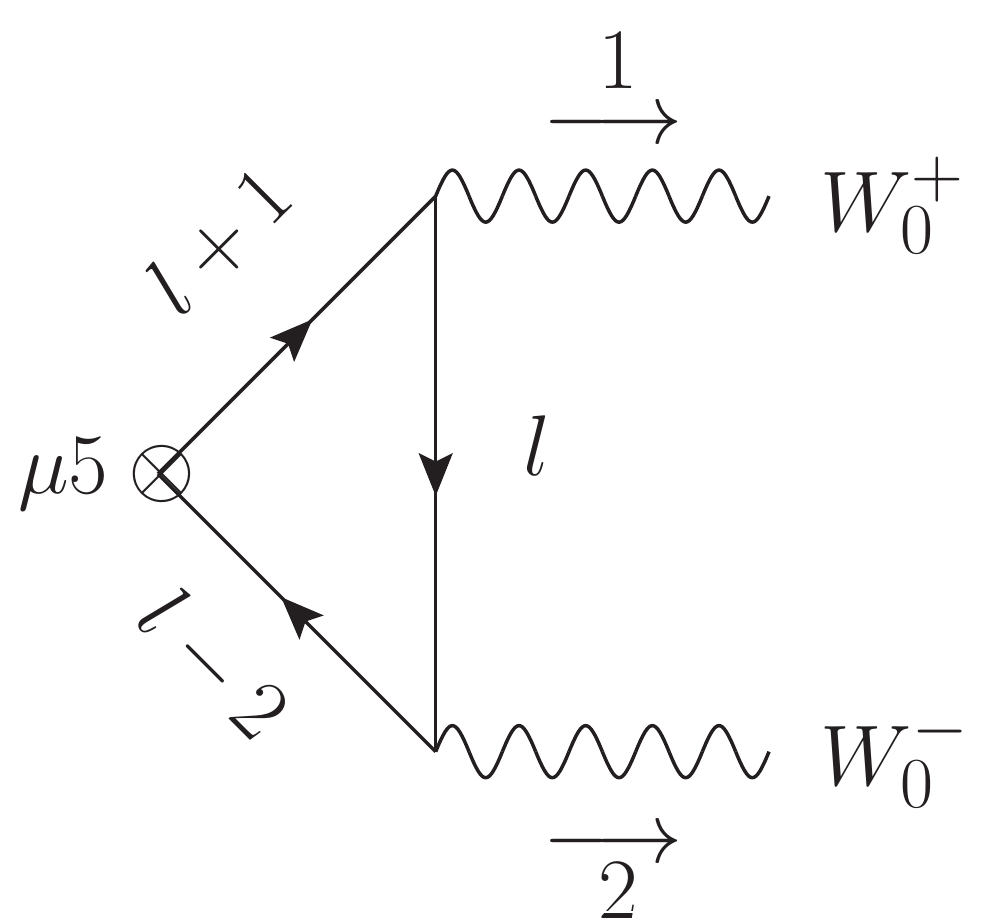} &\rightarrow\,\,
	\includegraphics[scale=0.26,valign=c]{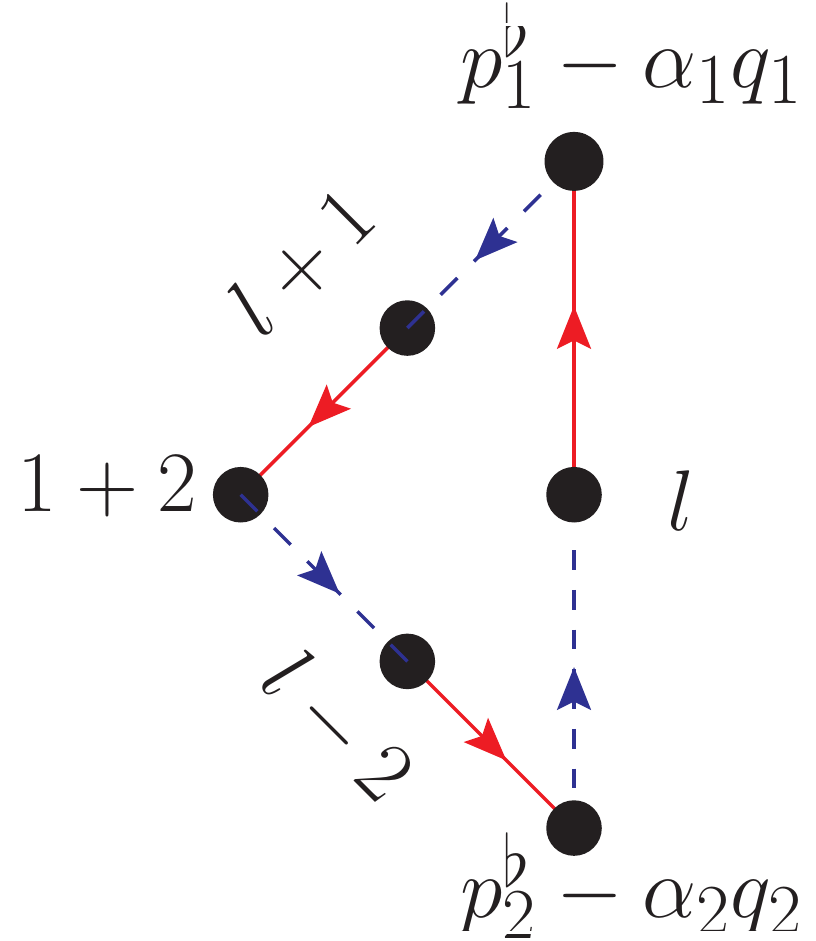}
	+ \mu^2 \includegraphics[scale=0.26,valign=c]{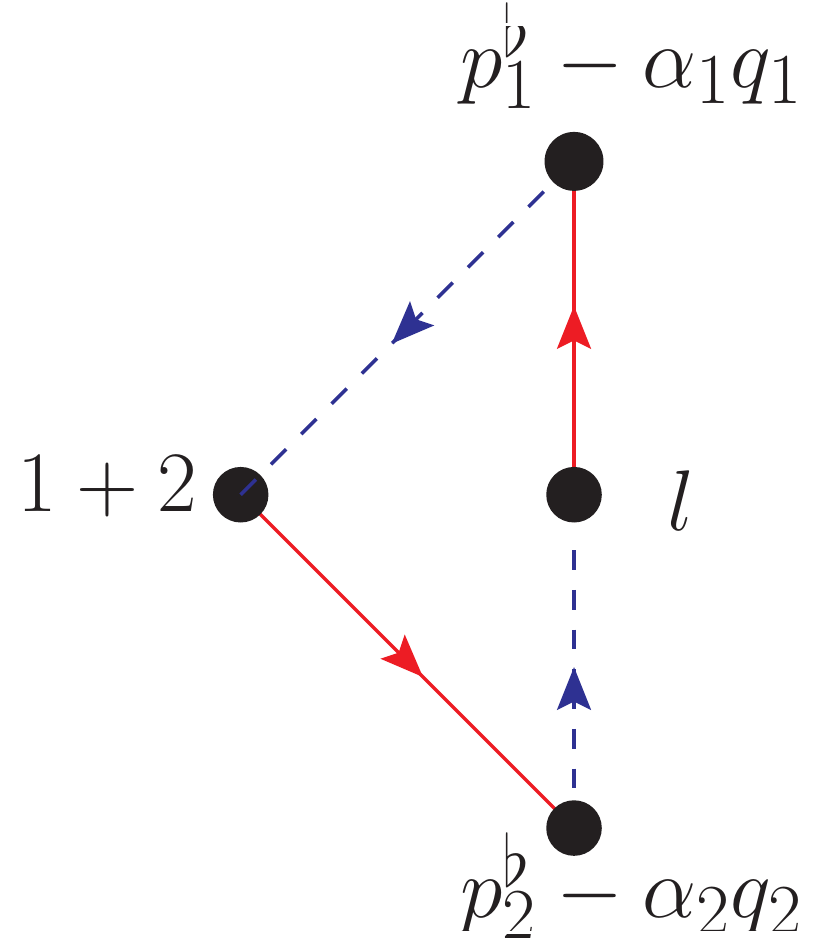}~,
\end{align}
where both terms have closed chirality-flow strings.
(The chiral projectors ensure that these are the only two Lorentz structures in the calculation.)
These closed flow loops can either be written as sums of spinor products or as traces of Pauli matrices.

Finally, we comment on the fermion self-energy in a theory with a spontaneously broken symmetry like the Standard Model.
In the Feynman gauge, as shown in \eqref{eq:fermion self energy flow},
the contribution from a boson of mass $m$ to the self-energy is
\begin{align}
	\includegraphics[scale=0.3,valign=c]{Jaxodraw/fermionSelfEnergyPhotonLR} &\sim \,
	g^2\lint \frac{\raisebox{-0.2\height}{\includegraphics[scale=0.3]{Jaxodraw/fermionSelfEnergyFlowLR}}}{
	(l^2-m^2)(l+p)^2}~,
	\label{eq:fermion self energy Z}
\end{align}
where we label the generic coupling factors as $g^2$ and ignore factors of $-1$, $i$ and $\sqrt{2}$. 
Note that it is obvious in chirality flow that this has exactly the same structure as the contribution from a Goldstone boson of mass $m_G = m$
\begin{align}
	\includegraphics[scale=0.3,valign=c]{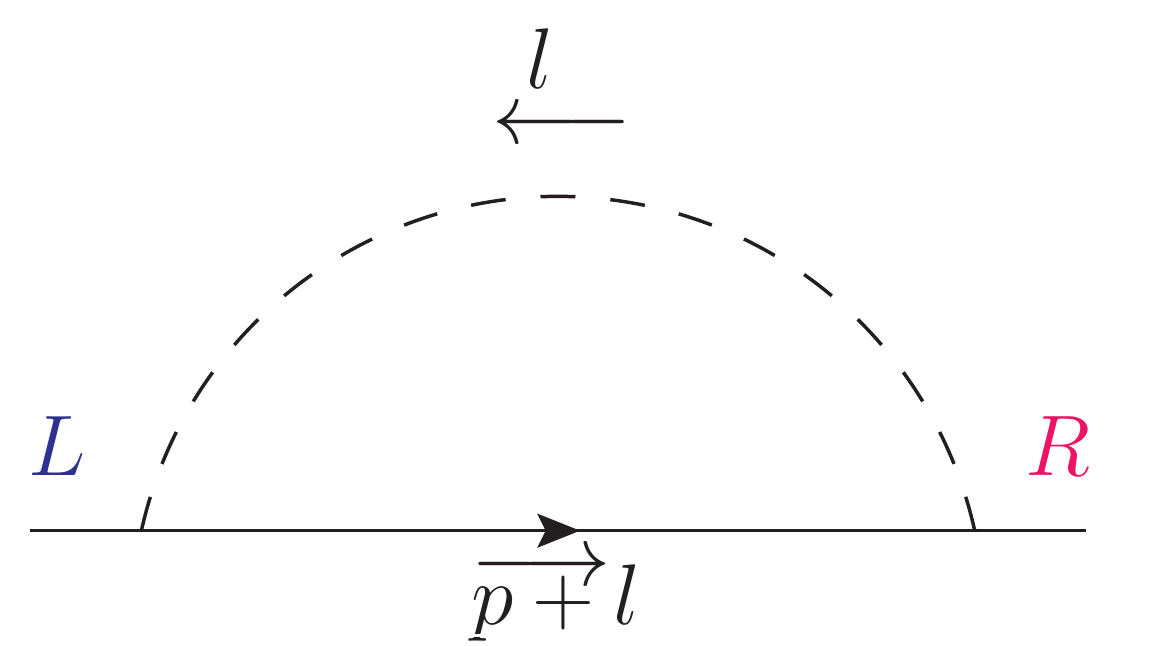} &\sim \,
	(g')^2\lint \frac{\raisebox{-0.2\height}{\includegraphics[scale=0.3]{Jaxodraw/fermionSelfEnergyFlowLR}}}{
		(l^2-m^2)(l+p)^2}~,
	\label{eq:fermion self energy chi}
\end{align}
where $g' \neq g$ is the coupling of the Goldstone to the fermion.
This type of transparency is one of the
nice features of chirality flow at both tree and one-loop level.

Further, in a general $\Rxi$ gauge, we can use relations such as
\eqref{eq:example tensor rank reductions} to relate the extra terms to
the Goldstone contribution.  
We envisage that the cancellation of gauge-parameter dependence can thus
be made more manifest by using chirality flow.

\section{Conclusion}
\label{conclusion}

In this paper we have studied the chirality-flow formalism at one-loop order.
We conclude that many of the simplifications seen at tree-level can be retained,
at least in the four-dimensional formulation of the 4$d$ helicity scheme.
In particular, the Lorentz algebra can be elegantly simplified, and
Feynman diagrams can be made to vanish by picking adequate reference
vectors for external gauge bosons and massive spinors.

Beyond this, we find that the tensor reduction procedure is also simplified for one of two reasons.
Either, because we reduce the number of required coefficients in the tensor decomposition,
or because we reduce the rank of the tensor integral. 
The former may happen either directly due to the Weyl equation or when applying the Fierz identity,
while the latter occurs when multiple momentum-dots containing the loop momentum are contracted together.

\section*{Acknowledgments}
AL and MS acknowledges support by the Swedish Research Council (contract
number 2016-05996, as well as the European Union’s Horizon 2020
research and innovation programme (grant agreement No 668679).  The authors
have in part also been supported by the European Union’s Horizon
2020 research and innovation programme as part of the Marie
Sklodowska-Curie Innovative Training Network MCnetITN3 (grant
agreement no. 722104).  

\appendix
\section{Additional chirality-flow rules}
\label{sec:additional rules}

In this section we collect the chirality-flow rules required for this paper which are not stated in the main text.
Additional chirality-flow rules can be found in \cite{Lifson:2020pai,Alnefjord:2020xqr}.

We begin with some algebra relations required to prove \eqsrefd{eq:fermion self energy taus}{eq:even jellyfish example}.
The vector indices of the Pauli matrices can be contracted using
\begin{align}
	\tau_\mu^{\dal\be}\taubar^\mu_{\ga\deta}
	&=\delta_{\ga}^{\ \be}\delta_{\ \deta}^{\dal}\,,
	&
	\taubar^\mu_{\al\dbe}\taubar_{\mu,\ga\deta}
	&=\eps_{\al\ga}\eps_{\dbe\deta}\,,
	&
	\tau^{\mu,\dal\be}\tau_{\mu}^{\dga\eta}
	&=\eps^{\dal\dga}\eps^{\be\eta}\,,
		\label{eq:taufierz}
\end{align}
while a $\tau$ can be turned into a $\taubar$ or vice versa using
\begin{align}
	\taubar^\mu_{\al\dbe} 
	&= 
	\epsilon_{\al\ga}\epsilon_{\dbe\deta}\tau^{\mu,\deta\ga}~,
	&
	\tau^{\mu,\dal\be} 
	&= 
	\epsilon^{\dal\dga}\epsilon^{\be\eta}\taubar^\mu_{\eta\dga}~,
	\label{eq:tau taubar}
\end{align}
where the index positions are crucial, because
\begin{align}
	\eps_{\al\be}\eps^{\be\ga} &=\delta_{\al}^{\ \ga}~,
	&
	\eps^{\dal\dbe}\eps_{\dbe\dga} &= \delta^{\dal}_{\ \dga}~,
	&
	\eps_{\al\be} &= - \eps_{\be\al}~,
	&
	\eps_{\dbe\dga} &= -\eps_{\dga\dbe}~.
	\label{eq:eps relations}
\end{align}

Next, we recall the equations behind arrow flips. 
A string of chirality-flow lines can have all of its arrows flipped using one of
\begin{align}
		\lanSp{i} \taubar^{\mu_1} \dots \taubar^{\mu_{2n+1}} \rsqSp{j} &= 
	 \lsqSp{j} \tau^{\mu_{2n+1}} \dots \tau^{\mu_{1}} \ranSp{i}~,
	\nonumber \\
	\lanSp{i} \taubar^{\mu_1} \dots \tau^{\mu_{2n}} \ranSp{j} &= 
	- \lanSp{j} \taubar^{\mu_{2n}} \dots \tau^{\mu_{1}} \ranSp{i}~,
	\nonumber \\
	\lsqSp{i} \tau^{\mu_1} \dots \taubar^{\mu_{2n}} \rsqSp{j} &= 
	-\lsqSp{j} \tau^{\mu_{2n}} \dots \taubar^{\mu_{1}} \rsqSp{i} ~, 
	\label{eq:arrow swaps} 
\end{align}
where we note that the Pauli matrices stand in a sequence $\tau\taubar\tau\dots$ or $\taubar\tau\taubar\dots$.
In practice, when swapping the chirality-flow arrows on a string of flow lines, 
if the endpoints of the string have the same chirality (line type) then a minus sign is required,
while if the endpoints are of opposite chirality (line type) then no minus sign is needed.

In the spinor helicity formalism, for example when converting the axial anomaly from 
\eqref{eq:axial anomaly Peskin} to \eqref{eq:axial anomaly as spinor products},
it can be useful to write all objects as (sums of) outer products of spinors.
For example, every momentum can be written as 
\begin{align}
	\sla{p} \equiv p^\mu\si_\mu &\overset{p_i^2=0}{=} \sum_i \rsqSp{i}\lanSp{i}~,
	&
	\slabar{p} \equiv p^\mu\sibar_\mu &\overset{p_i^2=0}{=} \sum_i \ranSp{i}\lsqSp{i}~,
	\label{eq:mom outer products}
\end{align}
where we used that $p = \sum_i p_i$ with $p_i^2 = 0$
to write a massive momentum $p$ as a sum of massless ones
(see \eqref{eq:massive momentum decomposition}),
and the massless polarization vectors in \eqref{eq:pol vecs massless}
can be written as
\begin{align}
	\sla{\eps}_{\blue{L}}(p_i,r) &= \frac{\rsqSp{i}\lanSp{r}}{\lan r i \ran}
	&
	&\text{or}
	&
	\slabar{\eps}_{\blue{L}}(p_i,r) &=\frac{\ranSp{r}\lsqSp{i}}{\lan r i \ran} ~,\nonumber\\
	\sla{\eps}_{\red{R}}(p_i,r) &=\frac{\rsqSp{r}\lanSp{i}}{\sql i r \sqr}
	&
	&\text{or}
	&
	\slabar{\eps}_{\red{R}}(p_i,r) &=\frac{\ranSp{i}\lsqSp{r}}{\sql i r \sqr} ~.
	\label{eq:pol vecs massless outer prods}
\end{align}

In our discussion of non-abelian theories, we require the polarization vectors of outgoing massive particles
\begin{align}
	\left(\eps^\mu_{+}(p^\flat,q)\right)^*
	&\longrightarrow
	&
	&\frac{1}{\cAngle{qp^\flat}}
	\!\raisebox{-0.25\height}{\includegraphics[scale=0.355]{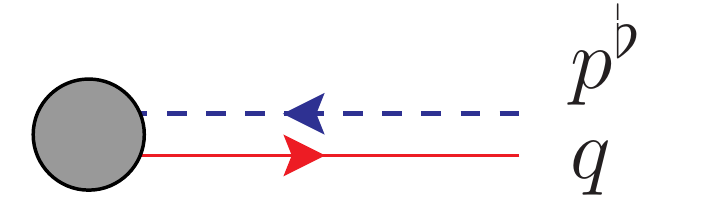}}
	&
	&\text{or}
	& 
	&\frac{1}{\cAngle{qp^\flat}}
	\!\raisebox{-0.25\height}{\includegraphics[scale=0.355]{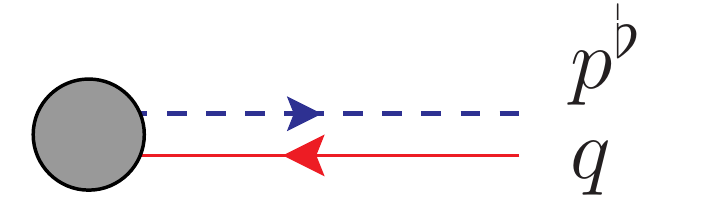}}
	\!\!\!\!, \nonumber
	\\
	\left(\eps^\mu_{-}(p^\flat,q)\right)^*
	&\longrightarrow
	&
	&\frac{1}{\cSquare{p^\flat q}}
	\raisebox{-0.25\height}{\includegraphics[scale=0.355]{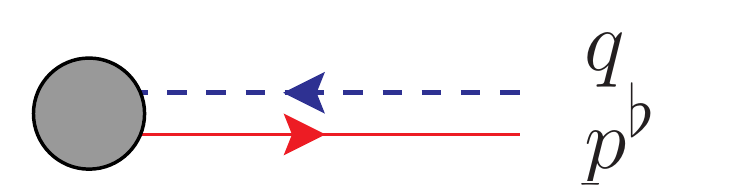}}
	&
	&\text{or}
	&
	&\frac{1}{\cSquare{p^\flat q}}
	\raisebox{-0.25\height}{\includegraphics[scale=0.355]{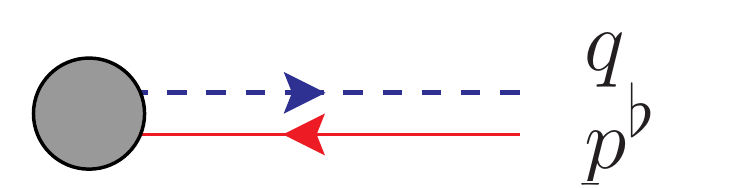}}
	\!\!\!\!,
	\nonumber \\
	\left(\eps^\mu_{0}(p^\flat,q)\right)^*
	&\longrightarrow
	&
	&\frac{1}{m\sqrt{2}}
	\!\raisebox{-0.25\height}{\includegraphics[scale=0.355]{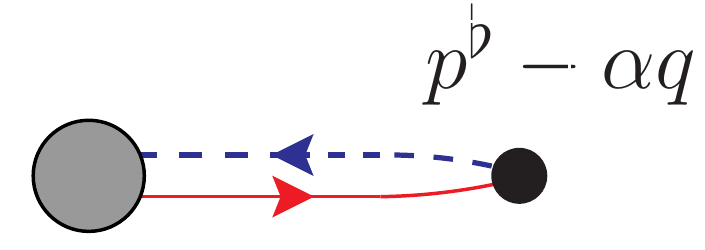}}
	&
	&\text{or}
	& 
	&\frac{1}{m\sqrt{2}}
	\!\raisebox{-0.25\height}{\includegraphics[scale=0.355]{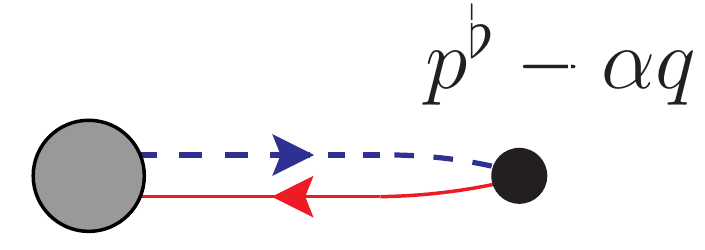}}
	\!\!\!\!,	
	\label{eq:pol vecs massive flow}
\end{align}
as well as the four-gluon vertex 
\begin{align}
	\raisebox{-0.425\height}{\includegraphics[scale=0.275]{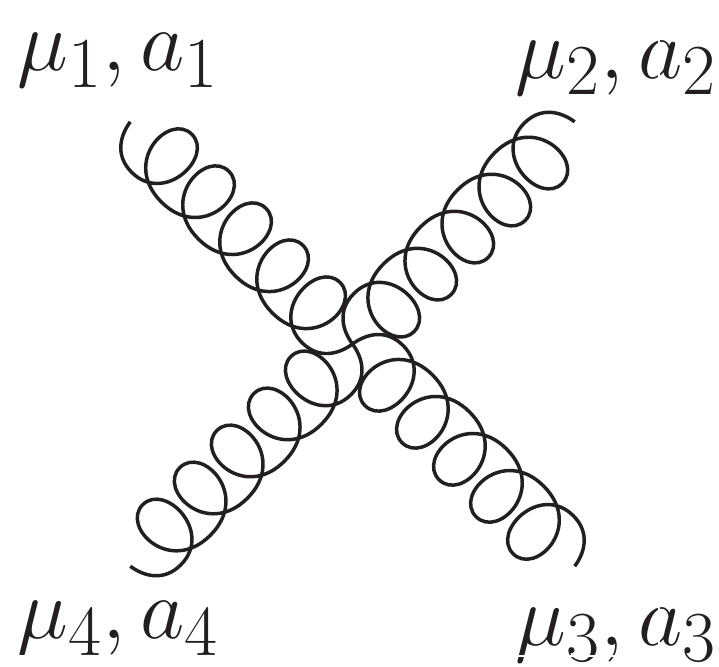}}
	&=\;\;
	i\left(\!\frac{g_s}{\sqrt{2}}\right)^{\!\!2}\!\!
	\;\sum\limits_{Z(2,3,4)}
	if^{a_1a_2b}if^{ba_3a_4}
	\;\Big(
	g^{\mu_1\mu_3}g^{\mu_2\mu_4}
	-g^{\mu_1\mu_4}g^{\mu_2\mu_3}
	\Big)\;,
	\nonumber\\
	&\rightarrow \;\;
	i\left(\!\frac{g_s}{\sqrt{2}}\right)^{\!\!2}\!\!
	\;\sum\limits_{Z(2,3,4)}
	if^{a_1a_2b}if^{ba_3a_4}
	\left(       \includegraphics[scale=0.225,valign=c]{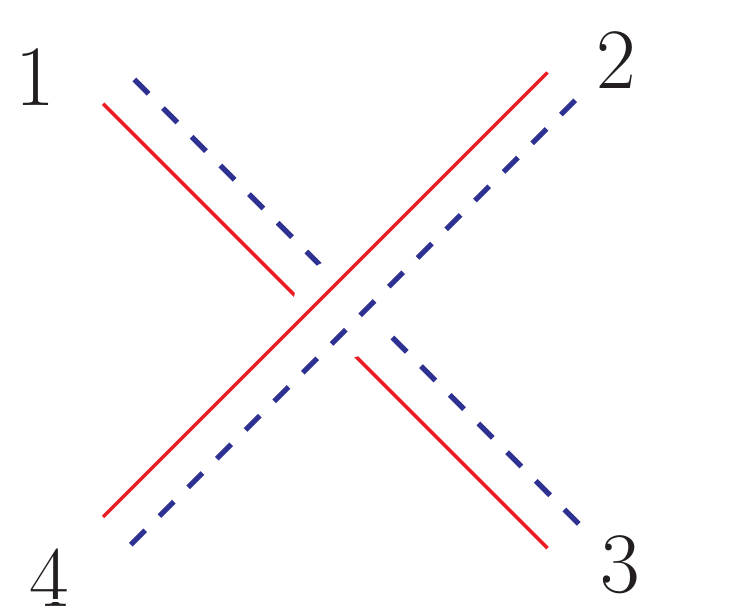}
	-
	\includegraphics[scale=0.225,valign=c]{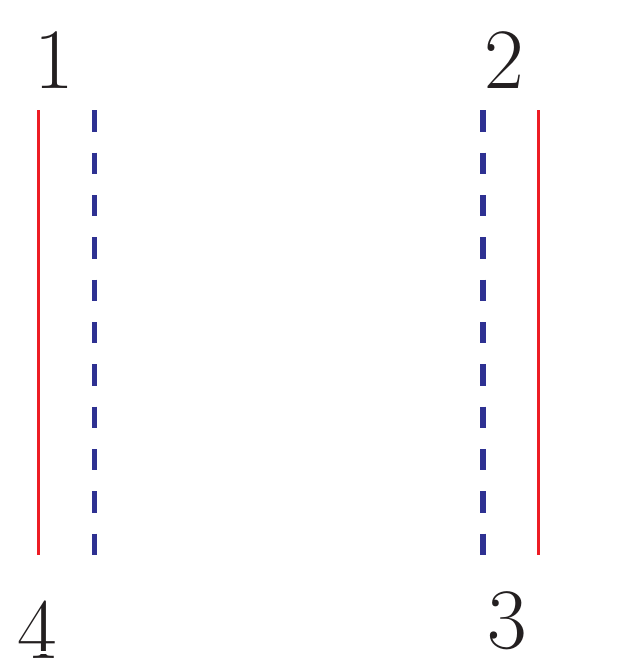}
	 \right)~,
	\label{eq:four_gluon_vertex}
\end{align}
where $Z(2,3,4)$ denotes the set of cyclic permutations of the integers $2,3,4$,
and in both cases the arrow directions which give a continuous flow in a given diagram are chosen.

\bibliographystyle{JHEP}  
\bibliography{loops} 

\end{document}